\documentclass[lettersize,journal]{IEEEtran}

\usepackage{amsmath,amsfonts, amsthm}
\usepackage{algpseudocodex}
\algnewcommand\algorithmicforeach{\textbf{for each}}
\algdef{S}[FOR]{ForEach}[1]{\algorithmicforeach\ #1\ \algorithmicdo}
\usepackage{algorithmicx}
\usepackage{amssymb}
\usepackage{algorithm}
\usepackage{array}
\usepackage{textcomp}
\usepackage{stfloats}
\usepackage{url}
\usepackage{verbatim}
\usepackage{graphicx}
\PassOptionsToPackage{hyphens}{url}
\usepackage{hyperref}
\usepackage{cite}
\usepackage{titlesec}
\titlespacing*{\subsubsection}{0pt}{*1}{*0}

\usepackage{booktabs}

\usepackage{graphicx}
\graphicspath{{images/}}
\DeclareGraphicsExtensions{.pdf,.jpg,.png}
\usepackage{wrapfig}
\usepackage[font=small,skip=2pt]{caption}
\usepackage{subcaption}
\usepackage{textcomp}
\usepackage{mathtools}

\newtheorem{theorem}{Theorem}

\newtheorem{corollary}{Corollary}[theorem]
\newcommand{\pctext}[2]{\text{\parbox{#1}{\centering #2}}}
\usepackage{multirow}

\newif\ifextended
\extendedtrue

\usepackage{color,soul}
\usepackage{adjustbox}
\usepackage{enumitem}
\usepackage{listings}
\usepackage{mwe}
\usepackage{xspace}
\usepackage{color,soul}

\lstdefinestyle{mystyle}{
    breaklines=true,
    basicstyle=\ttfamily \scriptsize,
    columns=fullflexible,
    frame=single,
    showstringspaces=false,
    captionpos=b,
    literate={time}{{time}}1 {date}{{date}}1 {table}{{table}}1 {action}{{action}}1 {at}{{at}}1
}

\usepackage{listings} %
\lstMakeShortInline[columns=fixed]|

\newcommand{\oursystem}{Sieve\xspace}

\newcommand\etal{\textit{et al.\ }}

\newcommand{\vDatabase}{\ensuremath{\mathcal{D}}}
\newcommand{\vRelationSet}[1]{\ensuremath{\mathcal{R}}_{#1}}
\newcommand{\vRelation}[2]{\textit{r}^{#1}_{#2}}
\newcommand{\vTupleSet}[1]{\ensuremath{\mathcal{T}}_{#1}}
\newcommand{\vTuple}[2]{\textit{t}^{#1}_{#2}}
\newcommand{\vQuery}[2]{\textit{Q}^{#1}_{#2}}

\newcommand{\vIndexSet}[1]{\ensuremath{\mathcal{I}}_{#1}}
\newcommand{\vIndex}[1]{\textit{i}_{#1}}

\newcommand{\vUserSet}[1]{\ensuremath{\mathcal{U}}_{#1}}
\newcommand{\vUser}[2]{\textit{u}^{#1}_{#2}}

\newcommand{\vGroupMethod}[1]{\textit{group(}{#1}\textit{)}}
\newcommand{\vProfileMethod}[1]{\textit{profile(}{#1}\textit{)}}

\newcommand{\vPolicySet}[1]{\ensuremath{\mathcal{P}}_{#1}}
\newcommand{\vPolicy}[2]{\textit{p}^{#1}_{#2}}

\newcommand{\vPolicyTuple}[3]{$\langle$#1, #2, #3$\rangle$}

\newcommand{\vTupleExpression}[3]{$\langle$#1, #2, #3$\rangle$}

\newcommand{\vTupleExpressionRange}[5]{$\langle$#1, #2, #3, #4, #5$\rangle$}

\newcommand{\vPolicyExpression}[1]{\textit{\ensuremath{\mathcal{E}}(}{#1}\textit{)}}

\newcommand{\vPolicyGuardedExpression}[1]{\textit{\ensuremath{\mathcal{G}}(}{#1}\textit{)}}

\newcommand{\vCostMethod}[1]{\textit{cost(}{#1}\textit{)}}

\newcommand{\vGuard}[2]{\textit{G}^{#1}_{#2}}

\newcommand{\vCandidateGuardSet}[2]{\ensuremath{\mathcal{CG}}^{#1}_{#2}}

\newcommand{\vSelectivityMethod}[1]{\rho({#1})}
\newcommand{\vReadCost}{c_r}
\newcommand{\vEvalCost}{c_e}
\newcommand{\vShortCircuit}{\alpha}
\newcommand{\vSetCardinality}[1]{\lvert {#1} \rvert}

\newcommand{\vMergeMethod}[2]{\theta({#1},{#2})}

\newcommand{\vBenefitMethod}[1]{\textit{benefit(}{#1}\textit{)}}
\newcommand{\vReadCostMethod}[1]{\textit{read\_cost(}{#1}\textit{)}}
\newcommand{\vUtilityMethod}[1]{\textit{utility(}{#1}\textit{)}}

\newcommand{\vObjectConditions}[2]{{\tt OC}^{#1}_{#2}}
\newcommand{\vObjectCondition}[2]{{\tt oc}^{#1}_{#2}}
\newcommand{\vQuerierConditions}[2]{{\tt QC}^{#1}_{#2}}
\newcommand{\vQuerierConditionsAnd}[2]{{\tt qc}^{#1}_{#2}}
\newcommand{\vQuerierCondition}[2]{{\tt qc}^{#1}.{#2}}

\newcommand{\vPolicyAction}[2]{{\tt AC}^{#1}_{#2}}

\newcommand{\vQueryMetadatas}[1]{{\tt QM(\vQuery{}{#1})}}
\newcommand{\vQueryMetadata}[2]{{\tt QM(\vQuery{}{#1})}{.#2}}

\newcommand{\vBaselineP}{$Baseline_{P}$}
\newcommand{\vBaselineI}{$Baseline_{I}$}
\newcommand{\vBaselineU}{$Baseline_{U}$}

\newcommand{\GEs}{Guarded Expressions\xspace}

\newcommand{\squishlist}{
	\begin{list}{$\bullet$}
		{
			\setlength{\itemsep}{0pt}
			\setlength{\parsep}{3pt}
			\setlength{\topsep}{3pt}
			\setlength{\partopsep}{0pt}
			\setlength{\leftmargin}{1.5em}
			\setlength{\labelwidth}{1em}
			\setlength{\labelsep}{0.5em} } }

\newcommand{\squishend}{
	\end{list}  }

\begin{document}

\title{Scalable Enforcement of Fine Grained Access Control Policies in Relational Database Management Systems}

\author{Anadi Shakya, 
Primal Pappachan, 
David Maier,
Roberto Yus,
Sharad Mehrotra,
and Johann-Christoph Freytag
\thanks{Anadi Shakya, Primal Pappachan, and David Maier are with the Department of Computer Science, Portland State University, Portland, Oregon 97201, United States of America (email: ashakya@pdx.edu; primal@pdx.edu; maier@pdx.edu).}
\thanks{Roberto Yus is with the Department of Computer Science and Electrical Engineering, University of Maryland, Baltimore County, Baltimore, Maryland 21250, United States of America (email: ryus@umbc.edu).}
\thanks{Sharad Mehrotra is with the Department of Computer Science, University of California, Irvine, Irvine, California 92697, United States of America (email: sharad@ics.uci.edu).}
\thanks{Johann-Christoph Freytag with the Institut für Informatik at Humboldt-Universität zu Berlin 10099, Germany.  (email: freytag@informatik.hu-berlin.de ).}
}

\markboth{Journal of \protect\LaTeX\ Class Files,~Vol.~14, No.~8, August~2021}%
{Shell \MakeLowercase{\textit{et al.}}: A Sample Article Using IEEEtran.cls for IEEE Journals}

\maketitle

\begin{abstract}
The proliferation of smart technologies and evolving privacy regulations, such as GDPR and CPRA, has necessitated the management of Fine-Grained Access Control (FGAC) policies in Database Management Systems (DBMS). Existing approaches for enforcing FGAC policies do not scale to thousands of policies, leading to degraded query performance and reduced system effectiveness. In this paper, we present \oursystem, a middleware solution for relational DBMSs that combines advanced query rewriting techniques with caching mechanisms to optimize FGAC policy enforcement. \oursystem rewrites a query with \textit{guarded expressions}, which group and filter policies, and can efficiently utilize underlying indices in the DBMS. Additionally, \oursystem integrates a caching mechanism using an effective replacement strategy and a refresh mechanism to adapt \oursystem to dynamic workloads. 
Our experiments, conducted on two DBMSs using real and synthetic datasets, demonstrate \oursystem's ability to scale to large datasets and policy corpora, maintaining low query latency and system load and boosting policy evaluation performance by up to 2--10$\times$ on 200--1,200 policies. The caching extension further improves query performance by 6-22\% under dynamic workloads, particularly with larger cache sizes. These results highlight \oursystem's applicability for real-time access control in smart environments, supporting the efficient and scalable management of user preferences and privacy policies.
\end{abstract}

\begin{IEEEkeywords}
IoT environments, Privacy, Access control, Query optimization, Caching, Middleware
\end{IEEEkeywords}

\IEEEpubidadjcol

\section{Introduction}
\label{sect:intro}

\IEEEPARstart{S}{mart} devices today capture and store large volumes of personal data for many purposes, such as providing personalized services, health monitoring, building management, and ad placements. Continuous data capture through sensors embedded in physical spaces has significant privacy implications~\cite{FARAHANI2018659, Acar2020, Kroger2022}. Regulations, such as the European General Data Protection Regulation (GDPR)~\cite{gdpr}, the California Privacy Rights Act (CPRA)~\cite{CPRA}, and the Oregon Consumer Privacy Act (OCPA)~\cite{OCPA}, impose strict requirements on how organizations manage user data. These requirements include purpose limitation, data minimization, and restricted retention of personal data. 

Access control policies have emerged as a key enabler for implementing compliance with various privacy regulations. For example, in Database Management Systems (DBMSs), access control supports provisions like \textit{Article 25} of the GDPR, which mandates \textit{Data Protection by Design and by Default}~\cite{Shastri2019, michelakaki2023unlocking, DBLP:journals/pvldb/ShastriBWKC20}. As smart spaces and Internet of Things (IoT) ecosystems become ubiquitous, where sensors continuously monitor individuals,  organizations have to create and enforce many fine-grained access control policies. Supporting such fine-grained policies is essential for user control but presents significant challenges for DBMSs. Beyond regulatory compliance, FGAC can also promote data sharing and reuse. For example, users may be more willing to share personal data (such as for medical research) if they can specify exactly who can access it, when, and for what purpose---instead of facing coarse, \textit{all-or-nothing} sharing options.

The increasing demand for expressive and scalable access control brings into focus the broader challenge of data governance. Data Governance, which involves specifying and enforcing data usage policies, has been identified as one of the major challenges of modern data management, as highlighted in the 2022 Seattle Database Report~\cite{SeattleDBReport2022}. A key requirement in data governance is efficiently enforcing access-control policies during query execution, especially as the number of policies grows. This scalability issue is particularly critical in smart environments. A motivating example, detailed in Section~\ref{sect:caseStudy}, illustrates how processing even simple analytical queries may require checking hundreds to thousands of access-control policies. Current DBMSs struggle with real-time enforcement, making this problem increasingly urgent. Although our focus is motivated by smart spaces and IoT, this need extends to other domains governed by data-protection regulations.

Today, DBMSs enforce Fine-Grained Access Control (FGAC) using one of two mechanisms~\cite{bertino2011access}: (1)~\textit{Policy as schema}, where access control is managed through authorization views~\cite{rizvi2004extending}; and (2)~\textit{Policy as data}, where policies are stored in tables and appended as predicates to query execution~\cite{agrawal2002hippocratic, byun2005purpose}. Both approaches rely on query rewriting~\cite{stonebraker1974access}, but their performance significantly degrades as the number of policies increases. In realistic deployments, such as IoT-enabled smart spaces or campus environments, users often define FGAC policies based on time, location, and purpose. As described in our example in Section~\ref{sect:caseStudy}, this can quickly result in hundreds or thousands of active policies per user. Under the ``policy as data" approach, each policy is translated into multiple query predicates. For example, enforcing 1,000 fine-grained access control policies can add up to 3,000 predicates in the \texttt{WHERE} clause. Today's DBMSs struggle to evaluate these complex expressions in Disjunctive Normal Form (DNF)~\cite{kim} and, therefore, resort to full table scan and real-time query execution, becoming prohibitively expensive.

To address these challenges, we introduce \oursystem\footnote{We call our method Sieve because it filters out irrelevant tuples and policies early in the process, allowing only the necessary data and access conditions to pass through for evaluation. This selective filtering improves efficiency and reduces overhead.}\cite{SIEVEPappachanYMF20}, a middleware solution that improves the enforcement of FGAC in DBMSs by optimizing query execution with \textit{guarded expressions} (GEs). These GEs represent a rewritten form of queries, effectively combining FGAC policies with query hints and User Defined Functions (UDFs) to support efficient, on-demand query evaluation over large policy sets.
The middleware implementation ensures compatibility with diverse relational DBMSs, making it adaptable to various IoT use cases~\cite{iotbenchmark}. This adaptability enables the exploration of novel ideas without being constrained by the limitations of prior work focusing on specific systems~\cite{DBLP:journals/dase/ColomboF16}. 

\oursystem employs two key strategies to reduce policy enforcement overhead: (1) Policies are pre-processed into GEs, which minimize the number of tuples checked during query execution. This approach is inspired by the predicate simplification techniques of Chaudhuri \etal~\cite{chaudhuri2003factorizing}; (2) Only relevant policies are dynamically selected based on query context, such as the querier metadata (e.g., purpose). This relevance filtering is achieved through a policy-check operator implemented as a UDF, ensuring efficiency by avoiding unnecessary policy evaluations. These strategies are combined into a unified framework that adaptively selects the optimal enforcement mechanism based on query and policy characteristics. Our real IoT dataset evaluation demonstrates significant performance gains over traditional query-rewriting approaches.


While \oursystem is highly effective in static scenarios, where both access-control policies and queries remain largely the same over time, dynamic environments introduce additional overhead. For example, in a smart campus setting, a professor may query the location of students in their class for various purposes (e.g., attendance, office hours) at different times during the year (e.g., the beginning vs. the end of the semester).
In such scenarios, where a querier repeatedly issues the same or similar queries, the system incurs unnecessary overhead from retrieving policies and regenerating the same guarded expressions, especially when the underlying policies have not changed.
As shown in Figure~\ref{fig:guardgen.png}, the cost of GE regeneration can exceed 200ms for large policy sets. While this may seem acceptable in isolation, it becomes a performance bottleneck when accumulated across large and frequent workloads. 




We extend \oursystem with a caching mechanism that stores previously generated GEs instead of generating them from scratch. 
Unlike conventional caching approaches that store query results, \oursystem caches GEs, which serve as intermediate query rewrites to enforce fine-grained access control. This approach allows \oursystem to reuse a cached GE across different queries from the same querier. In static workloads, where the policy set remains unchanged, \oursystem achieves high reuse of cached GEs. For instance, different queries issued by the same querier for the same purpose can reuse the same GE if governed by the same set of policies.

\oursystem, extended with caching, is particularly useful in dynamic scenarios where the same querier issues multiple queries over time. For example, two different queries issued by the same querier for the same purpose may still use the same GE if governed by the same set of policies. By identifying such common access-control contexts, \oursystem supports GE reuse beyond exact query matches, improving efficiency in both static and dynamic environments. Even when policies change, caching can still reduce overhead by reusing GEs associated with the same querier and access-control metadata. To support workloads with frequent policy insertions, deletions, or updates, we incorporate two strategies. The replacement strategy applies the clock algorithm~\cite{yang2023fifo} to evict GEs with low reuse potential. The refresh strategy determines whether to incrementally update a cached GE or regenerate it based on changes in the applicable policy set. These techniques allow \oursystem to maintain correct policy enforcement while avoiding unnecessary recomputation.

Experimental results, based on a workload generator that simulates static and dynamic IoT scenarios, demonstrate significant scalability and performance improvements, highlighting the utility of \oursystem under various conditions. The workload generator models dynamic environments with varying rates of policy insertion, query execution, and policy deletion to capture fluctuating access control activity.

\noindent The primary contributions of this work are as follows:
\begin{enumerate}
    \item We introduce \oursystem, a middleware solution that enforces FGAC policies efficiently. Our approach generates and reuses guarded expressions to minimize policy enforcement overhead during query execution to outperform traditional query rewriting methods. 
    \item We extend \oursystem with a caching mechanism designed for dynamic workloads, incorporating a clock-based replacement policy and a refresh strategy to maintain correctness while minimizing regeneration costs.
    \item We conduct experiments to evaluate the performance of \oursystem under static workloads, demonstrating its effectiveness in minimizing query execution costs. We further evaluate \oursystem with caching under dynamic conditions, showing significant performance improvements in query response times and system scalability. 
\end{enumerate}

\noindent\textbf{Outline of the paper.} 
Section~\ref{sect:problemSetting} presents a case study of a real IoT deployment, where a large set of access control policies is expected to be defined. 
Section~\ref{sect:sieveApproach} formalizes the query, policy model, and access control semantics of \oursystem. 
Section~\ref{sect:guardSel} presents an algorithmic solution to generate appropriate guarded expressions, the core mechanism of \oursystem. 
Section~\ref{sect:caching} introduces caching mechanisms for dynamic workload scenarios, including cache replacement and refresh strategies.
Section~\ref{sect:implementingSieve} describes the details of \oursystem generated query rewrites and various optimization techniques used.
Section~\ref{sect:expSieve} presents the experimental evaluation of \oursystem in static settings, demonstrating improvements in query evaluation efficiency across two datasets and two DBMSs. This is followed by an evaluation of \oursystem with caching, designed to handle diverse workloads of FGAC policies and queries in dynamic environments.
Section~\ref{sect:relWork} reviews relevant related works.
Finally, Section~\ref{sect:conclusion} presents conclusions and future work.

\noindent\textbf{Comparison to Conference version.} 
This version extends our prior conference publication \cite{SIEVEPappachanYMF20} by introducing the following new contributions: (1) A novel caching algorithm for improving query planning and policy retrieval time, which involves a replacement and a refresh strategy; (2) A workload generator that simulates diverse policy and query patterns to evaluate system performance in dynamic IoT scenarios; (3) Additional experiments assessing the performance and utility of caching across different settings, including whether \oursystem with caching improves query processing over \oursystem without it, the effectiveness of the cache, the impact of the replacement strategy on guard generation and execution time, the optimal cache size for maximizing performance, and how workload parameters such as policy deletions and querier popularity affect cache behavior; and (4) An expanded related work section with more details on existing approaches for caching on how they applies to our problem setting.

\section{Problem Setting}
\label{sect:problemSetting}

We present a smart campus setting where users specify a large number of FGAC policies to manage access to their collected data. Using this context, we review the challenges of managing and enforcing these policies in static and dynamic scenarios, highlighting the need for scalable solutions such as \oursystem.

\subsection{Smart Campus Case Study} 
\label{sect:caseStudy}

\begin{figure}[!htb]
    \centering
    \includegraphics[width=0.4\textwidth]{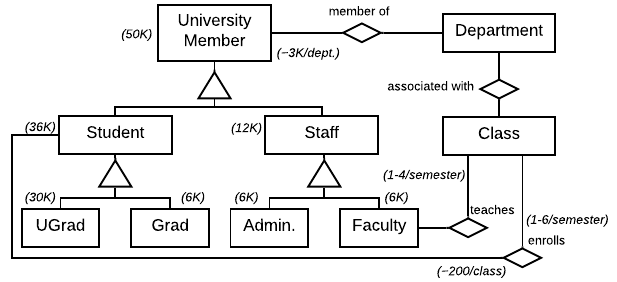}
    \caption{Users and their relationships in a Smart Campus Scenario.}
    \label{fig:entityLattice}
\end{figure}

We consider a motivating application wherein an academic campus supports variety of smart data services such as real-time queue size monitoring in different food courts, occupancy analysis to understand building usage (e.g., room occupancy as a function of time and events, determining how space organization impacts interactions amongst occupants, etc.), or automating class attendance and understanding correlations between attendance and grades~\cite{joon}.
While such solutions present interesting benefits, such as improving student performance~\cite{joon} and better space utilization, there are privacy challenges~\cite{pappachan2017towards} in the management of such data. This case study is based on our experience building a smart campus with various applications ranging from real-time services to offline analysis over the past decade. The deployed system, entitled TIPPERS~\cite{mehrotra2016tippers}, is in daily use in several buildings in our UC Irvine campus\footnote{More details about the system and the applications it supports can be found in the work by Roberto et al.~\cite{DBLP:journals/toit/YusBMV22}}. TIPPERS at our campus captures connectivity events (i.e., logs of the connection of devices to WiFi APs) that can be used, among other purposes, to analyze the location of individuals to provide them with services.


We use the UC Irvine campus, with the various users presented in Figure~\ref{fig:entityLattice} (along with the expected number of members in brackets), as a use case. Consider a professor on campus posing the following analytical query to evaluate the correlation between regular attendance in her class and student performance at the end of the semester. Specifically, the professor retrieves the grades of students enrolled in CS101 along with the number of distinct days they were present at the WiFi access point `1200' between 9--10 AM from September 25 to December 12, 2024, using WiFi logs as a proxy for attendance: 


\begin{lstlisting}[style=mystyle, caption={Query for analyzing attendance and performance}, label={lst:query}]
StudentPerf(WifiDataset, Enrollment, Grades)=
(SELECT G.student, G.grade, 
        count(distinct W.ts_date) AS attended
 FROM WiFiDataset W, Enrollment E, Grades G
 WHERE E.class = "CS101" AND
        E.student = W.owner AND
        G.student = W.owner AND
        W.ts_time between "9am" AND "10am" AND
        W.ts_date between "9/25/24" AND "12/12/24" AND
        W.wifiAP = "1200"
 GROUP BY G.student, G.grade)
\end{lstlisting}


Let us assume that within the students in the professor's class, there exist different privacy profiles (as studied in the mobile world by Lin et al.~\cite{lin2014privacy}). Adapting the distribution of users by profile to our domain, we can assume that~20\% of the students might have a common default policy (``unconcerned" group),~18\% may want to define their precise policies (``advanced users"), and the rest will depend on the situation (for which we consider, conservatively,~2/3~to be ``unconcerned" and~1/3~``advanced")\footnote{Examples of advanced access control policies are included in Section~\ref{sect:modeling}}. Simplifying this and applying it to a class of 200 students, we have 120 unconcerned users who will adopt the default policy and 80 advanced users who will define their own policies. With the conservative assumption that there are two default policies per default user and at least 4 specific policies per advanced user, we have 560 policies. Typically, advanced users define more policies than this conservative assumption, so we would add two additional policies per group, increasing the number of policies to 880 or 1.2K (with three additional policies per group). For example, a default policy may allow location sharing during class hours, while an advanced policy may restrict access to specific days, times, and locations for a particular professor.

Given the above policies for a single class, if students take 1-6 classes and faculty teach 1-4 classes per semester, a query to analyze students attendance listed above with performance over classes a professor taught over the year would be 3.3K (560 policies/class * 2 classes/quarter * 3 quarters/year) to 7.2K (considering our 1.2K policies/class estimation). We only focused on a single data type captured in this analysis (i.e., connectivity data) with two conditions per policy (e.g., time and location), and policies defined by a given user at the group-level (and not at the individual-level, which will even further increase the number of policies).

In addition to the many FGAC policies that need to be evaluated at query time, dynamic IoT scenarios introduce additional challenges. First, individuals may occasionally insert, update, or delete new policies in the DBMS as they change their privacy preferences.
For instance, students may define or update their class sharing preferences at the start of a semester. 
Second, the same querier may repeatedly pose the exact or similar queries to the DBMS. For example, professors may issue multiple queries to retrieve attendance records to determine participation points after each class activity, or when project submissions are due, students may frequently query the locations of their project teammates to coordinate meetings. 
If the same query or a similar one reappears and the relevant policies have not changed, the system can reuse the earlier guarded expression rather than regenerating it from scratch. This reuse avoids redundant computation. Typically, policy updates are more frequent at the start of the semester, whereas query frequency peaks later. We recognize this pattern and introduce caching mechanisms to improve efficiency in such dynamic workloads.

\section{\oursystem approach to FGAC}
\label{sect:sieveApproach}

We describe the three fundamental policy-driven data processing entities: data, query, and policies. For policies, we delve deeper and explain what each attribute represents. Then, using these three components, we describe the access control semantics used in this paper. We finish the section with a sketch of the approach followed by \oursystem to speed up policy enforcement. We have summarize frequently used notations in Table~\ref{table:notation_table}.

\subsection{Modeling Policy-Driven Data Processing}
\label{sect:modeling}

\begin{table*}[t]
\scriptsize
\centering
\begin{tabular}{m{3cm}m{12cm}}
  \textbf{Notation} & \textbf{Definition} \\ 
 \hline \\[-1em]
  $\vDatabase$  & Database \\ 
 \hline \\[-1em]
 $\vIndex{i} \in \vIndexSet{}$ & Index and set of indexes in $\vDatabase$ \\ 
 \hline \\[-1em]
 $\vRelation{}{i} \in \vRelationSet{}$ & Relation and set of relations in $\vDatabase$ \\ 
 \hline \\[-1em]
 $\vUser{}{k} \in \vUserSet{}$  & User and set of users in $\vDatabase$ \\ 
 \hline \\[-1em]
 $\vTuple{}{j} \in \vTupleSet{}$; $\vTupleSet{\vRelation{}{i}}$; $\vTupleSet{\vUser{}{k}}$;  $\vTupleSet{\vQuery{}{i}}$; $\vTupleSet{\vPolicy{}{l}}$ & Tuple and set of tuples: in $\vDatabase$; in $\vRelation{}{i}$; owned by $\vUser{}{k}$; required to compute $\vQuery{}{i}$; controlled by $\vPolicy{}{l}$ \\
 \hline \\[-1em]
 $\vGroupMethod{\vUser{}{k}}$  & Groups $\vUser{}{k}$ is part of \\ 
 \hline \\[-1em]
 $\vQuery{}{i}$; $\vQueryMetadatas{i}$ & Query; Metadata of $\vQuery{}{i}$ \\ 
 \hline \\[-1em]
 $\vPolicy{}{l} \in \vPolicySet{}$; $\vPolicySet{\vQueryMetadatas{i}}$ & Access control policy and set of policies in $\vDatabase$; set of policies related to a query given its metadata \\
 \hline \\[-1em]
 $\vObjectCondition{l}{i} \in \vObjectConditions{l}{}$;$\vQuerierConditionsAnd{l}{i} \in \vQuerierConditions{l}{}$;$\vPolicyAction{l}{}$ & Object conditions; querier conditions; action of $\vPolicy{}{l}$ \\ 
 \hline \\[-1em]
  $\vPolicyExpression{\vPolicySet{}} = \vObjectConditions{1}{} \lor \cdots \lor \vObjectConditions{\vSetCardinality{\vPolicySet{}}}{}$ & {\em Policy expression} of $\vPolicySet{}$ \\ 
 \hline \\[-1em]
  $\vPolicyGuardedExpression{\vPolicySet{}}=\vGuard{}{1} \lor \cdots \lor \vGuard{}{n}$ & {\em Guarded policy expression} of $\vPolicySet{}$ (DNF of guarded expressions) \\ 
 \hline \\[-1em]
  $\vGuard{}{i} = \vObjectCondition{i}{g} \land \vPolicySet{\vGuard{}{i}}$ & {\em Guarded expression} which consists of conjunctive expression of {\em guard} ($\vObjectCondition{i}{g}$) and a set of policies for which ($\vObjectCondition{i}{g}$) is a common factor. We refer to these policies as a {\em policy partition} ($\vPolicySet{\vGuard{}{i}}$) \\
 \hline \\[-1em]
  $\vCandidateGuardSet{}{}$ & {\em Candidate guards} for $\vPolicyExpression{\vPolicySet{}}$ \\
 \hline \\[-1em]
  $eval(exp, \vTuple{}{t})$ & function which evaluates a tuple $\vTuple{}{t}$ against a expression $exp$ \\
 \hline \\[-1em]
  $\Delta(\vPolicySet{\vGuard{}{i}}, \vQueryMetadata{i}{}, \vTuple{}{t})$ & policy operator \\
 \hline \\[-1em]
 $\vSelectivityMethod{pred}$  & estimated cardinality of a predicate \\
 \hline \\[-1em]
 $\vEvalCost$; $\vReadCost$  & cost of evaluating a tuple against the set of object conditions of a policy;  cost of reading a tuple from the disk  \\
 \hline \\[-1em]
 $\vShortCircuit$  & average number of policies that a tuple is checked against before it satisfies one  \\
 \end{tabular}
\caption{Frequently used notations.}
\label{table:notation_table}
\vspace{-0.5cm}
\end{table*}

\vspace{0.1cm}
\noindent
\textbf{Data Model.} Let us consider a database $\vDatabase$ consisting of a set of relations $\vRelationSet{}$, a set of data tuples $\vTupleSet{}$, a set of indexes $\vIndexSet{}$, and set of users $\vUserSet{}$. $\vTupleSet{\vRelation{}{i}}$ represents the set of tuples in the relation $\vRelation{}{i} \in \vRelationSet{}$. Users are organized in collections or {\em groups}, which are hierarchical (i.e., a group can be subsumed by another). For example, the group of undergraduate students is subsumed by the group of students. Each user can belong to multiple groups and we define the method $\vGroupMethod{\vUser{}{k}}$ which returns the set of groups $\vUser{}{k}$ is member of. Each data tuple $\vTuple{}{j} \in \vTupleSet{}$ belongs to a $\vUser{}{k} \in \vUserSet{}$ or a group whose access control policies restrict/grant access over that tuple to other users. We assume that for each data tuple $\vTuple{}{j} \in \vTupleSet{}$ there exists an owner $\vUser{}{k} \in \vUserSet{}$ who owns it, whose access control policies restrict/grant access over that tuple to other users (the ownership can be also shared by users within a group). This ownership is explicitly stated in the tuple by using the attribute $\vRelation{}{i}.owner$ that exists for all $\vRelation{}{i} \in \vRelationSet{}$ and that we assume is indexed (i.e., $\forall \; \vRelation{}{i} \in \vRelationSet{} \; \exists \; \vIndex{j} \in \vIndexSet{} \mid \vIndex{j}$ is an index over the attribute $\vRelation{}{i}.owner$). $\vTupleSet{\vUser{}{k}}$ represents the set of tuples owned by user $\vUser{}{k}$. 

\vspace{0.1cm}
\noindent
\textbf{Query Model.} 
The SELECT-FROM-WHERE query posed by a user $\vUser{}{k}$ is denoted by $\vQuery{}{i}$ and tuples in the relations in the FROM statement(s)  of query are denoted by $\vTupleSet{\vQuery{}{i}} = \bigcup\limits_{i=1}^{n} \vTupleSet{\vRelation{}{i}}$. In our model, we consider that queries have associated metadata $\vQueryMetadatas{i}$ which consists of information about the querier and the context of the query. This way, we assume that for any given query $\vQuery{}{i}$, $\vQueryMetadatas{i}{}$ contains the identity of the querier (i.e., $\vQueryMetadata{i}{querier}$) as well as the purpose of the query (i.e., $\vQueryMetadata{i}{purpose}$). In the example query in Section~\ref{sect:caseStudy}, $\vQueryMetadata{i}{querier}$=``Prof.Smith" and $\vQueryMetadata{i}{purpose}$=``Marking Attendance".

\vspace{0.1cm}
\noindent
\textbf{Access Control Policy Model.} A user specifies an access control policy (in the rest of the paper we will refer to it simply as policy) to allow or to restrict access to certain data she owns, to certain users/groups under certain conditions. Let $\vPolicySet{}$ be the set of policies defined over $\vDatabase$ such that $\vPolicy{}{l} \in \vPolicySet{}$ is defined by a user $\vUser{}{k}$ to control access to a set of data tuples in $\vRelation{}{i}$. Let that set of tuples be $\vTupleSet{\vPolicy{}{l}}$ such that $\vTupleSet{\vPolicy{}{l}} \subseteq \vTupleSet{\vUser{}{k}} \cap \; \vTupleSet{\vRelation{}{i}}$. We model such policy as $\vPolicy{}{l}=$\vPolicyTuple{$\vObjectConditions{l}{}$}{$\vQuerierConditions{l}{}$}{$\vPolicyAction{l}{}$}, where each element represents:

\noindent
$\bullet$
Object Conditions ($\vObjectConditions{l}{}$) are defined using a conjunctive boolean expression $\vObjectCondition{l}{1} \land \vObjectCondition{l}{2} \land ... \land \vObjectCondition{l}{n}$ which determines the access controlled data tuple(s).
Each {\em object condition} ($\vObjectCondition{l}{c}$) is a boolean expression \vTupleExpression{$attr$}{$op$}{$val$} where $attr$ is an attribute (or column) of $\vRelation{}{i}$, $op$ is a comparison operator (i.e., $=$, $!=$, $<$, $>$, $\geq$, $\leq$, |IN|, |NOT IN|, |ANY|, |ALL|), and $val$ can be either: (1) A constant or a range of constants or (2) A derived value(s) defined in terms of the expensive operator (e.g., a user defined function to perform face recognition) or query on $\vDatabase$ that will obtain such values when evaluated. To represent boolean expressions involving a range defined by two comparison operators (e.g., $4\leq a <20$) we use the notation \vTupleExpressionRange{$attr$}{$op1$}{$val1$}{$op2$}{$val2$} (e.g., \vTupleExpressionRange{$a$}{$\geq$}{$4$}{$<$}{$20$}). 
We assume that there exists exactly one $\vObjectCondition{l}{c} \in \vObjectConditions{l}{}$ such that $\vObjectCondition{l}{c} =$ \vTupleExpression{$\vRelation{}{i}.owner$}{=}{$\vUser{}{k}$} or $\vObjectCondition{l}{c} =$ \vTupleExpression{$\vRelation{}{i}.owner$}{=}{$\vGroupMethod{\vUser{}{k}}$}. We will refer to this object condition as $\vObjectCondition{l}{owner}$ in the rest of the paper. 

\noindent
$\bullet$
Querier Conditions ($\vQuerierConditions{l}{}$) identifies the metadata attributes of the query to which the access control policy applies. $\vQuerierConditions{l}{}$ is a conjunctive boolean expression $\vQuerierConditionsAnd{l}{1} \land \vQuerierConditionsAnd{l}{2} \land \cdots \land \vQuerierConditionsAnd{l}{m}$. 
Our model follows the well-studied Purpose-Based Access Control (Pur-BAC) model~\cite{byun2005purpose} to define the querier conditions.
Thus, we assume that each policy contains has at least two querier conditions such as $\vQuerierCondition{l}{querier}$ =  
\vTupleExpression{$\vQueryMetadata{i}{querier}$}{=}{$\vUser{}{k}$}
or $\vQuerierCondition{l}{querier} =$ \vTupleExpression{$\vQueryMetadata{i}{querier}$}{=}{$\vGroupMethod{\vUser{}{k}}$} (that defines either a user or group), and a $\vQuerierCondition{l}{purpose}$ = \vTupleExpression{$\vQueryMetadata{i}{purpose}$}{=}{\textit{purpose}} which models the intent/purpose of the querier (e.g., safety, commercial, social, convenience, specific applications on the scenario, or any~\cite{DBLP:conf/percom/0001K17}). 
Other pieces of querier context (such as the IP of the machine from where the querier posed the query, or the time of the day) can easily be added as querier conditions although in the rest of the paper we focus on the above mentioned querier conditions.

\noindent
$\bullet$
Policy Action ($\vPolicyAction{l}{}$) defines the enforcement operation, or {\em action}, which must be applied on any tuple $\vTuple{}{j} \in \vTupleSet{\vPolicy{}{l}}$.
We consider the default action, in the absence of an explicit-policy allowing access to data, to be \textit{deny}. Such a model is standard in systems that collect/manage user data. 
Hence, explicit access control actions associated with policies in our context are limited to \textit{allow}.  
If a user expresses a policy with a deny action (e.g., to limit the scope/coverage of an allow policy), we can factor in such a deny policy into the explicitly listed allow policies. For instance, given an explicit allow policy ``allow John access to my location'' and an overlapping deny policy ``deny everyone access to my location when in my office'', we can factor in the deny policy by replacing the original allow policy by ``allow John access to my location when I am in locations other than my office''. We therefore restrict our discussions to allow policies. 

Based on this policy model, we show two sample policies in the context of the motivating scenario explained before. First, we describe a policy with object conditions containing a constant value. This policy is defined by John to regulate access to his connectivity data to Prof. Smith only if he is located in the classroom and for the purpose of class attendance as follows: |$\langle$[W.owner = John $\land$ $\texttt{W.ts-time}$ $\geq$ 09:00 $\land$ $\texttt{W.ts-time}$ $\leq$ 10:00 $\land$ W.wifiAP = 1200], [Prof. Smith $\land$ Marking Attendance], allow$\rangle$|. Second, we describe the same policy with an object condition derived from a query to express that John wants to allow access to his location data only when he is with Prof. Smith. The object condition is updated as: |$\langle$[W.owner = John $\land$ W.wifiAP = (SELECT W2.wifiAP FROM WifiDataset AS W2 WHERE $\texttt{W2.ts-time}$ = $\texttt{W.ts-time}$ AND W2.owner = "Prof.Smith")],[Prof. Smith $\land$ Marking Attendance],allow$\rangle$|

\vspace{0.1cm}
\noindent
\textbf{Access Control Semantics.} 

Let $\vPolicySet{}$ be the set of policies defined on relation $\vRelation{}{k}$ that control access to $\vQuery{}{i}$. $\vPolicyExpression{\vPolicySet{}}$ is the Disjunctive Normal Form (DNF) expression of $\vPolicySet{}$ such that $\vPolicyExpression{\vPolicySet{}} = \vObjectConditions{1}{} \lor \cdots \lor \vObjectConditions{\vSetCardinality{\vPolicySet{}}}{}$ where $\vObjectConditions{l}{}$ is conjunctive expression of object conditions from $\vPolicy{}{l} \in \vPolicySet{}$. After appending $\vPolicyExpression{\vPolicySet{}}$ to $\vQuery{}{i}$ we obtain: |SELECT * FROM $\vRelation{}{j}$ MINUS SELECT * FROM $\vRelation{}{k}$ WHERE $\vPolicyExpression{\vPolicySet{}}$|. Consider a tuple $\vTuple{}{k} \in \vTupleSet{k}$ which has policy $\vPolicy{}{l} \in \vPolicySet{}$ that denies access $\vQuery{}{i}$ to $\vTuple{}{k}$. If there exists a tuple $\vTuple{}{j} \in \vTupleSet{\vRelation{}{j}}$ such that $\vTuple{}{j} = \vTuple{}{k}$, then performing set difference operations before checking policies on $\vRelation{}{k}$ will result in a tuple set that does not include $\vTuple{}{j}$. On the other hand, if policies for $\vRelation{}{k}$ are checked first, then $\vTuple{}{k} \not\in \vTupleSet{\vQuery{}{i}}$ and therefore $\vTuple{}{j}$ will be in the query result. 

We define access control as the task of deriving $\vTupleSet{\vQuery{}{}}' \subseteq \vTupleSet{\vQuery{}{}}$ which is the projection of $\vDatabase$ on which $\vQuery{}{}$ can be executed for access control policies defined for it's querier. Thus $\forall$ $\vTuple{}{t} \in \vTupleSet{\vQuery{}{i}}$, $\vTuple{}{t} \in \vTupleSet{\vQuery{}{i}}'$ $\Leftrightarrow$ $eval(\vPolicyExpression{\vPolicySet{}}, \vTuple{}{t})=True$. The function $eval(\vPolicyExpression{\vPolicySet{}}, \vTuple{}{t})$ evaluates a tuple $\vTuple{}{t}$ against the policy expression $\vPolicyExpression{\vPolicySet{}}$ that applies to $\vQuery{}{i}$ as follows:
{\small
\begin{alignat*}{2}
  eval(\vPolicyExpression{\vPolicySet{}}, \vTuple{}{t})   &= \begin{cases}
        True { \pctext{2.2in}{if $\exists$ \; $\vPolicy{}{l} \in \vPolicySet{} \mid$ eval($\vObjectConditions{l}{}, \vTuple{}{t}$) = True}}%
        \\
        False { \pctext{2.2in}{otherwise}}%
        \\
        \end{cases}
\end{alignat*}
}%

\noindent where $eval(\vObjectConditions{l}{}, \vTuple{}{t})$ evaluates the tuple against the object conditions of $\vPolicy{}{l}$ as follows:
{\small
\begin{alignat*}{2}
  eval(\vObjectConditions{l}{}, \vTuple{}{t})   &= \begin{cases}
        True {\pctext{2.2in}{if $\forall \; \vObjectCondition{l}{c} \in \vObjectConditions{l}{} \mid \vTuple{}{t}$.attr = $\vObjectCondition{l}{c}$.attr $\implies$ eval($\vObjectCondition{l}{c}$.op,$\vObjectCondition{l}{c}$.val,$\vTuple{}{t}$.val) = True}}%
        \\
        False {\pctext{2.2in}{otherwise}}%
        \\
        \end{cases}
\end{alignat*}
}%

\noindent where $eval(\vObjectCondition{l}{c}.op,\vObjectCondition{l}{c}.val,\vTuple{}{t}.val)$ compares the object condition value ($\vObjectCondition{l}{c}.val$) to the corresponding tuple value ($\vTuple{}{t}.val$) that matches the attribute of the object condition, using the object condition operator. The expensive operator/query is evaluated to obtain the value if the latter is a derived value. 

Given the above semantics, the order of evaluating policies and query predicates is important for the correctness of results. Depending upon the query operations, evaluating policies after them is not guaranteed to produce correct results. This is trivially true in the case for aggregation or projection operations that remove certain attributes from a tuple. In queries with non-monotonic operations such as set difference, performing query operations before policy evaluation will result in inconsistent answers.

This access control semantics satisfies the sound and secure properties of the correctness criterion defined by~\cite{wang2007correctness}. If no policies are defined on $\vTuple{}{t}$ then the tuple is not included in $\vTupleSet{\vQuery{}{i}}'$ as our access control semantics is opt-out by default.


\subsection{Overview of \oursystem Approach}
\label{sect:ourApproach}


\begin{figure*}[!htb]
\centering
\subfloat[]{\includegraphics[width=3in]{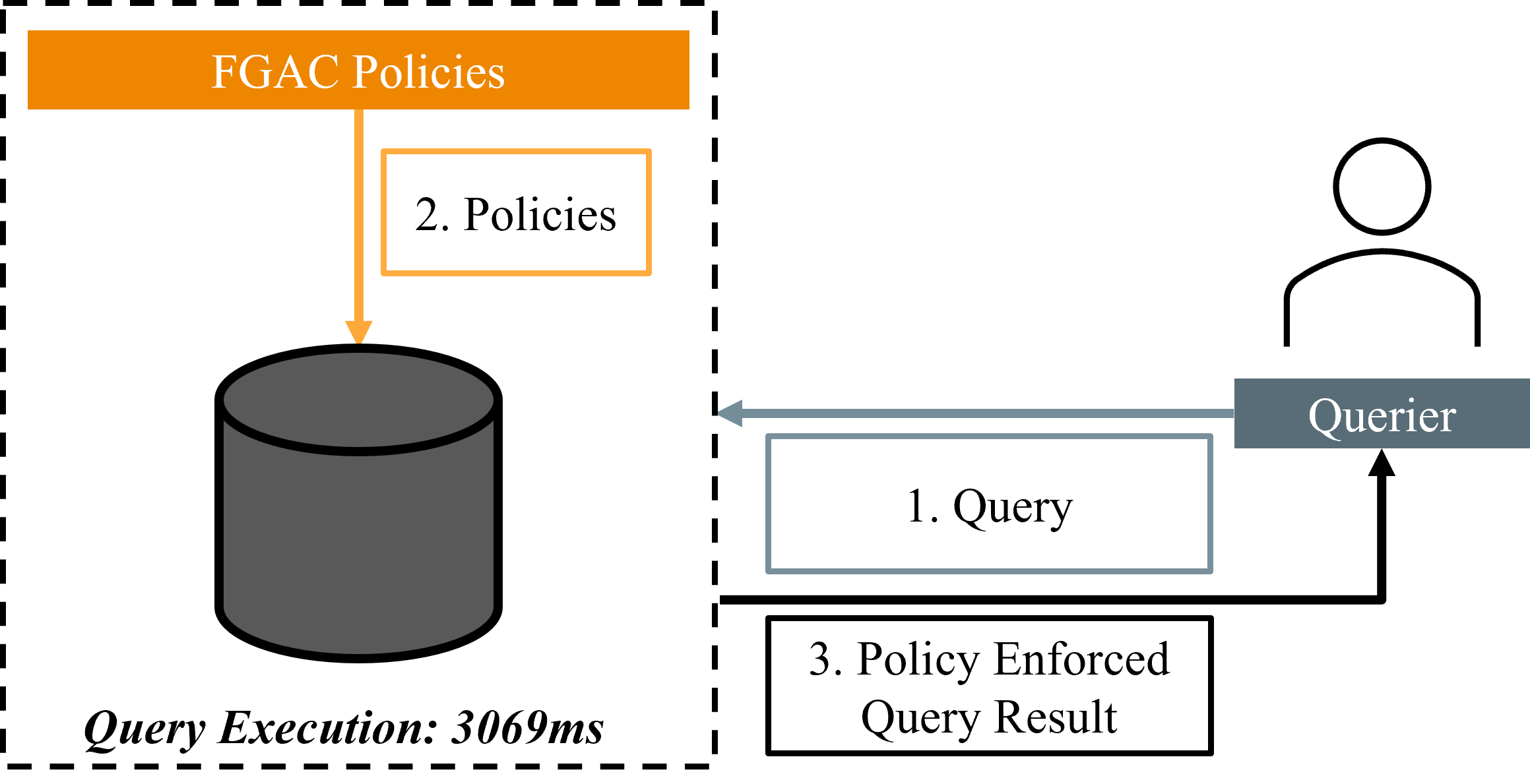}%
\label{fig:fgac_traditional}}
\hfil
\subfloat[]{\includegraphics[width=3in]{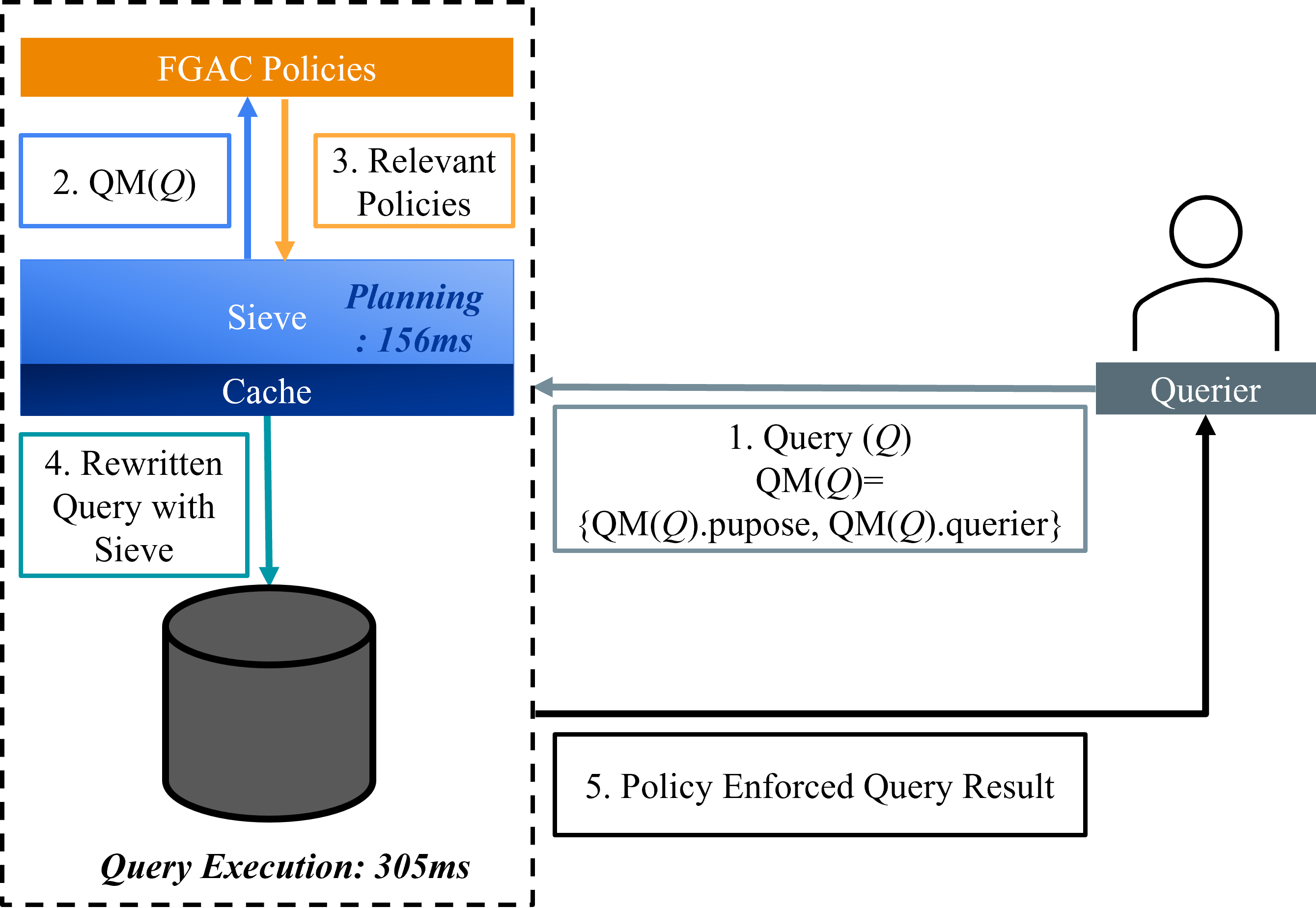}%
\label{fig:fgac_caching}}
\caption{FGAC enforcement comparison (a) Traditional FGAC checks all policies per query, causing high execution time (3069ms). (b) \oursystem uses query metadata (QM) to fetch only relevant policies and reuses planning via caching, reducing execution to 305ms and planning to 156ms.}
\label{fig:fgac_comparison}
\end{figure*}



Traditional FGAC enforcement requires checking access policies during query execution, resulting in significant delays. When a query is issued, the system retrieves all access control policies, regardless of their relevance to the querier, integrates them into the execution plan, and evaluates the modified query, as shown in Figure~\ref{fig:fgac_traditional}. This leads to unnecessary overhead, as irrelevant policies are also checked, and increases the number of tuples each policy is applied to, making query evaluation longer and more expensive.

\oursystem addresses these inefficiencies by introducing a planning phase which retrieves only the relevant queries based on query metadata and rewrites queries carefully to exploit the underlying indexes. This reduces the number of policies evaluated and tuples checked, improving performance during execution. With caching, \oursystem further reduces overhead by storing and reusing previously computed query plans. This helps accelerate both planning and execution phases, as shown in Figure~\ref{fig:fgac_caching}, making the system more efficient and responsive under different workloads.

We evaluate both approaches using a simulated environment with 16 queriers, each defining 200 access policies, totaling 3,200 policies, and each querier posing a single query, resulting in 16 queries. Each query is executed three times to ensure stability, and the average latency per query is reported. Under the traditional FGAC model, policies are appended to the query without reuse, leading to high per-query latency. In contrast, \oursystem with caching achieves up to $10\times$ faster performance by narrowing policy selection, reducing tuple evaluation, and reusing previously computed query plans.

To understand the observed performance gains, we delve into the core contributors to policy enforcement overhead. Specifically, the processing time for a query depends on two primary factors: the volume of policies requiring evaluation and the set of tuples subject to these policies.

Analytically, for a given query $\vQuery{}{}$, the evaluation time of the policy set $\vPolicySet{}$ across the tuple set $\vTupleSet{\vQuery{}{}}$ (represented as $eval(\vPolicyExpression{\vPolicySet{}}, \vTuple{}{}) \; \forall \; \vTuple{}{} \in \vTupleSet{\vQuery{}{}}$) is directly influenced by the cardinality of $\vPolicySet{}$ and $\vTupleSet{\vQuery{}{}}$. Consequently, reducing the policy evaluation overhead necessitates a two-pronged approach.

First, we can employ efficient, low-cost filters to rapidly eliminate tuples $\vTupleSet{\vQuery{}{}}$, thereby minimizing the number of tuples that undergo computationally expensive policy checks. Second, we can streamline the policy expressions themselves, reducing the complexity and length of the expressions applied to each tuple $\vTuple{}{}$ to determine its inclusion in the query result.

\oursystem implements these as follows:

\squishlist

    \item \textbf{Reducing Number of Policies.}
    Not all policies in $\vPolicySet{}$ are relevant to a specific query $\vQuery{}{i}$. We can first easily filter out those policies that are defined for different queriers/purposes given the query metadata $\vQueryMetadatas{i}$. For instance, when Prof. Smith poses a query for grading, only the policies defined for him and the faculty group for grading purpose are relevant out of all policies defined on campus.
    We denote the subset of policies which are relevant given the query metadata $\vQueryMetadatas{i}$ by $\vPolicySet{\vQueryMetadatas{i}} \subseteq \vPolicySet{}$ where $\vPolicy{}{l} \in \vPolicySet{\vQueryMetadatas{i}}$ iff  $\vQueryMetadata{i}{purpose} = \vQuerierCondition{l}{purpose} \land (\vQueryMetadata{i}{querier} = \vQuerierCondition{l}{querier} \lor \vQuerierCondition{l}{querier} \in group(\vQueryMetadata{i}{querier}))$. 
    In addition, for a given tuple $\vTuple{}{t} \in \vTupleSet{\vQuery{}{i}}$ we can further filter policies in $\vPolicySet{\vQueryMetadatas{i}}$ that we must check based on the values of attributes in $\vTuple{}{t}$.
For instance, we can further restrict the set of policies relevant for Prof. Smith's query by considering information of each tuple involved in the query such as its owner (i.e., {$\vTuple{}{t}.owner$}). This way, if the tuple belongs to John, only policies defined by John have to be checked from the previous set. This two-stage filtering process is sound and conservative: it ensures that all policies necessary for correct and complete enforcement are retained, while excluding those that are provably irrelevant. In particular, $\vPolicySet{\vQueryMetadatas{i}}$ serves as a superset of all policies that could potentially influence the query result. Any further reduction beyond this point may lead to under-enforcement, thereby compromising policy compliance guarantees.

    \item \textbf{Reducing Number of Tuples.}
    Even if the number of policies to check is minimized, the resulting expression $\vPolicyExpression{\vPolicySet{}}$ might still be computationally complex. 
    We might improve performance by filtering out tuples based on low-cost filters derived from $\vPolicyExpression{\vPolicySet{}}$. 
    Such processing can be even faster if such simplified expressions could leverage existing indexes $\vIndexSet{}$ over attributes in the database.
    We therefore rewrite the policy expression  $\vPolicyExpression{\vPolicySet{}}= \vObjectConditions{1}{} \lor \cdots \lor \vObjectConditions{\vSetCardinality{\vPolicySet{}}}{}$ as a {\em guarded policy expression} $\vPolicyGuardedExpression{\vPolicySet{}}$ which is a disjunction of {\em guarded expressions} $\vPolicyGuardedExpression{\vPolicySet{}} = \vGuard{}{1} \lor \cdots \lor \vGuard{}{n}$. Each $\vGuard{}{i}$ consists of a {\em guard} $\vObjectCondition{i}{g}$ and a {\em policy partition} $\vPolicySet{\vGuard{}{i}}$ where $\vPolicySet{\vGuard{}{i}} \subseteq \vPolicySet{}$. Note that $\vPolicySet{\vGuard{}{i}}$ partitions the set of policies, i.e., $\vPolicySet{\vGuard{}{i}} \cap \vPolicySet{\vGuard{}{j}} = \emptyset \; \forall \; \vGuard{}{i}, \vGuard{}{j} \in \vPolicyGuardedExpression{\vPolicySet{}}$. Also, all policies in $\vPolicySet{}$ are covered by one of the guarded expressions, i.e.,  $\forall \; \vPolicy{}{i} \in \vPolicySet{} \; (\exists \; \vGuard{}{i} \in \vGuard{}{} \mbox{\ such that\ } \vPolicy{}{i} \in \vPolicySet{\vGuard{}{i}})$.  We will represent the guarded expression $\vGuard{}{i} = \vObjectCondition{i}{g} \land \vPolicySet{\vGuard{}{i}}$ where $\vPolicySet{\vGuard{}{i}}$ is the set of policies but for simplicity of expression we will use it as an expression where there is a disjunction between policies.
        
    The {\em guard} term $\vObjectCondition{i}{g}$ is an object condition 
    that can support efficient filtering by exploiting an index. In particular, 
    it satisfies the following properties: 
        
    \squishlist
        \item $\vObjectCondition{i}{g}$ is a simple predicate over an attribute
        (e.g., $ts-time > 9am$) consisting of an attribute name, a comparison operator, and a constant value. Also, the attribute in $\vObjectCondition{i}{g}$ has an index on it (i.e., $\vObjectCondition{i}{g}.attr \in \vIndexSet{}$).
        \item The guard  $\vObjectCondition{i}{g}$  can serve as a filter for all the policies in  the partition $\vPolicySet{\vGuard{}{i}}$ (i.e.,
        $\forall \, \vPolicy{}{l} \in \vPolicySet{\vGuard{}{i}} \; \exists \; \vObjectCondition{l}{j} \in \vObjectConditions{l}{} \mid \vObjectCondition{l}{j} \implies \vObjectCondition{i}{g}$).
    \squishend
    
\squishend

For example, consider the policy expression of all the policies defined by students to grant the professor access to their data in different situations. Let us consider that many such policies grant access when the student is connected to the WiFi AP of the classroom. For instance, in addition to John's previous policy, let us consider that Mary defines the policy |$\langle$[W.owner = Mary $\land$ W.wifiAP = 1200], [Prof. Smith $\land$ Marking Attendance], allow$\rangle$|. This way, such predicate (i.e., wifiAP=1200) could be used as a guard that will group those policies, along with others that share that predicate, to create the following expression:
|wifiAP=1200 AND ((owner=John AND $\texttt{ts-time}$ between 9am AND 10am OR (owner=Mary) OR ...)|

\oursystem adaptively selects a query execution strategy when a query is posed leveraging the above ideas. First, given $\vQuery{}{i}$, \oursystem filters out policies based on $\vQueryMetadatas{i}$. Then, using the resulting set of policies, it replaces any relation $\vRelation{}{j} \in \vQuery{}{i}$ by a projection that satisfies policies in $\vPolicySet{\vQueryMetadatas{i}}$ that are defined over $\vRelation{}{j}$. It does so by using the guarded expression  $\vPolicyGuardedExpression{\vPolicySet{\vRelation{}{j}}}$ constructed as a query |SELECT * FROM $\vRelation{}{j}$ WHERE $\vPolicyGuardedExpression{\vPolicySet{\vRelation{}{j}}}$|

By using $\vPolicyGuardedExpression{\vPolicySet{\vRelation{}{j}}}$ and its guards $\vObjectCondition{i}{g}$, we can efficiently filter out a high number of tuples and only evaluate the relevant tuples against the more complex policy partitions $\vPolicySet{\vGuard{}{i}}$. The generation of $\vPolicyGuardedExpression{\vPolicySet{\vRelation{}{j}}}$ can be performed either offline before queries arrive or even online as the algorithm (see Section~\ref{sect:guardSel}) is efficient for large numbers of policies (as we will show in our experiments).


A tuple that satisfies the guard $\vObjectCondition{i}{g}$ is then checked against $\vPolicyExpression{\vPolicySet{\vGuard{}{i}}} = \vObjectConditions{1}{} \lor \cdots \lor \vObjectConditions{\vSetCardinality{\vPolicySet{\vGuard{}{i}}}}{}$. As it is a DNF expression, in the worst case (a tuple that does not satisfy any policy) it must be evaluated against each $\vObjectConditions{j}{} \in \vPolicyExpression{\vPolicySet{\vGuard{}{i}}}$. To further reduce runtime overhead, \oursystem optionally employs a tuple-aware filtering operator, $\Delta(\vPolicySet{\vGuard{}{i}}, \vQueryMetadatas{i}, \vTuple{}{t})$, which refines the policy set to $\hat{\vPolicySet{\vGuard{}{i}}}$ based on query metadata and tuple context. Evaluation is then performed using $eval(\hat{\vPolicySet{\vGuard{}{i}}}, \vTuple{}{t})$, and the guarded expression $\vGuard{}{i} = \vObjectCondition{i}{g} \land \vPolicySet{\vGuard{}{i}}$ becomes $\vGuard{}{i} = \vObjectCondition{i}{g} \land \Delta(\vPolicySet{\vGuard{}{i}}, \vQueryMetadatas{i}, \vTuple{}{t})$. \oursystem determines whether to apply $\Delta$ for a given $\vGuard{}{i} \in \vPolicyGuardedExpression{\vPolicySet{\vRelation{}{j}}}$ by estimating its cost relative to directly evaluating $\vPolicyExpression{\vPolicySet{\vGuard{}{i}}}$. Due to space constraints, implementation details of the $\Delta$ operator are provided in the Appendix.


To enhance scalability, \oursystem integrates caching to reduce the overhead of regenerating guarded expressions. 
When a query $\vQuery{}{i}$ from querier $\vQueryMetadata{i}{querier}$ is received, \oursystem checks whether a guarded expression associated with the current metadata and policy set exists in the cache. If found, $\vGuard{}{i}$ is retrieved and used to rewrite $\vQuery{}{i}$, avoiding recomputation. Otherwise, a new $\vGuard{}{i}$ is generated, stored in the cache, and used for query rewriting.



Cache management in \oursystem employs a replacement policy inspired by the Clock algorithm~\cite{CLOCKcorbato1968paging}. While the Clock algorithm provides an efficient FIFO-based eviction strategy, we adapt it to maintain a cache of $\langle \vQueryMetadata{i}{querier}, \vGuard{}{i} \rangle$ pairs, each associated with a \textit{use} bit. This design enables \oursystem to track recently accessed GEs and prioritise their retention. When the cache reaches capacity, the replacement policy scans for the first entry with \textit{use-bit} $=0$ and evicts it, ensuring that frequently accessed GEs remain available while optimising resource utilisation. This approach effectively balances simplicity with performance, making it well-suited for dynamic FGAC workloads where efficient cache management is essential for scalability.

A refresh strategy updates cached GEs when new policies are inserted. The system checks for mergeability if new policies $\vPolicySet{new}$ exist for $\vQueryMetadata{i}{querier}$ after $\vGuard{old}{i}$ was cached. If mergeable, a new $\vGuard{new}{i} = \vObjectCondition{i}{g} \land \vPolicySet{\vGuard{new}{i}}$ is regenerated over all policies, including the new ones of $\vQueryMetadata{i}{querier}$. If merging is not feasible, the cached $\vGuard{old}{i}$ is incrementally updated by incorporating the new policies and generating a new guarded expression over them, resulting in $\vGuard{}{i} \cup \vGuard{new}{i}$. This ensures queries always use the latest policies without unnecessary recomputation.


The main challenges in implementing \oursystem are: (1) Selecting appropriate guards and generating the guarded expression; (2) Dynamically selecting a strategy and constructing a query that can be executed in an existing DBMS using the selected strategy. We explain our algorithm to generate guarded expressions for a set of policies in Section~\ref{sect:guardSel}. This generation may occur offline when policy changes are infrequent. Otherwise, it can be performed dynamically when a policy update occurs or at query time to accommodate more dynamic workloads as seen in \oursystem with caching.
An additional set of challenges arises in efficiently managing caching: (3) Ensuring cache consistency so that cached \GEs remain valid despite policy updates; and (4) Selecting an appropriate replacement strategy to optimize memory usage while maintaining frequently accessed \GEs. We describe our caching approach, including replacement and refresh strategies, in Section~\ref{sect:caching}. We later explain how \oursystem can be implemented in existing DBMSs and how it selects an appropriate strategy depending on the query and the set of policies~\ref{sect:implementingSieve}.


\section{Creating Guarded Expressions}
\label{sect:guardSel}

Our goal is to translate a policy expression  $\vPolicyExpression{\vPolicySet{}}= \vObjectConditions{1}{} \lor \cdots \lor \vObjectConditions{\vSetCardinality{\vPolicySet{}}}{}$ into a guarded policy expression $\vPolicyGuardedExpression{\vPolicySet{}} = \vGuard{}{1} \lor \cdots \lor \vGuard{}{n}$ such that the cost of evaluating $\vPolicyGuardedExpression{\vPolicySet{}}$ given $\vDatabase$ and $\vIndexSet{}$\footnote{For our purposes we will assume that the set of available indexes is known.} is minimized
\begin{equation}\label{eq:minCost}
\min cost (\vPolicyGuardedExpression{\vPolicySet{}},\vGuard{}{}) = \min \sum_{\vGuard{}{i} \in \vGuard{}{}} cost(\vGuard{}{i})   
\end{equation}

\noindent where $\vGuard{}{}$ is the set of all the guarded expressions in $\vPolicyGuardedExpression{\vPolicySet{}}$. A guarded expression $\vGuard{}{i}$ corresponds to $\vGuard{}{i}=\vObjectCondition{i}{g} \land \vPolicySet{\vGuard{}{i}}$ where $\vObjectCondition{i}{g}$ is an object condition on an indexed attribute. To model $cost(\vGuard{}{i})$ let us define first the cost of evaluating a tuple against a set of policies as
\begin{equation}\label{eq:costEvalTuplePolicies}
cost(eval(\vPolicyExpression{\vPolicySet{\vGuard{}{i}}}, \vTuple{}{t})) = \vShortCircuit.\vSetCardinality{\vPolicySet{\vGuard{}{i}}}.\vEvalCost
\end{equation}

\noindent where $\vShortCircuit$ represents the average number of policies in $\vPolicySet{\vGuard{}{i}}$ that the tuple $\vTuple{}{t}$ is checked against before it satisfies one (as the policies in $\vPolicyExpression{\vPolicySet{\vGuard{}{i}}}$ form a disjunctive expression\footnote{We assume that the execution of such disjunctive expression stops with the first policy condition evaluating to true and skipping the rest of the policy conditions.}), and $\vEvalCost$ represents the average cost of evaluating $\vTuple{}{t}$ against the set of object conditions for a policy $\vPolicy{}{l} \in \vPolicySet{\vGuard{}{i}}$ (i.e., $\vObjectConditions{l}{}$). The values of $\vShortCircuit$ and $\vEvalCost$ are determined experimentally using a set of sample policies and tuples. Hence, we model $cost(\vGuard{}{i})$ as
\begin{equation}\label{eq:costPartition}
cost(\vGuard{}{i}) = \vSelectivityMethod{\vObjectCondition{i}{g}}.(\vReadCost + cost(eval(\vPolicyExpression{\vPolicySet{\vGuard{}{i}}}, \vTuple{}{t})))
\end{equation}

\noindent where $\vSelectivityMethod{\vObjectCondition{i}{g}}$ denotes the estimated cardinality\footnote{Estimated using histograms maintained by the database.} of the guard $\vObjectCondition{i}{g}$ and $\vReadCost$ represents the cost of reading a tuple from the disk (the value of $\vReadCost$ is also obtained experimentally). Given this cost model, the number of policies in  $\vPolicySet{\vGuard{}{i}}$ and selectivity of $\vObjectCondition{i}{g}$ contribute to most of the cost when evaluating $\vGuard{}{i}$.

The first step in determining $\vPolicyGuardedExpression{\vPolicySet{}}$ is to generate all the {\em candidate guards} ($\vCandidateGuardSet{}{}$), given the object conditions from $\vPolicySet{}{}$, which satisfy the properties of guards as explained in Section~\ref{sect:ourApproach}. Different choices may exist for the same policy given $\vIndexSet{}$; the second step is to select a subset of guards from $\vCandidateGuardSet{}{}$ with the goal of minimizing the evaluation cost of $\vPolicyGuardedExpression{\vPolicySet{}}$.

\subsection{Generating Candidate Guards}

Each policy $\vPolicy{}{l} \in \vPolicySet{\vQuery{}{j}}$ is guaranteed to have at least one object condition (i.e., $\vObjectCondition{l}{owner}$), that trivially satisfies the properties of a guard as 1) $\vObjectCondition{l}{owner}.val$ is a constant and $\vObjectCondition{l}{owner}.attr \in \vIndexSet{}$; and 2) $\vObjectCondition{l}{owner} \in \vObjectConditions{l}{}$. Therefore, we first include all the $\vObjectCondition{l}{k}$ in $\vCandidateGuardSet{}{}$. Similarly, any $\vObjectCondition{l}{c}$ on an indexed attribute with a constant value, belonging to any policy $\vPolicy{}{l}$, can be added to the candidate set $\vCandidateGuardSet{}{}$. However, if only those were to be used as guards, then the size of their corresponding policy partitions $\vPolicySet{\vGuard{}{i}}$ might be small as only policies defined by the same person or policies with the exact same object condition (including, attribute, value, and operation) would be grouped by such a guard.
Exploiting the property that different policies might have common object conditions reduces the number of $\vGuard{}{i}$ and increases the size of their policy partitions $\vPolicySet{\vGuard{}{i}}$ thus decreasing the potential number of evaluations. 
Hence, we create additional candidate guards by {\em merging} range object conditions of different policies on the same attribute, but with different constant values (e.g., if the conditions of two policies on attribute $a$ are $3<a<10$ and $4<a<15$, respectively, the condition $3<a<15$ could be created as a guard, by merging the two conditions, to group both policies in its policy partition). The following theorem limits object conditions that should be considered for this merge based on their overlap.

\begin{theorem}{\label{the:mergeTheorem}} 
Given two candidate guards $\vObjectCondition{x}{c}$ = ($attr^{x}_{1}, op^{x}_{1}$, $val^{x}_{1}, op^{x}_{2}, val^{x}_{2}$) $\in \vObjectConditions{x}{}$, $\vObjectCondition{y}{c} = (attr^{y}_{1}, op^{y}_{1}, val^{y}_{1}, op^{y}_{2}, val^{y}_{2}) \in \vObjectConditions{y}{}$ such that $attr^{x}_{1} = attr^{y}_{1}$ and $attr^{x}_{1} \in \vIndexSet{}$, it is not beneficial to generate a guard by merging them as $\vObjectCondition{x \oplus y}{c}$ = ($attr^{x}_{1}$, $op^{x}_{1}$, $val^{x \oplus y}_1, op^{y}_{2}, val^{x \oplus y}_2$) with $val^{x \oplus y}_1=min(val^{x}_{1},val^{y}_{1})$ and $val^{x \oplus y}_2=max(val^{x}_{2},val^{y}_{2})$ iff $[val^{x}_{1}, val^{x}_{2}] \cap [val^{y}_{1}, val^{y}_{2}]=\phi$.
\end{theorem} 

\textit{Proof:} By Equation~\ref{eq:costPartition}, and considering a guarded expression that contains only a single policy $\vPolicy{}{l}$, the evaluation cost using $\vObjectCondition{l}{c}$ as guard is given by
\begin{equation}\label{eq:costPolicy}
\vCostMethod{\vPolicy{}{l}} = \vSelectivityMethod{\vObjectCondition{l}{c}}.(\vReadCost + \vEvalCost)
\end{equation}

Given two policies $\vPolicy{}{x}$ and $\vPolicy{}{y}$ with candidate guards $\vObjectCondition{x}{c}$ and $\vObjectCondition{y}{c}$\footnote{For simplification of notation in this proof we use $\vObjectCondition{x/y}{c}$ to denote the values in the range $[val^{x/y}_{1}, val^{x/y}_2]$.} such that $\vObjectCondition{x}{c} \cap \vObjectCondition{y}{c} = \emptyset$, it is trivial to see that the cost of evaluating their merge is always greater than evaluating them separately. W.l.o.g., let us consider that $min(val^{x}_{1},val^{y}_{1})=val^{x}_{1}$ and $max(val^{x}_{2},val^{y}_{2})=val^{y}_{2}$ hence the evaluation cost if they were to be merged would be
\begin{multline}\label{eq:costMergeEmptyIntersection}
\vCostMethod{\vPolicy{}{x}\oplus\vPolicy{}{y}} = \vSelectivityMethod{\vObjectCondition{x \oplus y}{c}}.(\vReadCost + \vEvalCost)= \\
(\vSelectivityMethod{\vObjectCondition{x}{c}}+\vSelectivityMethod{\vObjectCondition{y}{c}}).(\vReadCost + \vEvalCost)+\vSelectivityMethod{\vObjectCondition{e}{c}}.(\vReadCost + 2.\vEvalCost)
\end{multline}

\noindent where $\vObjectCondition{e}{c}= (attr^{x}_{1}, op^{x}_{1}, val^{x}_{1}, op^{y}_{2}, val^{y}_{2})$ and hence $\vSelectivityMethod{\vObjectCondition{e}{c}}>=0$, which makes $\vCostMethod{\vPolicy{}{x}\oplus\vPolicy{}{y}}>=\vCostMethod{\vPolicy{}{x}}+\vCostMethod{\vPolicy{}{y}}$. \qed

For situations where $[val^{x}_{1}, val^{x}_{2}] \cap [val^{y}_{1}, val^{y}_{2}] \neq \phi$ we can derive the condition that will make merging beneficial. As previously, let us consider w.l.o.g. that $min(val^{x}_{1},val^{y}_{1})=val^{x}_{1}$ and $max(val^{x}_{2},val^{y}_{2})=val^{y}_{2}$. If the candidate guards were to be merged the new cost of evaluation would be given by $\vCostMethod{\vPolicy{}{x}\oplus\vPolicy{}{y}} = \vSelectivityMethod{\vObjectCondition{x}{c} \cup \vObjectCondition{y}{c}}.(\vReadCost + 2.\vEvalCost)$ which, applying the inclusion-exclusion principle, becomes
\begin{multline}\label{eq:costMergeNotEmptyIntersection}
    \vCostMethod{\vPolicy{}{x}\oplus\vPolicy{}{y}} = (\vSelectivityMethod{\vObjectCondition{x}{c}} + \vSelectivityMethod{\vObjectCondition{y}{c}} - \\
    \vSelectivityMethod{\vObjectCondition{x}{c} \cap \vObjectCondition{y}{c}}).(\vReadCost + 2.\vEvalCost)
\end{multline}

Given that merging will be beneficial if $\vCostMethod{\vPolicy{}{x}\oplus\vPolicy{}{y}}<\vCostMethod{\vPolicy{}{x}}+\vCostMethod{\vPolicy{}{y}}$ and by Equations~\ref{eq:costPolicy} and~\ref{eq:costMergeNotEmptyIntersection} we have
\begin{equation}
    \begin{split}
    &  (\vSelectivityMethod{\vObjectCondition{x}{c}} + \vSelectivityMethod{\vObjectCondition{y}{c}} - \vSelectivityMethod{\vObjectCondition{x}{c} \cap \vObjectCondition{y}{c}}).(\vReadCost + 2.\vEvalCost) < \\
    & \vSelectivityMethod{\vObjectCondition{x}{c}}.(\vReadCost + \vEvalCost) + \vSelectivityMethod{\vObjectCondition{y}{c}}.(\vReadCost + \vEvalCost) \\
    & \vSelectivityMethod{\vObjectCondition{x}{c}}.\vEvalCost + \vSelectivityMethod{\vObjectCondition{y}{c}}.\vEvalCost-\vSelectivityMethod{\vObjectCondition{x}{c} \cap \vObjectCondition{y}{c}}(\vReadCost + 2.\vEvalCost)<0
    \end{split}
\end{equation}

\noindent Using inclusion exclusion principle if follows that
\begin{equation}{\label{eq:overlapCon}}
    \dfrac{\vSelectivityMethod{\vObjectCondition{x}{c} \cap \vObjectCondition{y}{c}}}{\vSelectivityMethod{\vObjectCondition{x}{c} \cup \vObjectCondition{y}{c}}} > \dfrac{\vEvalCost}{\vReadCost + \vEvalCost}
\end{equation}

\noindent which is the condition to be checked to merge those two overlapping candidates. Equation~\ref{eq:overlapCon} is checked by the function $\vMergeMethod{\vObjectCondition{x}{c}}{\vObjectCondition{y}{c}}$ which returns $\vObjectCondition{x \oplus y}{c}$ if merging $\vObjectCondition{x}{c}$ and $\vObjectCondition{y}{c}$ is beneficial and $\phi$ otherwise. 
$\vObjectCondition{x \oplus y}{c}$ is added to $\vCandidateGuardSet{}{}$ and we maintain in a mapping structure that both $\vPolicy{}{x}$ and $\vPolicy{}{y}$ are relevant to that candidate guard (this information will be used in the second step). 
Given Theorem~\ref{the:mergeTheorem}, $\vMergeMethod{\vObjectCondition{x}{c}}{\vObjectCondition{y}{c}}$ is computed only if $\vSetCardinality{\vObjectCondition{x}{c} \cap \vObjectCondition{y}{c}} \neq \phi$, we first order the candidate guards by their left range value in ascending order.
The number of checks to be done could still be high as a candidate guard could potentially merge with another transitively. For example, given a situation where $\vObjectCondition{x}{c} \cap \vObjectCondition{y}{c} \neq \phi$, $\vObjectCondition{y}{c} \cap \vObjectCondition{z}{c} \neq \phi$, and $\vObjectCondition{x}{c} \cap \vObjectCondition{z}{c} = \phi$, which might make $\vMergeMethod{\vObjectCondition{x}{c}}{\vObjectCondition{y}{c}} \neq \phi$, $\vMergeMethod{\vObjectCondition{y}{c}}{\vObjectCondition{z}{c}} \neq \phi$, and $\vMergeMethod{\vObjectCondition{x}{c}}{\vObjectCondition{z}{c}} = \phi$, it could be possible the transitive merge of $\vObjectCondition{x}{c}$ with $\vObjectCondition{y \oplus z}{c}$ is beneficial (i.e., $\vMergeMethod{\vObjectCondition{x}{c}}{\vObjectCondition{y \oplus z}{c}} = \phi$). 
We present a condition to limit the number of checks to be performed due to such transitive overlaps for a given a $\vCandidateGuardSet{}{}$ with candidate guards sorted in the ascending order of their left range values.

First, we show as a consequence of Theorem~\ref{the:mergeTheorem} that transitive merges will not be useful under the following condition.

\begin{corollary}{\label{theorem:pruningCon}}
Given two candidate guards $\vObjectCondition{x}{c}$ and $\vObjectCondition{y}{c}$, such that $\vObjectCondition{x}{c} \cap \vObjectCondition{y}{c} \neq \phi$ and whose merging is not beneficial (i.e., $\vMergeMethod{\vObjectCondition{x}{c}}{\vObjectCondition{y}{c}} = \phi$), and given another candidate guard $\vObjectCondition{y \oplus z}{c}$, generated after merging $\vObjectCondition{y}{c}$ and $\vObjectCondition{z}{c}$, the transitive merge of $\vObjectCondition{x}{c}$ and $\vObjectCondition{y \oplus z}{c}$ will not be beneficial (i.e., $\vMergeMethod{\vObjectCondition{x}{c}}{\vObjectCondition{y \oplus z}{c}} = \phi$) if $\vObjectCondition{x}{c} \cap \vObjectCondition{z}{c} = \phi$.
\end{corollary}

Let us consider $\vObjectCondition{x}{c} \cap \vObjectCondition{z}{c} = \phi$.
By Equation~\ref{eq:costPolicy} and Equation~\ref{eq:costMergeNotEmptyIntersection}, we calculate the cost of such a merge by
\begin{multline}
\vCostMethod{\vPolicy{}{x}\oplus(\vPolicy{}{y\oplus z})} =  (\vSelectivityMethod{\vObjectCondition{x}{c}}+\vSelectivityMethod{\vObjectCondition{y}{c}} + \vSelectivityMethod{\vObjectCondition{z}{c}} + \\ \vSelectivityMethod{\vObjectCondition{x}{c} \cap \vObjectCondition{y}{c}} + \vSelectivityMethod{\vObjectCondition{y}{c} \cap \vObjectCondition{z}{c}}).(\vReadCost + 3.\vEvalCost) 
\end{multline}

\noindent which makes $\vCostMethod{\vPolicy{}{x}\oplus(\vPolicy{}{y}\oplus\vPolicy{}{z})}>\vCostMethod{\vPolicy{}{x}}+\vCostMethod{\vPolicy{}{y}}+\vCostMethod{\vPolicy{}{z}}$ and hence $\vMergeMethod{\vObjectCondition{x}{c}}{\vObjectCondition{y \oplus z}{c}} = \phi$.

In addition, in the situation described in Corollary~\ref{theorem:pruningCon}, we can show that there is no need to merge $\vObjectCondition{x}{c}$ with any other candidate following $\vObjectCondition{z}{c}$.

\begin{corollary}
Given the situation explained in Corollary~\ref{theorem:pruningCon}, let us define  $\hat{\vCandidateGuardSet{}{}}=\vCandidateGuardSet{}{}\setminus\{\vObjectCondition{x}{c},\vObjectCondition{y}{c},\vObjectCondition{z}{c}\}$. For any $\vObjectCondition{w}{c} \in \hat{\vCandidateGuardSet{}{}}$, the transitive merge with $\vObjectCondition{x}{c}$ is not beneficial (i.e., $\vMergeMethod{\vObjectCondition{x}{c}}{\vObjectCondition{y \oplus z \oplus w}{c}} = \phi \; \forall \; \vObjectCondition{w}{c} \in \hat{\vCandidateGuardSet{}{}}$).
\end{corollary}

As the candidate guards are sorted by their left ranges and $\vObjectCondition{x}{c} \cap \vObjectCondition{z}{c} = \phi$, we also have $\vObjectCondition{x}{c} \cap \vObjectCondition{w}{c} = \phi$. Therefore, as shown in Corollary~\ref{theorem:pruningCon}, the transitive merge with $\vObjectCondition{x}{c}$, $\vObjectCondition{y}{c}$, $\vObjectCondition{z}{c}$, and $\vObjectCondition{w}{c}$ will not be beneficial ($\vMergeMethod{\vObjectCondition{x}{c}}{\vObjectCondition{y \oplus z \oplus w}{c}} = \phi$).

To summarize, the steps for generating $\vCandidateGuardSet{}{}$ from a set of policies $\vPolicySet{}$ are then as follows: 1) For all $\vPolicy{l}{} \in \vPolicySet{}$ collect object conditions that satisfy guard properties by their attribute; 2) For each such collection sort range object conditions by their left range; 3) For the first candidate guard ($\vObjectCondition{1}{c}$), verify whether the next candidate guard ($\vObjectCondition{2}{c}$) is such that $\vMergeMethod{\vObjectCondition{1}{c}}{\vObjectCondition{2}{c}} \neq \phi$. In that case, merge both candidate guards to generate $\vObjectCondition{1 \oplus 2}{c}$ which is added to $\vCandidateGuardSet{}{}$ ($\vPolicy{}{1}$ and $\vPolicy{}{2}$ get associated to the new merged candidate). Otherwise, if $\vMergeMethod{\vObjectCondition{1}{c}}{\vObjectCondition{2}{c}} = \phi$, then we check $\vObjectCondition{1}{c}$ with the following candidate guards until the condition in Corollary~\ref{theorem:pruningCon} is satisfied and move to the next candidate guard when it does and repeat the process.

\subsection{Selecting Cost Optimal Guards}

We next select the subset of guards  $\vGuard{}{} \in \vCandidateGuardSet{}{}$ that minimizes the cost according to Equation~\ref{eq:minCost}. The guard selection problem can be formally stated as 
\begin{multline}\label{eq:minimizationProblem}
\min_{\vGuard{}{} \subseteq \vCandidateGuardSet{}{}} \vCostMethod{\vGuard{}{}}
    = \\ \sum_{\vGuard{}{i}\in \vGuard{}{}} \vCostMethod{\vGuard{}{i}} \; \forall \; \vPolicy{}{i} \in \vPolicySet{} \;
    \exists \; \vGuard{}{i} \in \vGuard{}{} \mid \vPolicy{}{i} \in \vPolicySet{\vGuard{}{i}}    
\end{multline}

The problem of selecting $\vGuard{}{}$ from $\vCandidateGuardSet{}{}$ such that every policy in $\vPolicySet{}$ is covered exactly once (as this would limit the extra checkings) can be shown to be NP-hard by reducing weighted Set-Cover problem to it. In the weighted Set-Cover problem, we have a set of elements $E = {e_1, \cdots, e_n}$ and a set of subsets over $E$ denoted by $S = {S_1, \cdots, S_m}$ with each set $S_i \in S$ having a weight $w_i$ associated with it. The goal of set cover problem is to select $\min_{\hat{S} \subseteq S} \sum{S_i.w_i} \mid S_i \in \hat{S}$  and $E = \bigcup_{S_i \in \hat{S}} S_i$. From our guard selection problem we have $E$ and $S$ equivalent to $\vPolicySet{}$ and $\vCandidateGuardSet{}{}$ respectively. We assign $e_i$ to $S_i$ when the corresponding $\vPolicy{}{i}$ is assigned to $\vGuard{}{i}$.
The weight function $w_i$ set to $\vCostMethod{\vGuard{}{i}}$ where the evaluation cost of $\vPolicyGuardedExpression{\vGuard{}{i}}$ is set to zero. So we have $w_i = \vSetCardinality{\vObjectConditions{i}{c}}.\vReadCost$. If a polynomial time algorithm existed to solve this problem, then it would solve set-cover problem too.

The evaluation cost of a policy depends on the guard it is assigned to. We define a utility heuristic\footnote{Similar to the one used by~\cite{hellerstein1998optimization} for optimizing queries with expensive predicates.} which ranks the guards by their benefit per unit read cost. Without a guard, $\vPolicyExpression{\vPolicySet{}}$ will be evaluated by a linear scan followed by the checking of $\vPolicyExpression{\vPolicySet{}}$ as filter on top of $\vTuple{}{t} \in \vTupleSet{\vQuery{}{i}}$. The guard $\vObjectCondition{i}{c} \in \vGuard{}{}$ reduces the number of tuples that have to be checked against each $\vPolicyExpression{\vPolicySet{\vGuard{}{i}}}$. The benefit of a guard captures this difference by $\vBenefitMethod{\vGuard{}{i}} = \vEvalCost.\vSetCardinality{\vPolicySet{\vGuard{}{i}}}.(\vSetCardinality{\vRelation{}{i}} - \vSelectivityMethod{\vObjectCondition{i}{c}})$. Using this benefit method, and the read cost of evaluating $\vGuard{}{i}$ defined earlier, we define the utility of $\vGuard{}{i}$ as $\vUtilityMethod{\vGuard{}{i}}= \frac{\vBenefitMethod{\vGuard{}{i}}}{\vReadCostMethod{\vGuard{}{i}}}$.

Algorithm~\ref{alg:guardSelection} uses this heuristic to select the best possible guards to minimize the cost of policy evaluation. First, it iterates over $\vCandidateGuardSet{}{}$ and stores each guarded expression $\vGuard{}{i} \in \vCandidateGuardSet{}{}$ (comprised of a guard $\vObjectCondition{g}{i}$ and a policy partition $\vPolicySet{\vGuard{}{i}}$) in a priority queue in descending order of their utility. 
Next, the priority queue is polled for the $\vGuard{}{i}$ with the highest utility. If $\vPolicySet{\vGuard{}{i}}$ intersects with another $\vPolicySet{\vGuard{}{j}} \in \vCandidateGuardSet{}{}$, $\vPolicySet{\vGuard{}{j}}$ is updated to remove the intersection of policies and $\vUtilityMethod{\vGuard{}{j}}$ is recomputed after which the new $\vGuard{}{j}$ is reinserted into priority queue in the order of its utility. The result is thus the subset of candidates guards $\vGuard{}{}$ that maximizes the benefit and covers all the policies in $\vPolicySet{}$, that is, minimizes  $cost(\vPolicyGuardedExpression{\vPolicySet{}},\vGuard{}{})$ in Equation~\ref{eq:minimizationProblem}.

\begin{algorithm}
\scriptsize
\caption{Selection of guards}
\label{alg:guardSelection}
\begin{algorithmic}[1]
    \Function{GuardSelection}{$\vCandidateGuardSet{}{}$}
        \For{$i \gets 1$ to $\vSetCardinality{\vCandidateGuardSet{}{}}$}
            \State $C[i] \gets$ \Call{cost}{$\vGuard{}{i}$}
            \State $U[i] \gets$ \Call{utility}{$\vGuard{}{i}$}
        \EndFor
        \State $Q \gets \varnothing$
        \For{$i \gets 1$ to $\vSetCardinality{\vCandidateGuardSet{}{}}$}
            \State \Call{PriorityInsert}{$Q$, $\vGuard{}{i}$, $U[i]$}
        \EndFor
        \While{$Q$ is not empty}
            \State $\vGuard{}{max} \gets$ \Call{Extract-Maximum}{$Q$}
            \State $\vGuard{}{} \gets \vGuard{}{max}$
            \For{$\vGuard{}{i}$ in $Q$}
                \If{$\vPolicySet{\vGuard{}{i}} \cap \vPolicySet{\vGuard{}{max}} \neq \varnothing$}
                    \State $\vPolicySet{\vGuard{}{i}} \gets \vPolicySet{\vGuard{}{i}} \setminus \vPolicySet{\vGuard{}{max}}$
                    \State \Call{Remove}{$Q, \vGuard{}{i}$}
                    \If{$\vPolicySet{\vGuard{}{i}} \neq \varnothing$} 
                        \State $B \gets$ \Call{benefit}{$\vGuard{}{i}$}
                        \State $U[i] \gets \dfrac{B}{C[i]}$
                        \State \Call{PriorityInsert}{$Q$, $\vGuard{}{i}$, $U[i]$}
                    \EndIf
                \EndIf
            \EndFor
        \EndWhile
        \State \Return $\vGuard{}{}$
    \EndFunction
\end{algorithmic}
\end{algorithm}


\section{Caching Approach in Sieve}\label{sect:caching}


While \oursystem optimizes policy enforcement through guarded expressions (GEs), its reliance on generating a new GE for every query poses challenges in dynamic environments. These include scenarios where policies and queries frequently change, such as in smart campuses. Consider the beginning of an academic term, where a surge of policy insertions occurs as students define access preferences for attendance tracking or space usage analysis. Faculty may repeatedly query the database throughout the term, while policies may evolve dynamically as students add or drop classes. In such cases, regenerating GEs for every query results in significant computational overhead, increasing query response times.

\begin{figure}[!hb]
\centering
\includegraphics[width=\linewidth]{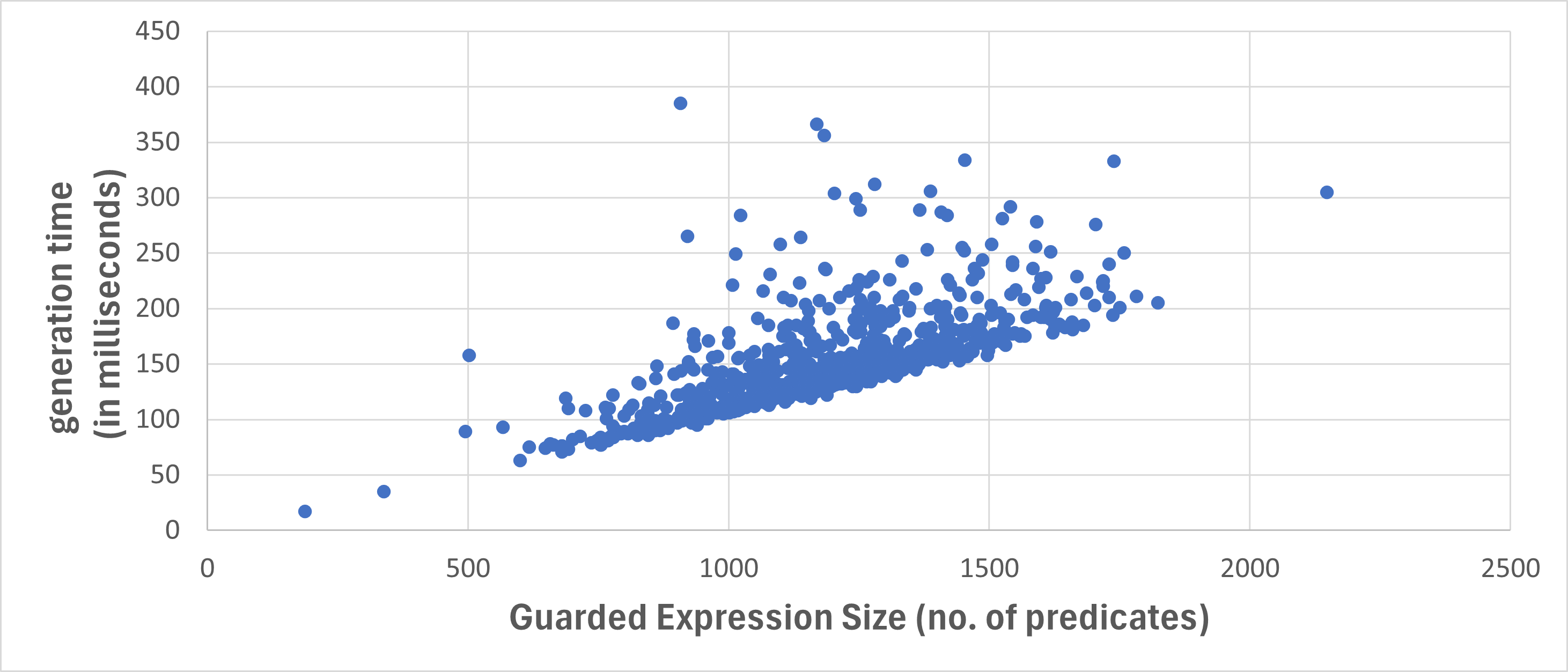}
\caption{Guarded Expression size vs. generation time.}
\label{fig:guardgen.png}
\end{figure}

Our observations reveal that many queries from the same querier share similar or identical access control requirements. This insight suggests an opportunity to reuse previously generated GEs for subsequent queries. To validate this, we conducted an experiment to analyze the relationship between GE size and generation time. Figure~\ref{fig:guardgen.png} illustrates that GE generation time increases sharply with GE size, making frequent regeneration computationally expensive in large-scale deployments. 

We propose a caching mechanism that stores and reuses GEs associated with a querier for future queries to address these challenges. This approach eliminates redundant computations by leveraging two strategies:
\begin{itemize}
    \item \textbf{Replacement Strategy:} A Clock-based algorithm~\cite{yang2023fifo} manages limited cache memory efficiently, ensuring that the least relevant GEs are evicted when constrained space. This strategy is further explained in Section~\ref{sec:replacement-policy}.
    \item \textbf{Refresh Strategy:} Ensures the correctness of cached GEs by deciding whether to regenerate or update an existing GE when policies change. This strategy evaluates the trade-off between the computational cost of regeneration and the potential query execution benefits. The strategy is further discussed in detail in Section~\ref{sec:refresh-strategy}
\end{itemize}

Caching reduces the frequency of expensive GE regenerations and improves query response times, making \oursystem scalable in dynamic workloads. By integrating these enhancements, \oursystem can handle real-world scenarios such as attendance tracking and space usage monitoring with greater efficiency and compliance.




\begin{figure*}[!htb]
\centering
\subfloat[]{\includegraphics[width=3.5in]{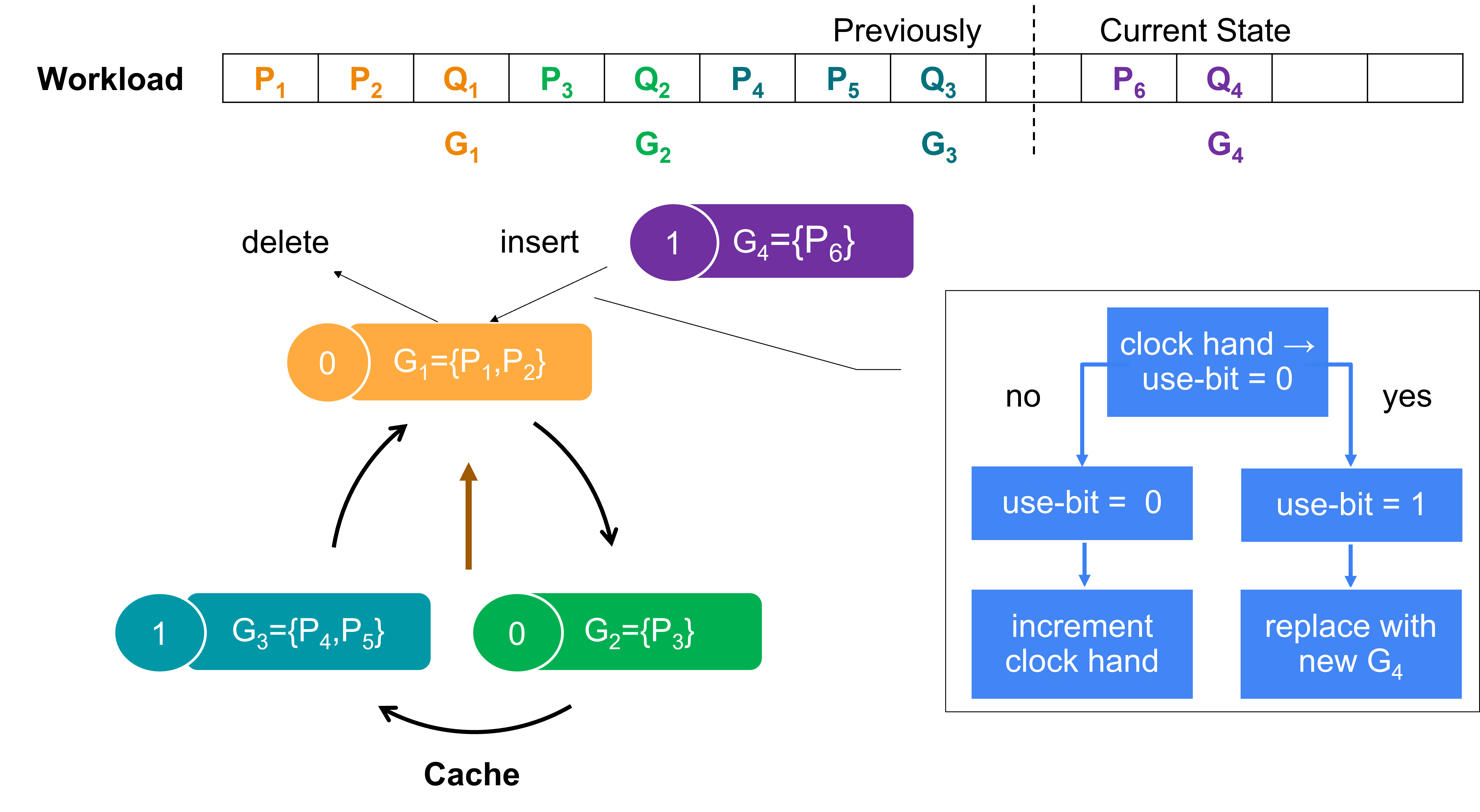}%
\label{fig:replacement}}
\hfil
\subfloat[]{\includegraphics[width=3.5in]{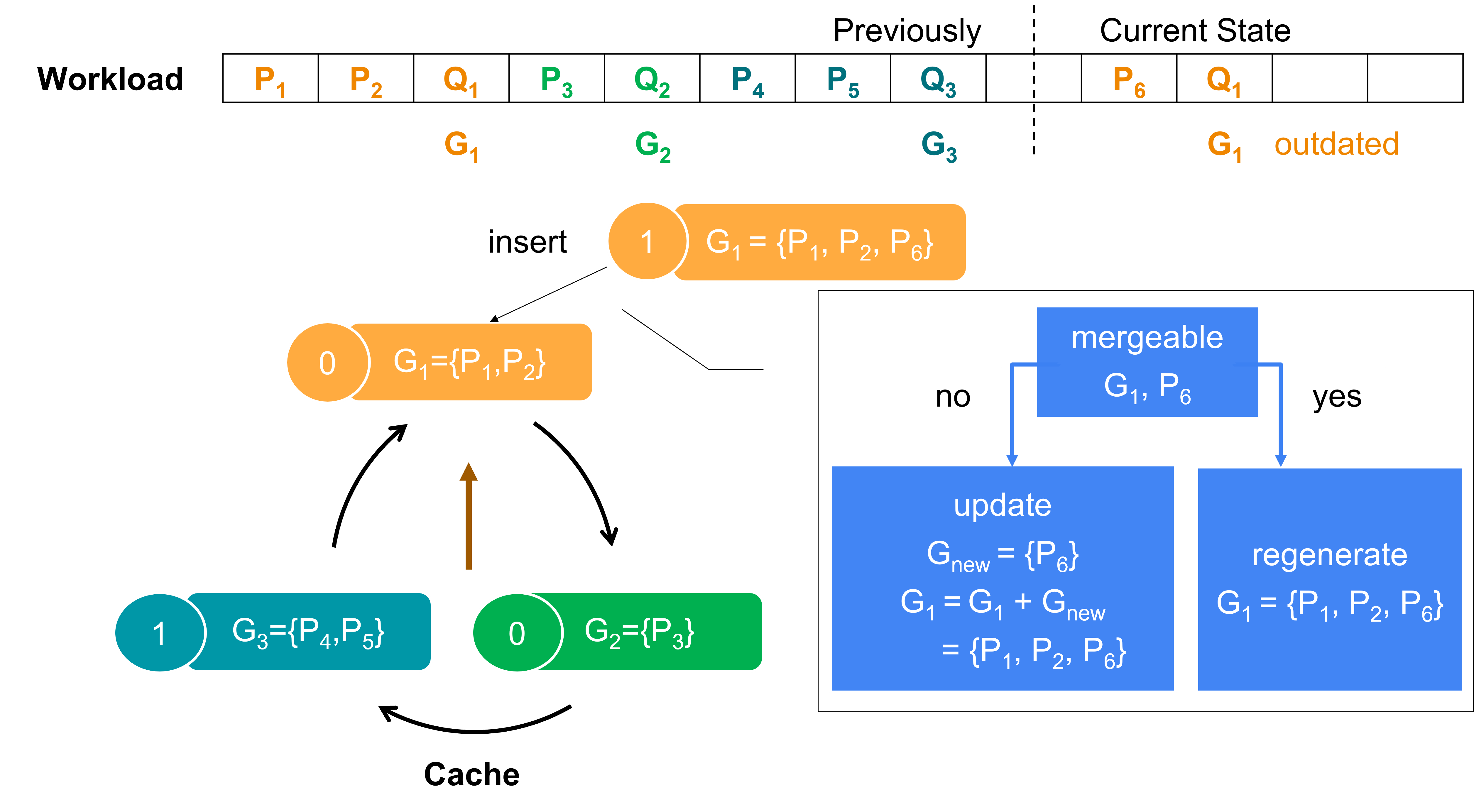}%
\label{fig:refresh}}
\caption{Caching Mechanism: (a) Cache Replacement Strategy; (b) Cache Refresh Strategy.}
\label{fig:caching-imp-figure}
\end{figure*}

\subsection{Replacement Strategy}\label{sec:replacement-policy}

We illustrate how caching in \oursystem works using Figure~\ref{fig:replacement}. The system operates with a circular cache containing limited slots, where each entry consists of a querier-GE pair associated with a set of policies and a \textit{use-bit} to track recent usage. The figure depicts a workload consisting of policy insertions and queries, where each inserted policy may apply to an immediately following query. For example, policies $\vPolicy{}{1}$ and $\vPolicy{}{2}$ are specified for a querier who later submits $\vQuery{}{1}$, generating $\vGuard{}{1}$. Similarly, subsequent queries ($\vQuery{}{2}$,$\vQuery{}{3}$) correspond to different policies, resulting in additional GEs ($\vGuard{}{2}$, $\vGuard{}{3}$), which are stored in the cache. Each time a GE is retrieved, its \textit{use-bit} is set to 1, indicating recent use.

As new queries arrive, the system stores generated GEs in the cache until they reach capacity. At this point, the replacement policy is triggered. As illustrated in Figure~\ref{fig:replacement}, \oursystem employs a Clock-based eviction strategy to determine which GE should be replaced. The algorithm scans the cache entries circularly, checking the \textit{use-bit} of each GE. If the \textit{use-bit} is 0, the entry is evicted and replaced with a new GE. If the \textit{use-bit} is 1, it is reset to 0, and the scan continues until an entry with a 0-bit is found. This approach ensures that frequently used GEs remain in the cache while replacing less relevant ones.


We evaluate our replacement strategy with different workload configurations, varying cache sizes and query-to-policy ratios, as described in Section~\ref{sect:expCaching}. Specifically, we analyze how smaller cache sizes impact the miss rate compared to larger ones. A limited cache capacity leads to higher misses due to reduced storage, potentially degrading performance. Conversely, a larger cache accommodates more GEs, improving hit rates and query response times. These experiments assess the effectiveness of our caching mechanism in optimizing enforcement costs under different workload dynamics.

\subsection{Refresh Strategy}\label{sec:refresh-strategy}

When new policies are inserted for a querier after their corresponding GE has already been cached, a refresh strategy is needed to ensure the cached entry remains up-to-date. In Figure~\ref{fig:refresh}, we illustrate this process. Suppose a new policy, $\vPolicy{}{6}$, is added for a querier whose query $\vQuery{}{1}$ previously generated a cached GE, $\vGuard{}{1} = \{\vPolicy{}{1}, \vPolicy{}{2}\}$. Since $\vGuard{}{1}$ does not include $\vPolicy{}{6}$, it is now considered outdated. To maintain query correctness while minimizing recomputation costs, our caching approach offers two possible refresh strategies:

\begin{itemize}
    \item \textbf{Regenerate:} A completely new GE is created, incorporating all relevant policies:
    \begin{equation}
        \vGuard{new}{1} = \{\vPolicy{}{1}, \vPolicy{}{2}, \vPolicy{}{6}\}
    \end{equation}
    The outdated $\vGuard{}{1}$ is fully replaced with $\vGuard{new}{1}$ in the cache. This approach ensures that future queries operate on a fresh, optimized GE without fragmentation.

    \item \textbf{Update:} Instead of regenerating from scratch, we generate an incremental GE:
    \begin{equation}
        \vGuard{new}{} = \{\vPolicy{}{6}\}
    \end{equation}
    The old GE is then merged with $\vPolicy{new}{}$:
    \begin{equation}
        \vGuard{}{1} = \vGuard{}{1} + \vGuard{new}{} = \{\vPolicy{}{1}, \vPolicy{}{2}, \vPolicy{}{6}\}
    \end{equation}
    This strategy minimizes recomputation by appending only the new policies to the cached entry.
\end{itemize}

To decide between regeneration and updating, we assess the mergeability of the newly inserted policy with the existing cached GE. Mergeability determines whether the guard in the cached GE overlaps with the predicates in the new policy. Specifically, if the guard in $\vGuard{}{1}$ is already present in $\vPolicy{}{6}$, we can regenerate the GE while ensuring it accurately represents the latest access control conditions. This avoids excessive fragmentation and results in a more compact, optimized GE.

Mergeable guards enable the system to extend cached entries efficiently without introducing redundancy, thereby reducing query execution overhead. When guard conditions are disjoint, updating the cached entry may not be the most effective strategy, and complete regeneration can become a more suitable option. While regeneration incurs a higher initial cost, it simplifies policy evaluation by ensuring that queries operate on an optimized, non-fragmented GE. Conversely, updating is computationally cheaper but may result in longer and more complex guards, which can degrade query performance due to the increased number of predicates.

In our experiments in Section~\ref{sect:expCaching}, we analyze the trade-offs between regeneration and updating to determine which strategy performs better under different workload conditions. By evaluating various cache sizes and policy update frequencies, we quantify the impact of mergeability on query response times and overall system efficiency. The choice between the two refresh strategies depends on how frequently policies change and whether merging prevents excessive guard growth or necessitates a fresh regeneration.

Algorithm~\ref{alg:caching} keeps the cache up to date by efficiently managing the retrieval and insertion of GEs. The cache is refreshed only when a GE is accessed—i.e., when a query invokes it. Given a query $\vQuery{}{}$, the algorithm extracts the querier identity from $\vQueryMetadata{}{querier}$ and checks for a corresponding GE in the cache. If found, \oursystem verifies whether any new policies have been defined for the querier since the GE’s last update. If not, the GE is reused as-is, and its \emph{use-bit} is set to 1 to indicate recent access. This use-bit is solely for the replacement policy to track recency of use. If new policies exist, they are evaluated for mergeability with the cached GE. Mergeable policies trigger regeneration; otherwise, the GE is incrementally updated. In either case, the new GE replaces the old one in the cache, and the use-bit is set. If no cached GE is found, a new one is generated from all applicable policies and inserted into the cache. This process ensures the cache remains relevant, adapting to policy changes while minimizing query execution overhead.

\begin{algorithm}
\scriptsize
    \caption{Sieve with caching.}
    \label{alg:caching}
    \begin{algorithmic}[1]
        \State \textbf{Input:} Query $\vQuery{}{i}$, Guarded Expression Cache $GCacheMap$
        \State \textbf{Output:} Guarded Expression $\vGuard{}{i}$

        \State $\vQueryMetadata{i}{querier} \gets \text{getQuerier}(\vQuery{}{i})$
        
        \If {$\vQueryMetadata{i}{querier}$ exists in $GCacheMap$}
            \State $\vGuard{}{i} \gets GCacheMap[\vQueryMetadata{i}{querier}]$
            \State $timestamp_{\vGuard{}{i}} \gets \text{getTimestamp}(\vGuard{}{i})$
            \State $\vPolicySet{\vQueryMetadata{i}{}}^{new} \gets \text{fetchPolicies}(\vQueryMetadata{i}{querier}, timestamp_{\vGuard{}{i}})$

            \If {$\vPolicySet{\vQueryMetadata{i}{}}^{new} = \varnothing$}
                \State \text{setUseBit}($GCacheMap$, $\vQueryMetadata{i}{querier}$)
                \State \Return $\vGuard{}{i}$
            \Else
                \If {\text{mergeable}($\vGuard{}{i}$, $\vPolicySet{\vQueryMetadata{i}{}}^{new}$)}
                    \State $\vPolicySet{\vQueryMetadata{i}{querier}} \gets \text{fetchAllPolicies}(\vQueryMetadata{i}{querier}, 0])$
                    \State $\vGuard{new}{i} \gets \text{regenerateG}(\vQueryMetadata{i}{querier}, \vPolicySet{\vQueryMetadata{i}{querier}})$
                \Else
                    \State $\vGuard{new}{i} \gets \text{updateG}(\vGuard{}{i}, \vPolicySet{\vQueryMetadata{i}{}}^{new})$
                \EndIf
                \State \text{replaceAndSet}($GCacheMap$, $\vGuard{new}{i}$)
                \State \Return $\vGuard{new}{i}$
            \EndIf
        \Else
            \State $\vPolicySet{\vQueryMetadata{i}{}} \gets \text{fetchPolicies}(\vQueryMetadata{i}{querier}, 0)$
            \State $\vGuard{new}{i} \gets \text{generateG}(\vQueryMetadata{i}{querier}, \vPolicySet{\vQueryMetadata{i}{querier}})$
            \State \text{addToCache}($GCacheMap$, $\vGuard{new}{i}$)
            \State \Return $\vGuard{new}{i}$
        \EndIf
    \end{algorithmic}
\end{algorithm}

\vspace{-0.6cm}
\section{Implementing \oursystem}
\label{sect:implementingSieve}

\oursystem is a general-purpose middleware that intercepts queries posed to a database, optimally rewrites them, and submits the queries to the underlying database on which it is layered for execution. \oursystem rewrites queries such that the rewritten queries can be executed efficiently to produce query results that are compliant with the policies. \oursystem's rewriting is based on: (a) decreasing the policies that have to be check per tuple and (b) reducing the number of tuples that have to be checked against policy expressions. In implementing this, \oursystem exploits the extensibility options of databases such as support for UDFs and index usage hints. The implementation of \oursystem with connectors for both MySQL and PostgreSQL is available at \url{https://github.com/DIPrLab/Sieve}.

\oursystem is a general-purpose middleware that intercepts queries posed to a database, optimally rewrites them, and submits the queries to the underlying database on which it is layered for execution. \oursystem rewrites queries such that the rewritten queries can be executed efficiently to produce query results that are compliant with the policies. \oursystem's rewriting is based on: (a) decreasing the policies that have to be checked per tuple and (b) reducing the number of tuples that have to be checked against policy expressions. In implementing this, \oursystem exploits the extensibility options of databases such as support for UDFs and index usage hints. The implementation of \oursystem with connectors for both MySQL and PostgreSQL is available at \url{https://github.com/DIPrLab/Sieve}. \textit{For brevity, detailed implementation aspects of the $\Delta$ operator and its integration with guards have been moved to the Appendix.}

\ifextended
\subsection{Persistence of Policies and Guards}
To store policies associated with all the relations in the database, \oursystem uses two additional relations, the policy table (referred to as $\vRelation{}{P}$), which stores the set of policies, and the object conditions table (referred to as $\vRelation{}{OC}$), which stores conditions associated with the policies. The structure of $\vRelation{}{P}$ corresponds to |$\langle$id, owner, querier, $\texttt{associated-table}$, purpose, $\texttt{action}$, $\texttt{ts-inserted-at}$$\rangle$|, where |$\texttt{associated-table}$| is the relation $\vRelation{}{i}$ for which the policy is defined and |$\texttt{ts-inserted-at}$| is the timestamp at policy insertion. The schema of $\vRelation{}{OC}$ corresponds to |$\langle$policy-id, attr, op, val$\rangle$| where |policy-id| is a foreign key to $\vRelation{}{P}$ and the rest of attributes represent the condition $\vObjectCondition{l}{c}$=\vTupleExpression{$attr$}{$op$}{$val$}. 
We emphasize that the value |val| in $\vRelation{}{OC}$ might correspond to a complex SQL condition in case of nested policies. For instance, the two sample policies defined in Section~\ref{sect:modeling} regulate access to student connectivity data for Prof. Smith; they are persisted as tuples |$\langle$1, John, Prof.Smith, $\texttt{WiFiDataset}$, Attendance Control, Allow, 2020-01-01 00:00:01$\rangle$| and |$\langle$2, John, Prof.Smith, $\texttt{WiFiDataset}$, Attendance Control, Allow, 2020-01-01 00:00:01$\rangle$| in $\vRelation{}{P}$ and with the tuples |$\langle$1, 1, wifiAP, $=$, 1200$\rangle$|, |$\langle$2, 1, $\texttt{ts-time}$, $\geq$, 09:00$\rangle$|, |$\langle$3, 1, $\texttt{ts-time}$, $\leq$, 10:00$\rangle$|, |$\langle$4, 2, wifiAP, $=$, SELECT W2.wifiAP FROM $\texttt{WiFiDataset}$ AS W2 WHERE W2.owner = "Prof.Smith" and $\texttt{W2.ts-time}$ = $\texttt{W.ts-time}$$\rangle$| in $\vRelation{}{OC}$.

A guarded policy expression $\vPolicyGuardedExpression{\vPolicySet{}}$ generated, per user and purpose, is stored in $\vRelation{}{GE}$ with the schema |$\langle$id, querier, $\texttt{associated-table}$, purpose, $\texttt{action}$, $\texttt{outdated}$, $\texttt{ts-inserted-at}$$\rangle$|. Guarded policy expressions are not continuously updated based on incoming policies as this would be unnecessary if their specific queriers do not pose any query. We use the |$\texttt{outdated}$| attribute, which is a boolean flag, to describe whether the guarded expression includes all the policies belonging to the querier. If at query time, the |$\texttt{outdated}$| attribute associated to the guarded policy expression for the specific querier/purpose (as specified in the query metadata $\vQueryMetadata{i}{querier}, \vQueryMetadata{i}{purpose}$) is found to be true, then that guarded policy expression is regenerated. After the guarded expression is regenerated for a querier, it is stored in the table with |outdated| set to false. 
Guard regeneration comes with an overhead. However, in our experience, the corresponding overhead is much less than the execution cost of queries. As a result, we generate guards during query execution using triggers in case the current guards are outdated. 
Guarded expressions $\vGuard{}{i}$ associated with a guarded policy expression $\vPolicyGuardedExpression{}$ are stored in two relations:  $\vRelation{}{GG}$=|$\langle$id, guard-expression-id, attr, op, val$\rangle$| to store the guard (i.e., $\vObjectCondition{i}{g}$=\vTupleExpression{$attr$}{$op$}{$val$}) and $\vRelation{}{GP}$=|$\langle$guard-id, policy-id$\rangle$| to store the policy partition (i.e., $\vPolicySet{\vGuard{}{i}}$).
\fi

\subsection{Implementing Policy Guarded Expression}
\label{sect:implementingGuards}
 
Our goal is to evaluate policies for query $\vQuery{}{i}$ by replacing any relation $\vRelation{}{j} \in \vQuery{}{i}$ by a projection of $\vRelation{}{j}$ that satisfies the guarded policy expression $\vPolicyGuardedExpression{\vPolicySet{\vRelation{}{j}}}$ where $\vPolicySet{\vRelation{}{j}}$ is the set of policies defined for the specific querier, purpose, and relation. To this end, we first use the |WITH| clause for each relation $\vRelation{}{j} \in \vQuery{}{i}$ that selects tuples in $\vRelation{}{j}$  satisfying the guarded policy expression\footnote{Using the above strategy the policy check needs to be only done once in the WITH clause even if the relation appears multiple times in the query.}. The rewritten query replaces every occurrence of $\vRelation{}{j}$ with the corresponding $\hat{\vRelation{}{j}}$. 

\begin{lstlisting}[style=mystyle]
  WITH $\hat{\vRelation{}{j}}$ AS (
    SELECT * FROM $\vRelation{}{j}$ WHERE $\vGuard{}{1}$ OR $\vGuard{}{2}$ OR $\cdots$ OR $\vGuard{}{n}$)
\end{lstlisting}

\oursystem utilizes extensibility features (e.g., index usage hints\footnote{\url{https://dev.mysql.com/doc/refman/8.0/en/index-hints.html}}, optimizer explain\footnote{\url{https://www.postgresql.org/docs/13/sql-explain.html}}, UDFs) offered by DBMSs that allows it to suggest index plans to the underlying optimizer. Since such features vary across DBMSs, guiding optimizers requires a platform dependent connector that can rewrite the query appropriately. In systems such as MySQL, Oracle, DB2, and SQL Server that support index usage hints, \oursystem can rewrite the query to explicitly force indexes on guards. For example, in MySQL using |FORCE INDEX| hints, which tell the optimizer that a table scan is very expensive and should only be used if the DBMS cannot use the suggested index to find rows in the table, the rewritten query will be as follows:

\begin{lstlisting}[style=mystyle]
  WITH $\hat{\vRelation{}{j}}$ AS (
    SELECT * FROM $\vRelation{}{j}$ [FORCE INDEX ($\vObjectCondition{1}{g}$)] WHERE $\vGuard{}{1}$ UNION 
    SELECT * FROM $\vRelation{}{j}$ [FORCE INDEX ($\vObjectCondition{2}{g}$)] WHERE $\vGuard{}{2}$ UNION$\cdots$ 
    SELECT * FROM $\vRelation{}{j}$ [FORCE INDEX ($\vObjectCondition{n}{g}$)] WHERE $\vGuard{}{n}$)
\end{lstlisting}

Some systems, like PostgreSQL, do not support index hints explicitly. In such cases, \oursystem still does the above rewrite but depends upon the underlying optimizer to select appropriate indexes.

\subsection{Exploiting Selection Predicates in Queries}
\label{sect:exploitingQuery}

So far, we only considered exploiting guarded policy expressions to optimize the overhead of policy checks while executing a query $\vQuery{}{i}$. 
We could further exploit selection predicates defined over relation $\vRelation{}{j}$ that appears in $\vQuery{}{i}$, especially if such predicates are highly selective, in reducing the cost of policy checking. The rewrite strategy discussed above, that is used to replace $\vRelation{}{j}$ into $\hat{\vRelation{}{j}}$, can be modified to include such selective query predicates in addition to the guarded policy expression.
Such a modification provides the optimizer with a choice on whether to use the index on the guards or to use the query predicate to filter the tuples in the relation on which we apply the policy checks. 

Instead of relying solely on the optimizer to select an execution plan\footnote{Optimizers might choose suboptimal plans when query predicates are as complex as the guarded policy expressions.}, \oursystem provides a cost-based hint to guide the optimizer by evaluating alternative execution strategies. In particular, \oursystem considers the following three options:

\begin{enumerate}
    \item $LinearScan$: A full scan of the relation, followed by the evaluation of the guarded policy expression.
    \item $IndexQuery$: An index scan based on the query predicate, followed by evaluation of the guarded policy expression.
    \item $IndexGuards$: An index scan based on the guards, followed by evaluation of the resulting policy partitions.
\end{enumerate}

\noindent
In each of these strategies, guarded expressions are used to generate the filtered relation $\hat{\vRelation{}{j}}$, although the underlying access methods differ.

To determine cost of each strategy, \oursystem first runs the |EXPLAIN| of query $\vQuery{}{i}$ which returns a high-level view of the query plan including, usually, for each relation in the query the particular access strategy (table scan or a specific index) the optimizer plans to use and estimated selectivity of the predicate on that attribute ($p$). Then, \oursystem estimates an upper bound of the cost for each strategy focusing on the cost of accessing data. $\vCostMethod{IndexGuards}$ is computed as $\sum_{\vGuard{}{i} \in \vGuard{}{}}\vSelectivityMethod{\vGuard{}{i}}.\vReadCost$, where $\vSelectivityMethod{\vGuard{}{i}}$ is the cardinality of the guard $\vObjectCondition{1}{g}$. If the optimizer selects to perform index scan on a query predicate $p$, then $\vCostMethod{IndexQuery} = \vSelectivityMethod{p}.\vReadCost$ otherwise $\vCostMethod{IndexQuery}=\infty$.   
\oursystem chooses between $IndexGuards$ and $IndexQuery$ based on which strategy is less costly. It then compares the better of the two strategies to $LinearScan$ choosing the latter if the random access due to index scan is expected to be more costly than the sequential access of linear of scan.
To implement the selected strategy, \oursystem rewrites the query (including the appropriate |WITH| clause(s) as explained in Section~\ref{sect:implementingGuards}) to append: 
An index hint (e.g., |FORCE INDEX| statement in MySQL) for each guard $\vGuard{}{i}$ to the |FROM| clause within the |WITH| clause as we showed previously (in the case of $IndexGuards$ strategy); or an index hint for the attribute of $p$ (for $IndexQuery$); or a hint to suggest the optimizer to igonore all indexes (e.g., |USE INDEX()| in MySQL) (for $LinearScan$). 

\subsection{Sample Query Rewriting in \oursystem}

Let us consider the query in Section~\ref{sect:caseStudy} to study the tradeoff between student performance and attendance to classes. In that case, \oursystem might rewrite the query as follows depending on the available policies and DBMS:
\begin{lstlisting}[style=mystyle]
WITH WiFiDatasetPol AS (
 SELECT * FROM WiFiDataset as W FORCE INDEX($\vObjectCondition{1}{g} \cdots \vObjectCondition{n}{g}$)
 WHERE ($\vObjectCondition{1}{g}$ AND W.ts-date between "9/25/19" AND "12/12/19" AND ($\vObjectCondition{1}{1}$ AND $\cdots$ AND $\vObjectCondition{1}{n}$))
        OR $\cdots$ OR 
       ($\vObjectCondition{n}{g}$ AND W.ts-date between "9/25/19" AND "12/12/19" AND delta(32,"Prof.Smith", "Analysis","owner","ts-date", "ts-time", "wifiAP")=true)
) StudentPerf(WifiDatasetPol, Enrollment, Grades)
 \end{lstlisting}
 

As the query has only one table with associated policies in its |FROM| clause (i.e., |$\texttt{WiFiDataset}$| table), the rewritten query contains one |WITH| clause, generated as explained in Section~\ref{sect:implementingGuards}. This clause creates |$\texttt{WiFiDatasetPol}$|, which is now used in the original query to replace the |$\texttt{WiFiDataset}$| table. The |WITH| clause includes the set of guards generated for the querier (``Prof. Smith'') and his purpose (``Analysis'') given the policies in the database. The query predicate on date (|$\texttt{ts-date}$ between "9/25/19" AND "12/12/19"|) was included along with each guard as outlined in Section~\ref{sect:exploitingQuery}. As \oursystem selected the $IndexGuards$ strategy, the |WITH| clause forces the usage of guards as indexes (through the |FORCE INDEX| command) as explained in Section~\ref{sect:exploitingQuery}. Finally, for one specific guarded expression ($\vGuard{}{n}$), \oursystem selected the $guard+\Delta$ strategy, where the policy partition is replaced by a call to the UDF that implements the $\Delta$ operator. The general UDF described was extended to retrieve only the relevant policies associated with that guard. Further details on the $guard+\Delta$ strategy and its implementation are provided in the Appendix.

\section{Experimental Evaluation} \label{sect:expSieve}

In this section, we conduct a detailed experimental evaluation of \oursystem in different settings. First, we measure \oursystem's query evaluation performance in static settings, focusing on its efficiency in enforcing FGAC policies in section~\ref{sec:query-eval-perf}. Second, we assess the performance of \oursystem enhanced with caching in dynamic settings by testing its ability to handle diverse workloads of FGAC policies and queries in section~\ref{sect:expCaching}. In each subsection, we begin by detailing our experimental setup, which includes the dataset, workload including FGAC policy and query templates, and DBMS configuration.


\subsection{\oursystem Performance in Static Settings}
\label{sec:query-eval-perf}



We evaluate \oursystem performance with a static workload to assess its efficiency in enforcing FGAC polices. \oursystem optimizes query execution by integrating guarded expressions, reducing overhead compared to traditional query rewriting methods. To understand its effectiveness, we conduct a series of experiments to evaluate the effectiveness of \oursystem in static settings by answering the following questions:
\begin{itemize}
    \item What is the cost of generating a guarded expression? \textit{(addressed in Experiment~1)}
    \item How efficient is \oursystem in real-time query evaluation in static settings? \textit{(addressed in Experiment~2)}
    \item How well does \oursystem generalize across database systems? \textit{(addressed in Experiment~3)}
    \item How does Sieve scale with very large policy workloads? \textit{(addressed in Experiment~4)}
\end{itemize}

We determine the best configurations for parameter-related experiments for optimizing guarded expression generation and policy enforcement. We analyze Sieve’s impact on query evaluation across different workloads and database environments for system-level experiments. In addition, we include a separate experiment in the Appendix that evaluates design trade-offs in applying the $\Delta$ operator and choosing between query- and guard-based indexing strategies. This structured evaluation clearly explains Sieve’s performance under various static conditions.

\subsubsection*{Experimental settings:}
\textbf{Datasets:} 
We used the {\em TIPPERS} dataset~\cite{mehrotra2016tippers} consisting of connectivity logs generated by the 64 WiFi Access Points (APs) at the Computer Science building at UC Irvine for three months. These logs are generated when a WiFi-enabled device (e.g., a smartphone or tablet) connects to one of the WiFi APs and contain the hashed identification of the device's MAC, the AP's MAC, and a timestamp. The dataset comprises 3.9M~events corresponding to 36K~different devices (the signal of some of the WiFi APs bleeds outside the building, and passerby devices/people are also observed). This information can be used to derive the occupancy levels in different parts of the building and provide diverse location-based services (see Section~\ref{sect:caseStudy}) since device MACs can be used to identify individuals. Since location information is privacy-sensitive, limiting access to this data based on individuals' preferences is essential. 
\ifextended
Table~\ref{tippers-schema} shows the schema of the different tables in the {\em TIPPERS} dataset. \texttt{WiFi\_Dataset} stores the logs generated at each \texttt{WiFi\_AP} when the devices of a \texttt{User} connects to them. \texttt{User\_Group} and \texttt{User\_Group\_Membership} keeps track of the groups and their members respectively.

\begin{table}[!htb]
\scriptsize
\centering
\caption{TIPPERS data schema.} \label{tippers-schema}
\begin{tabular}{l l l}  
      Table           & Columns  &  Data type \\ \hline
\multirow{3}{*}{Users} & id & int  \\  \cline{2-3} 
                  & device & varchar \\  \cline{2-3} 
                  & office &  int \\ \hline
\multirow{3}{*}{User\_Groups} & id  & int \\ \cline{2-3} 
                  & name & varchar  \\ \cline{2-3} 
                  & owner & varchar  \\ \hline
\multirow{2}{*}{User\_Group\_Membership} & user\_group\_id &  int \\ \cline{2-3} 
                  & user\_id & int  \\ \hline
\multirow{3}{*}{Location} & id  & int  \\ \cline{2-3} 
                  & name & varchar  \\ \cline{2-3} 
                  & type & varchar \\  \hline
\multirow{5}{*}{WiFi\_Dataset} & id  & int  \\ \cline{2-3} 
                  & wifiAP & int  \\ \cline{2-3} 
                  & owner  & int \\ \cline{2-3} 
                  & ts-time & time \\ \cline{2-3} 
                  & ts-date & date
\end{tabular}
\end{table}

\fi

We also used a synthetic dataset containing WiFi connectivity events in a shopping mall for scalability experiments with an even larger number of policies. We refer to this dataset as {\em Mall}. We generated the {\em Mall} dataset using the IoT data generation tool in~\cite{iotbenchmark} to generate synthetic trajectories of people in a space (we used the floorplan of a mall extracted from the Web) and sensor data based on those. The dataset contains 1.7M~events from 2,651 different devices representing customers. 
\ifextended
Table~\ref{mall-schema} shows the schema of the tables in the {\em Mall} dataset.  

\begin{table}[!htb]
\scriptsize
\centering
\caption{Mall data schema.} \label{mall-schema}
\begin{tabular}{l l l}  
      Table           & Columns  &  Data type \\ \hline
\multirow{3}{*}{Users} & id & int  \\  \cline{2-3} 
                  & device & varchar \\  \cline{2-3} 
                  & interest &  varchar \\ \hline
\multirow{3}{*}{Shop} & id  & int  \\ \cline{2-3} 
                  & name & varchar  \\ \cline{2-3} 
                  & type & varchar \\  \hline
\multirow{5}{*}{WiFi\_Connectivity} & id  & int  \\ \cline{2-3} 
                  & shop\_id & int  \\ \cline{2-3} 
                  & owner  & int \\ \cline{2-3} 
                  & obs\_time & time \\ \cline{2-3} 
                  & obs\_date & date
\end{tabular}
\end{table}

\fi

\textbf{Queries:} We used a set of query templates based on the recent IoT SmartBench benchmark~\cite{iotbenchmark}, which includes a mix of analytical and real-time tasks and target queries about (group of) individuals. 
Specifically, query templates $Q_1$ - Retrieve the devices connected for a list of locations during a time period (e.g., for location surveillance); $Q_2$ - Retrieve devices connected for a list of given MAC addresses during a time period (e.g., for device surveillance); $Q_3$ - Number of devices from a given group or profile of users in a given location (e.g., for analytic purposes). 
\ifextended
The SQL version of the queries is thus:
\begin{lstlisting}
Q1=(SELECT * FROM WiFi_Dataset AS W 
    WHERE W.wifiAP IN ($[ap]$) W.ts-time AND BETWEEN $t1$ AND $t2$
    AND W.ts-date BETWEEN $d1$ AND $d2$)
Q2=(SELECT * FROM WiFi_Dataset AS W 
    WHERE W.owner in ($[devices]$) 
    AND W.ts-time BETWEEN $t1$ AND $t2$ 
    AND W.ts-date BETWEEN $d1$ AND $d2$)
Q3=(SELECT * FROM WiFi_Dataset AS W, 
    User_Group_Membership AS UG 
    WHERE UG.user_group_id = $group\_id$ 
    AND UG.user_id = W.owner 
    AND W.ts-time BETWEEN $t1$ AND $t2$ 
    AND W.ts-date BETWEEN $d1$ AND $d2$)
\end{lstlisting}
\fi

Based on these templates, we generated queries at three different selectivities (low, medium, high) by modifying configuration parameters (locations, users, time periods).
Below, when we refer to a particular query type (i.e., $Q_1$, $Q_2$, or $Q_3$), we will refer to the set of queries generated for such type.

\textbf{Policy Generation:} The {\em TIPPERS} dataset, collected for a limited duration with special permission from the University for the purpose of research, does not include user-defined policies. We, therefore, generated a set of synthetic policies. As part of the {\em TIPPERS} project, we conducted several town hall meetings and online surveys to understand the privacy preferences of users about sharing their WiFi-based location data. 
The surveys, as well as prior research~\cite{DBLP:conf/percom/0001K17,lin2014privacy}, indicate that users express their privacy preferences based on different user profiles (e.g., students or faculty) or groups (e.g., my coworkers, classmates, friends, etc.). Thus, we used a profile-based approach to generate policies specifying which events belonging to an individual can be accessed by a given querier (based on their profile) for a specific purpose in a given context (e.g., location, time).

We classified devices in the {\em TIPPERS} dataset as belonging to users with different profiles (denoted by $\vProfileMethod{\vUser{}{k}}$ for User~$\vUser{}{k}$) based on the total time spent in the building and connectivity patterns. Devices that rarely connect to APs in the building (i.e., less than~5\% of the days) are classified as \textit{visitors}. The non-visitor devices are then classified based on the type of rooms they spent most time in: \textit{staff} (staff offices), \textit{undergraduate students} (classrooms), \textit{graduate students} (labs), and \textit{faculty} (professor offices). As a result, we classified 31,796 visitors, 1,029 staff, 388 faculty, 1,795 undergraduate, and 1,428 graduate students from a total of 36,436 unique devices in the dataset. Our classification is consistent with the expected numbers for the population of the monitored building. We also grouped users into groups based on the affinity of their devices to rooms in the building, which is defined in terms of time spent in each region per day. Thus, each device is assigned to a group with maximum affinity. We generated 56 groups with an average of 108 devices per group. 

We define two kinds of policies based on whether they are for an unconcerned or advanced user as described in Section~\ref{sect:caseStudy}. Unconcerned users subscribe to the default policies set by the administrator, which allow access to their data based on user groups and profiles. Given the schema in Table~\ref{tippers-schema} and the unconcerned user $\vUser{}{k}$ we generate the following default policies:

\ifextended
\squishlist
    \item Data associated with $\vUser{}{k}$ collected during working hours can be accessed by members of $\vGroupMethod{\vUser{}{k}}$.
    \item Data associated with $\vUser{}{k}$ collected at any time can be accessed by overlapping members of $\vGroupMethod{\vUser{}{k}}$ and $\vProfileMethod{\vUser{}{k}}$. 
\squishend

On the other hand, an advanced user defines, on average, 40 policies, given the large number of control options (such as device, time, groups, profiles, and locations) in our setting. The policy dataset generated contains 869,470 policies, each defining 472 policies on average and appearing as a querier in 188 policies defined by others on average. The above policies are defined to allow access to data in different situations. Any other access that is not captured by the previous policies will be denied (based on the default opt-out semantics defined in Section~\ref{sect:modeling}). 

\ifextended

Table~\ref{policy-table} shows the schema and several sample policies generated for three different queriers. The first sample policy is generated for \textit{Prof. John Smith} to mark the \textit{Attendance}, and the policy is defined for user $120$. \textit{Prof. John Smith} can access the data only when the time falls in the $09:00:00\leq ts-time \leq 10:00:00$ when queried for device wifiAP is $1200$. For brevity, the \textit{inserted\_at} and \textit{action} columns are skipped. Table~\ref{oc-table} shows the corresponding object conditions which are part of two policies.

\begin{table}[!htb]
\scriptsize
\centering
\caption{Policy Table} \label{policy-table}
 \begin{tabular}{c l l l} 
 id & table & querier & purpose \\ [0.5ex] 
 \hline
 1 & WiFi\_Dataset & Prof.John Smith & Attendance   \\ 
 \hline
 2 & WiFi\_Dataset & Bob Belcher & Lunch Group  \\ 
 \hline
 3 & WiFi\_Dataset & Prof.John Smith & Attendance  \\ 
 \hline
 4 & WiFi\_Dataset & Liz Lemon & Project Group  \\ 
 \hline
 5 & WiFi\_Dataset & Prof.John Smith & Attendance 
\end{tabular}
\end{table}

\begin{table}[!htb]
\scriptsize
\centering
\caption{Policy Object Conditions Table} \label{oc-table}
 \begin{tabular}{c c l l c r} 
 id & policy\_id & attr\_type & attr & op & val \\ [0.5ex] 
 \hline
 1 & 1 & int & owner & $=$ & 120   \\ 
 \hline
 2 & 1 & time & ts-time & $\geq$ & 09:00:00     \\ 
 \hline
 3 & 1 & time & ts-time & $\leq$ & 10:00:00     \\ 
 \hline
 4 & 1 & int & wifiAP & $=$ & 1200     \\ 
 \hline
 5 & 2 & int & owner &  $=$ & 145  \\ 
 \hline
 6 & 2 & int & wifiAP & $=$ & 2300
\end{tabular}
\end{table}

\fi

For the \textit{Mall} dataset, the shops were categorized into six types based on the services they provide(e.g., arcade, movies). We also classified customers into frequent and occasional based on their shop visits. We then defined two types of policies for each customer, depending on the prior classification. Frequent customers allowed shops they visit the most to have access to their location during open hours. On the other hand, occasional customers shared their data only with specific shop types depending on if there were sales or discounts. Finally, if a customer expressed an interest in a particular shop category, we also generated policies which allowed access of their data to the shops in the category for a short period of time (e.g., lightning sales). This policy dataset generated on top of the {\em Mall} dataset contains 19,364 policies defined for 35 shops (queriers) in the mall, with 551 policies on average per shop.

\textbf{Database Systems:}  We ran the experiments on an individual machine (CentOS 7.6, Intel(R) Xeon(R) CPU E5-4640, 2799.902 Mhz, 20480 KB cache size) in a cluster with a shared total memory of 132 GB. 
We performed experiments on MySQL 8.0.3 with InnoDB, an open-source DBMS supporting index usage hints. We configured the buffer\_pool\_size to 4 GB. We also performed experiments on PostgreSQL 13.0 with shared\_buffers configured to 4 GB.

\vspace{0.1cm}
\subsubsection*{\textbf{Experiment~1:}~Cost~for generating Guarded Expressions and Effectiveness}

This experiment aims to study the cost of generating guarded expressions for a querier as the factor of the number of policies and the quality of generated guards. For analyzing the cost of guarded expression generation, we generate guarded expressions for all the users using the algorithm described in Section~\ref{sect:guardSel} and collect the generation times in a set. We sort these costs (in milliseconds) of generating guarded expressions for different users and average the value for every group of 50 users, showing the result in Figure~\ref{fig:guardGenerationTest}. 

\setlength{\intextsep}{2pt}%
\setlength{\columnsep}{12pt}%
\begin{wrapfigure}{R}{0.25\textwidth}
\centering
\hspace{6mm}
	\includegraphics[width=0.25\textwidth]{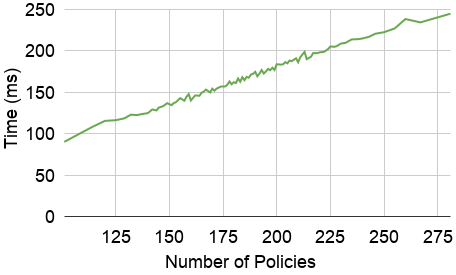}
\caption{Guard generation time.}
\label{fig:guardGenerationTest}
\end{wrapfigure}

The cost of guard generation increases linearly with the number of policies. 
As guarded expression generation also depends on the selectivity of policies and the number of candidate guards generated, which is also a factor of overlap between predicates, we sometimes observe a slight decrease in the time taken with increasing policies. The overhead of generating guarded expressions is minimal in static settings; for example, generating a guard for a querier with 160 associated policies (e.g., the student trying to locate classmates explained in Section~\ref{sect:caseStudy}) takes around 150ms. However, this regeneration cost can become a bottleneck in dynamic settings where policies are frequently updated and queries are repeatedly issued. 

\begin{center}
\begin{minipage}{\linewidth}
    \begin{minipage}[b]{0.54\linewidth}
        \vspace{0pt}
        \centering
        \scriptsize
        \begin{tabular}{lllll}
            & min  & avg  & max  & SD  \\ \hline
            $\vSetCardinality{\vPolicy{}{\vUser{}{k}}}$ & 31 & 187 & 359 & 38 \\\hline
            $\vSetCardinality{\vGuard{}{}}$ & 2 & 31 & 60 & 10  \\\hline
            $\vSetCardinality{\vPolicy{}{\vGuard{}{i}}}$ & 4  & 7 & 60 & 5  \\\hline
            $\vSelectivityMethod{\vGuard{}{i}}$ & 0.01\%  & 3\% & 24\%  & 2\%  \\\hline
            Savings  & 0.99  & 0.99 & 1  & $7\mathrm{e}{-4}$ \\ \\
        \end{tabular}
        \captionof{table}{Analysis of policies and generated guards.}
        \label{tab:guardAnalysis}
    \end{minipage}
    \hfill
    \begin{minipage}[b]{0.4\linewidth}
        \vspace{0pt}
        \centering
        \scriptsize
        \begin{tabular}{lrr}
            \multirow{2}{*}{$\vSelectivityMethod{\vGuard{}{}}$} & \multicolumn{2}{c}{$\vSetCardinality{\vGuard{}{}}$} \\
            \cmidrule(lr){2-3} & low   & high    \\ \hline
            low                & 227.2 & 537.0   \\ \hline
            high               & 469.0 & 1,406.7 \\ \hline \\ \\
        \end{tabular}
        \captionof{table}{Analysis of number of guards and total cardinality.}
        \label{tab:guardsNumbervsCardinality}
    \end{minipage}
\end{minipage}
\end{center}

We present the results of analyzing the policies and guarded expressions in Table~\ref{tab:guardAnalysis} and the analysis of the number of guards and total cardinality in Table~\ref{tab:guardsNumbervsCardinality}. Each user on average have defined 187 policies ($\vSetCardinality{\vPolicy{}{\vUser{}{k}}}$). These policies depend on their profiles (e.g., students) and group memberships. \oursystem creates an average of 31 guards per user with the mean partition cardinality (i.e., $\vSetCardinality{\vPolicy{}{\vGuard{}{i}}}$) as 7. The total cardinality of guards in the guarded expression is low (i.e., $\vSelectivityMethod{\vGuard{}{i}}$), which helps in filtering out tuples before performing policy evaluation. In cases with high cardinality guards (e.g., maximum of 24\%), \oursystem will not use force an index scan in that particular guard as explained in Section~\ref{sect:implementingSieve}. Savings are computed as the ratio of the difference between the total number of policy evaluations with and without using the guard and the number of policy evaluations. This was computed on a smaller sample of the entire dataset, and the results show that guards help in eliminating around 99\% of the policy checks compared to policy evaluation.

\vspace{0.1cm}
\subsubsection*{\textbf{Experiment~2:}~Query Evaluation Performance} 

We compare the performance of \oursystem (implemented as detailed in Section~\ref{sect:implementingSieve}) against three different baselines. In the first baseline, {\vBaselineP}, we append the policies that apply to the querier to the |WHERE| condition of the query.
Second, {\vBaselineI} performs an index scan per policy (forced using index usage hints) and combines the results using the |UNION|. Third, {\vBaselineU} is similar to {\vBaselineP}, but instead of using the policy expression, it uses a UDF defined on the relation to evaluate the policies. The UDF takes all the attributes of the tuple as input. {\vBaselineU} significantly reduces the number of policies to be evaluated per tuple and works well for low cardinality queries. However, UDF invocations are expensive; therefore, executing the UDF as late as possible might be preferable from the optimization perspective~\cite{hellerstein1998optimization}. To preserve the correctness of policy enforcement as defined in Section~\ref{sect:modeling}, UDF operations have to be performed before any non-monotonic query operations.

For each query type ($Q1$, $Q2$, $Q_3$), we generate a workload of queries with three different selectivity classes posed by five different queriers belonging to four different profiles. The values chosen for these three selectivity classes (low, medium, high) differed depending on the query type. We execute each query along with the access control mechanism~5~times and average the execution times. The experimental results below give the average performance per query. The time-out was set at 30 seconds. If a strategy is timed out for all queries of that group, we show the value $TO$. 
If a strategy timed out for some of the queries in a group but not all, the table shows the average performance only for those queries that were executed to completion; those time values are denoted as $t^{+}$.

\begin{table}[!ht]
\caption{Overall comparison of performance for $Q1$, $Q2$, and $Q3$ (in ms).}
\label{tab:performanceAllQueries}
\centering
\scriptsize
    \begin{tabular}{lcrrrr}
         &  $\vSelectivityMethod{\vQuery{}{}}$ & \vBaselineP & \vBaselineI & \vBaselineU & \oursystem \\ \hline
        \multirow{3}{*}{$Q1$} & low  & 1,668        & 906   & 9,122        & 418 \\ 
                              & mid  & 15,356       & 910   & 23,575$^{+}$ & 453 \\ 
                              & high & TO           & 937   & TO           & 523 \\ \hline
        \multirow{3}{*}{$Q2$} & low  & 860          & 916   & 7,787        & 407 \\ 
                              & mid  & 7,191        & 922   & 22,617$^{+}$ & 454 \\ 
                              & high & 29,765$^{+}$ & 962   & TO           & 475 \\ \hline  
        \multirow{3}{*}{$Q3$} & low  & 883          & 881   & 14,379        & 477 \\ 
                              & mid  & 2,217          & 2,209   & TO & 476 \\ 
                              & high & 3,502        & 3,543 & TO           & 521 \\
    \end{tabular}
\end{table}

Table~\ref{tab:performanceAllQueries} shows the average performance for the three query types. The performance of {\vBaselineP} and {\vBaselineU} degrades with increasing cardinality of the associated query as they rely on the query predicate for reading the tuples. The relative reduction in overhead for Q3 for {\vBaselineP} at high cardinalities is because the optimizer can use the low cardinality join condition to perform a nested index loop join. \oursystem and {\vBaselineI} performance stays the same across query cardinalities as they utilize the policy and guard predicates for reading the tuples and are therefore unaffected by the query cardinality. The increase in the speedup between these two sets of approaches clearly demonstrates that exploiting indices paid off. For {\vBaselineP}, the optimizer cannot exploit indices at high cardinalities and resorts to performing linear scans. In {\vBaselineU}, the cost of UDF invocation per tuple far outweighed any benefits from filtering policies. {\vBaselineI} generated by careful rewriting with an index scan per policy performs significantly better than the previous two baselines. The performance degradation of {\vBaselineI} for Q3 is due to the optimizer preferring to perform the nested loop join first instead of the index scans. Compared to all these baselines, {\oursystem} is significantly faster at all query cardinalities.

\vspace{0.1cm}
\subsubsection*{\textbf{Experiment~3:}~\oursystem on PostgreSQL} 

In the previous experiments, we used MySQL, which supports hints for index usage, thus enabling SIEVE to explicitly force the optimizer to choose guard indexes. However, other DBMSs, such as PostgreSQL, do not support index usage hints explicitly (as discussed in Section~\ref{sect:implementingGuards}). To study \oursystem's performance in such systems, we implemented a \oursystem connector to PostgreSQL using the same rewrite strategy but without index usage hints. To have a cumulative set of policies 
for evaluation, we chose 5 queriers with at least 300 policies in the dataset. For each querier, we divided their policies into 10 different sets of an increasing number of policies, starting with the smallest set of 75 policies. The order and the specific policies in these sets were varied 3 times by random sampling. The results in Figure~\ref{fig:exp4} show the average performance of different strategies for each set size averaged across queries and the samples for SELECT ALL queries.

The four strategies tested in this experiment are a best-performing baseline for MySQL (\vBaselineI (M)) from Experiment 3, the baseline in PostgreSQL (\vBaselineP (P)), and \oursystem in both MySQL and PostgreSQL (\oursystem(M) and \oursystem(P)). The results show that not only \oursystem outperforms the baseline in PostgreSQL, but also the speedup factor w.r.t. the baseline is even higher than in MySQL. Additionally, the speedup factor in PostgreSQL is the highest at the largest number of policies. Based on our analysis of the query plan chosen by PostgreSQL, it correctly chooses the guards for performing index scans (as intended by \oursystem) even without the index usage hints. In addition, PostgreSQL supports combining multiple index scans by preparing a bitmap in memory\footnote{\url{https://www.postgresql.org/docs/12/indexes-bitmap-scans.html}}. It used these bitmaps to \textit{OR} the guards' results whenever possible, and the only resultant table rows were visited and obtained from the disk. With a larger number of guards (for the larger number of policies), PostgreSQL was also able to filter out tuples more efficiently than using the policies. Thus, {\oursystem} benefits from a reduced number of disk reads (due to bitmap) and a smaller number of evaluations against the partition of the guarded expression.

\begin{center}
\begin{minipage}{\linewidth}
    \begin{minipage}[b]{0.49\linewidth}
        \vspace{0pt}
        \centering
        \includegraphics[width=\columnwidth]{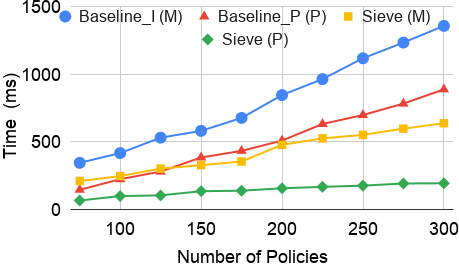}
        \captionof{figure}{\oursystem on MySQL and PostgreSQL.}
        \label{fig:exp4}
    \end{minipage}
    \hfill
    \begin{minipage}[b]{0.49\linewidth}
        \vspace{0pt}
        \centering
        \includegraphics[width=\columnwidth]{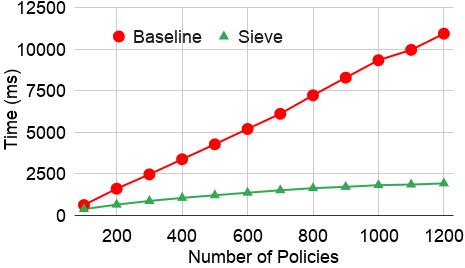}
        \captionof{figure}{Scalability comparison.}
        \label{fig:expMall}
    \end{minipage}
\end{minipage}
\end{center}

\vspace{0.1cm}
\subsubsection*{\textbf{Experiment~4:}~Scalability} 

The previous experiment shows that the speedup of \oursystem w.r.t. the baselines increases with an increasing number of policies, especially for PostgreSQL. We explore this aspect further on PostgreSQL using the {\em Mall} dataset, where generating a larger dataset of policies is possible as the underlying dataset contains a higher number of devices/customers. 
We used the same process as in Experiment 4 to generate a cumulative set of policies by choosing 5 queriers/shops with at least 1,200 policies defined for them. Figure~\ref{fig:expMall} reaffirms how the speedup of \oursystem compared against the baseline increases linearly, starting from a factor of 1.6 for 100 policies to a factor of 5.6 for 1,200 policies. We analyzed the query plan selected by the optimizer for the \oursystem rewritten queries. We observed that with a larger number of guards, PostgreSQL can utilize the bitmaps in memory to gain additional speedups from guarded expressions (as explained in Experiment 4). Also, this experiment shows that \oursystem outperforms the baseline for a different dataset, which shows the generality of our approach.

\subsection{\oursystem Performance in Dynamic Settings}
\label{sect:expCaching}

Our previous evaluation of \oursystem focused on query execution performance in static settings. In this second set of experiments, we evaluate \oursystem enhanced with caching and its ability to handle diverse FGAC policies and query workloads in dynamic environments.

We address the following research questions in our evaluation:
\begin{itemize}
    \item Which is the best replacement strategy for caching? \textit{(addressed in Experiment~5)}
    \item How does the querier's popularity affect cache efficiency? \textit{(addressed in Experiment~6)}
    \item How does window size affect cache efficiency? \textit{(addressed in Experiment~7)}
    \item Does caching improve end-to-end query processing performance? \textit{(addressed in Experiment~8)}
\end{itemize}

A realistic workload is essential for this evaluation. Therefore, we developed a workload generator that creates diverse workloads based on the TIPPERS dataset described earlier. These workloads allow us to simulate access control scenarios in dynamic IoT settings and measure the performance of \oursystem.

We classify our experiments into system-related and cache-effectiveness categories. System-related experiments assess caching’s impact on query execution and scalability. Cache-effectiveness experiments analyze configuration parameters such as replacement policy and window size. Additional experiments, including how different workloads and cache sizes affect performance, are included in the Appendix to provide a deeper view of performance under varying conditions. This structured evaluation demonstrates how caching enhances \oursystem’s ability to enforce FGAC policies in dynamic environments.

\subsubsection*{Experimental Settings:} 



\textbf{Dataset:} Our experiments utilize dynamically generated workloads that simulate real-world IoT environments. Each workload consists of policies and queries, structured to reflect access control requirements in smart-campus applications using the TIPPERS dataset~\cite{TIPPERS2016}.

Existing benchmarks, such as TPC-H, or IoT-specific ones, such as \cite{iotbenchmark}, do not include FGAC policies. These are widely used to benchmark database performance and focus primarily on evaluating query execution efficiency and throughput without addressing access control concerns. It does not incorporate any mechanisms to restrict or control data access at a granular level based on user roles, attributes, or contextual factors, which are essential in many real-world applications, especially those involving sensitive data.
While they evaluate database performance by focusing on query execution efficiency, they do not model access control constraints based on user roles, attributes, or contextual factors. To address this gap, we generate FGAC policies and queries tailored for two smart-campus scenarios:
\begin{enumerate}
    \item Marking Attendance: Faculty members query student locations to track attendance within predefined class hours.
    \item Monitoring Space Usage: Staff or faculty monitor space utilization by tracking user movements.
\end{enumerate}

\begin{figure}[!ht]
\centering
\includegraphics[width=\linewidth]{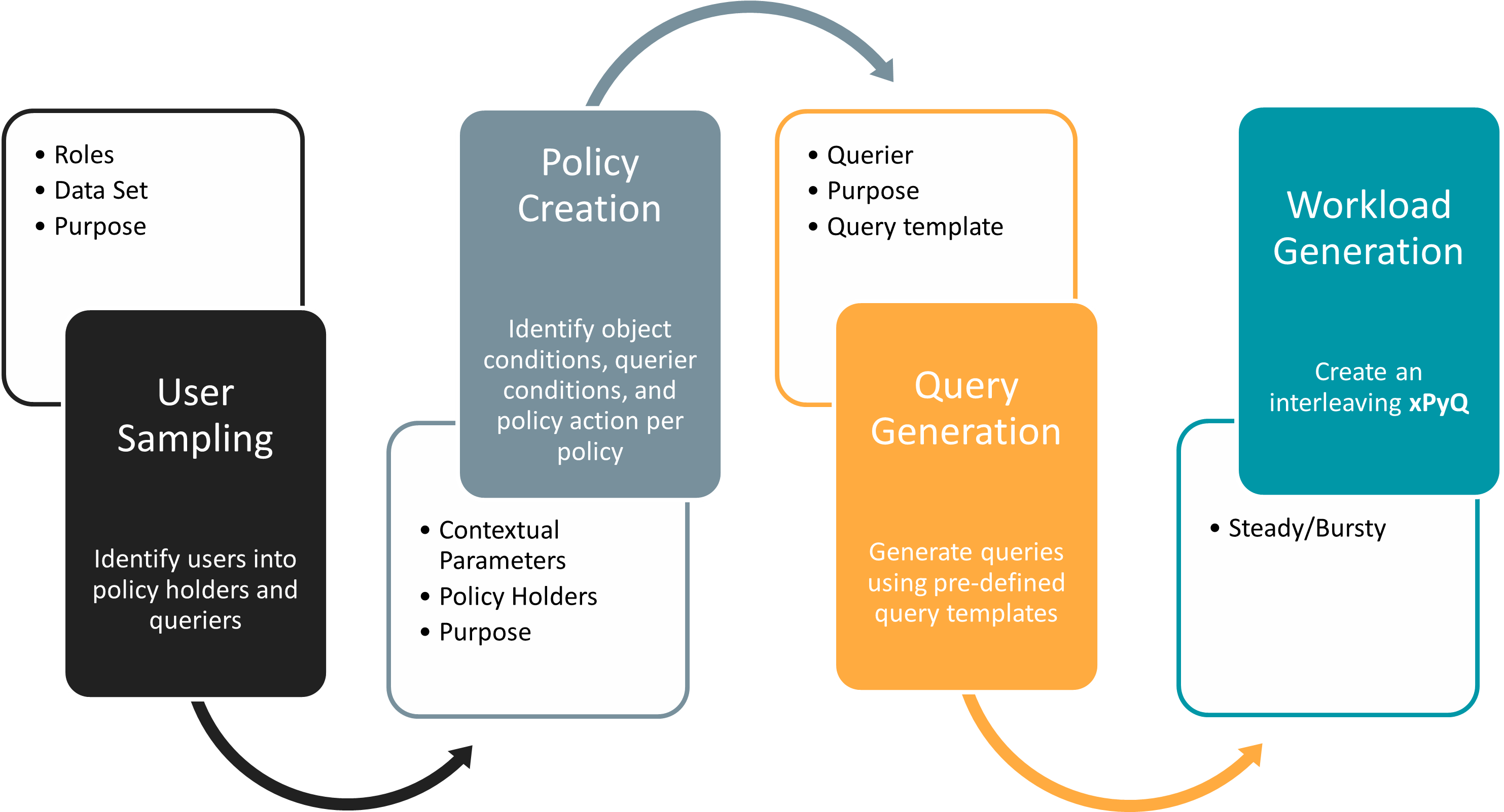}
\caption{Flow chart for generating the FGAC and query workload for Marking Attendance Scenario.}
\label{fig:flow-chart-AC}
\end{figure}

The workload generation process involves several steps to ensure that the workload accurately mimics real-world conditions. The flowchart for the \textit{marking attendance} scenario is presented in Figure \ref{fig:flow-chart-AC}. This flowchart can be used to generalize the workload generation process.

\begin{itemize}
    \item \textbf{Step 1: Sampling Users} – The process begins by sampling users to specify policyholders and query creators. For the \textit{marking attendance} scenario, the policyholders are sampled as students, while the queriers are sampled as faculty.
    
    \item \textbf{Step 2: Policy Creation} – The sampled policyholders are then passed to the policy creation process. Policies are created for each policyholder based on the contextual parameters of the scenario. For instance, in the \textit{marking attendance} scenario, policies are generated for students where the timestamp reflects class hours, the location is set to a classroom, and the querier is a faculty member.
    
    \item \textbf{Step 3: Query Creation} – Queries are generated for the sampled queriers using predefined query templates. In this case, queries for faculty are created based on the selected query templates relevant to the scenario.
    
    \item \textbf{Step 4: Workload Generation} – Finally, the generated policies and queries are passed to the workload generator. The generator selects either a bursty or steady state and creates an appropriate interleaving between policies and queries.
\end{itemize}

The complete workload generation methodology, including dataset usage, policy/query patterns, and dynamic simulation parameters, is described in the Appendix.

\noindent\textbf{Database System}: We ran the experiments on a dedicated machine (CentOS 7, Intel(R) Xeon(R) E5-2665, 2.4 GHz, 16 cores, 192 GB RAM, 10 TB local scratch). We performed experiments on MariaDB 5.5.68 with InnoDB, as it is an open-source DBMS that supports index usage hints. We configured the buffer\_pool\_size to 4 GB.


\subsubsection*{\textbf{Experiment~5}: Evaluation of Cache Refresh Strategies for Guard Regeneration}

In addition to caching, implementing an effective replacement strategy is crucial for managing data in the cache that has become stale. We conducted experiments on four different approaches to evaluate various strategies for replacing the guard expression (GE) using mergeability, as discussed in Section \ref{sec:refresh-strategy}.

\begin{enumerate}
    \item Baseline \(\mathbf{B_1}\), termed \textit{Always Regenerate}, regenerates the GE each time it is found to be outdated, similar to our proposed system with caching. 
    \item Baseline \(\mathbf{B_2}\), termed \textit{Max\_Consecutive\_Update\_Limit$<10$}, limits updates to the guards to ten instances, preventing the GE from becoming excessively long; after this limit is reached, the GE is regenerated. 
    \item Our Approach \(\mathbf{O_1}\), titled \textit{Mergeability Based on Regenerate/Update}, employs a dynamic approach to decide between regeneration and updates based on mergeability conditions, as detailed in Section \ref{sec:refresh-strategy}. This approach requires 100\% of the policies to be mergeable before regeneration. This ensures that we only merge when all the new policies can be merged into the existing GE, which encompasses the older policies.
    \item Our Approach \(\mathbf{O_2}\), titled \textit{Relaxed Mergeability Based on Regenerate/Update}, regenerates the GE if at least 50\% or more of the newly inserted policies can be merged with the existing GE. This approach relaxes $B_2$, which required 100\% of the policies to merge, thus potentially regenerating the GE earlier.

\end{enumerate}

\begin{figure}[!h]
\centering
\includegraphics[width=3.5in]{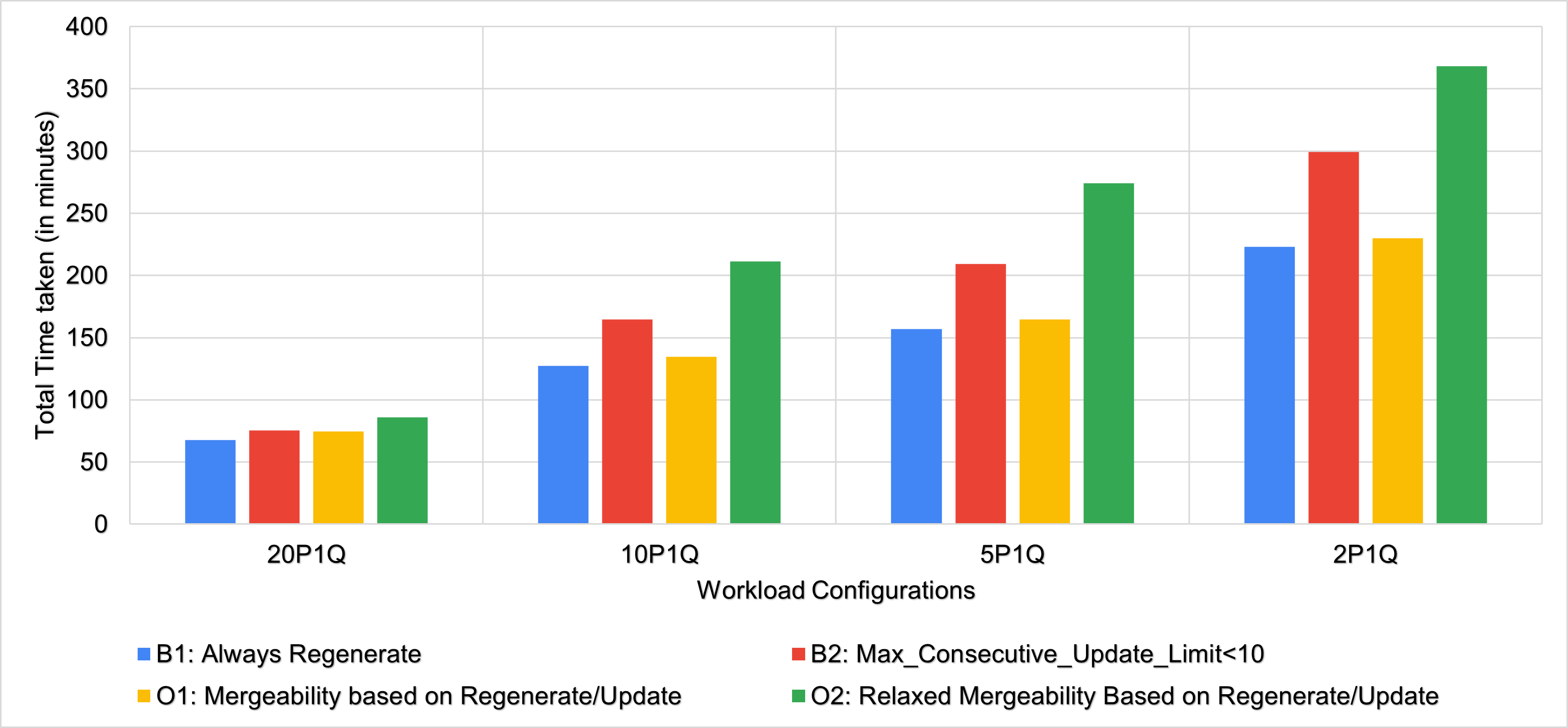}
\caption{Performance analysis of Cache Refresh Strategies - Steady State for \textit{marking attendance} scenario: Total Runtime Across Workloads.}
\label{fig:perf-merge}
\end{figure}

\begin{figure}[!h]
\centering
\includegraphics[width=2.2in]{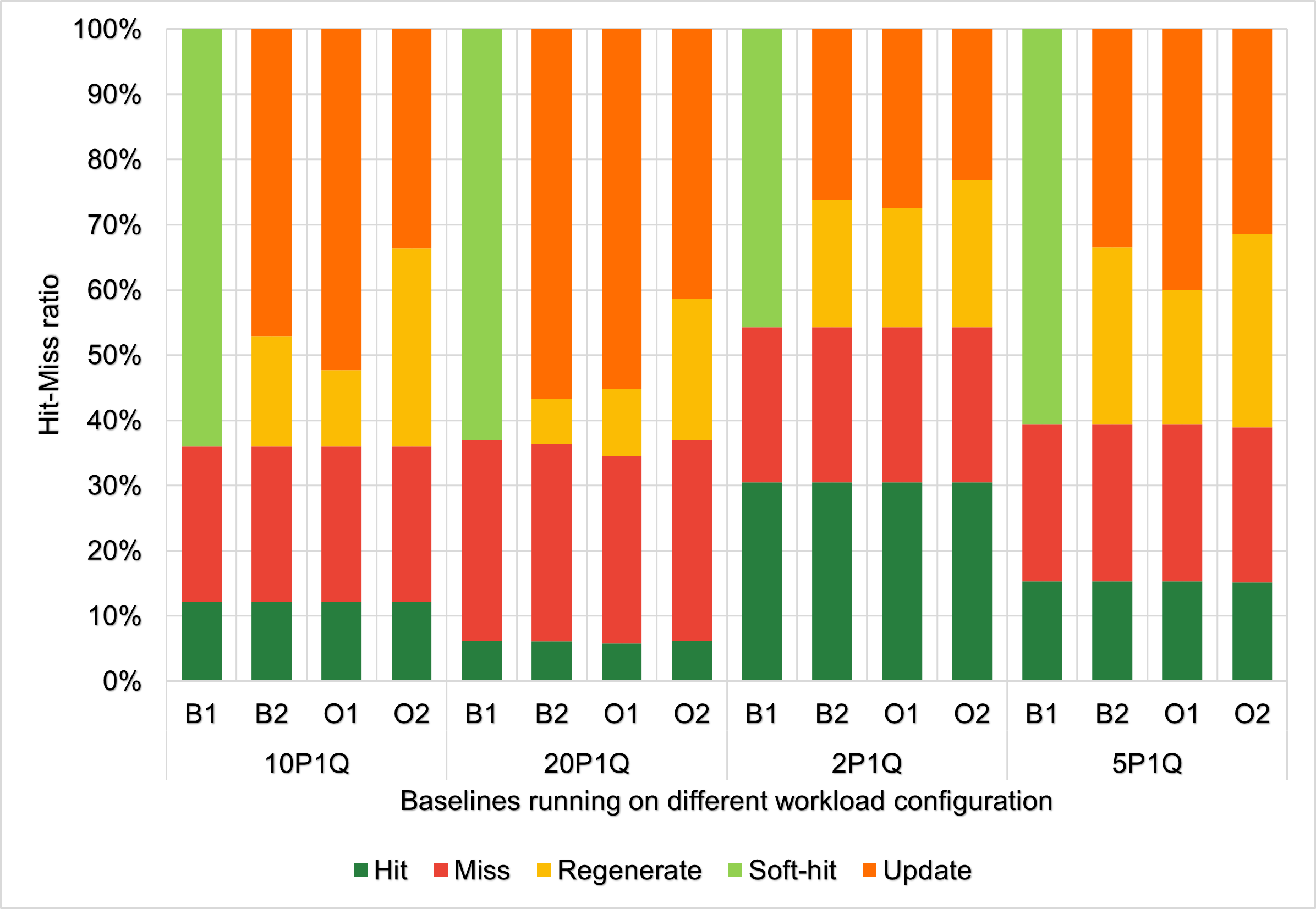}
\caption{Analyzing Cache Hit Rate with Varying Policy Configuration with Mergeability Approach - Steady State.}
\label{fig:merge}
\end{figure}

We generated performance graphs for four of these approaches to evaluate the different strategies for replacing the GE, as shown in Figure \ref{fig:perf-merge}. The x-axis of the graphs represents various workload configurations for varied policy insertion workload \textbf{xP1Q}, while the y-axis illustrates the total execution time in minutes. We use a fixed policy set of 31,520 entries and vary the query load across configurations (2P1Q–20P1Q), resulting in decreasing query counts from 15,760 to 1,576.




Our experiments focus on a marking attendance scenario with a cache size of 80\%, equivalent to 310 cache slots. The results indicate that Our Approach \(\mathbf{O_1}\) performs closest to the Baseline \(\mathbf{B_1}\) due to its dynamic approach of selecting between regeneration and updates based on mergeability conditions. In contrast, \(\mathbf{O_1}\)incurs significant overhead from constant regeneration, while Baseline \(\mathbf{B_1}\) triggers premature regenerations after a fixed number of updates, reducing overall efficiency. Our approach \(\mathbf{O_2}\) offers more flexibility but regenerates too frequently when partial updates suffice, thus adding unnecessary overhead. Furthermore, \(\mathbf{B_2}\) system may select sub-optimal guards when updates are preferred over regenerations, increasing query processing times as the policy expressions become less optimized.

\begin{figure*}[ht]
\centering
\subfloat[]{\includegraphics[width=2in]{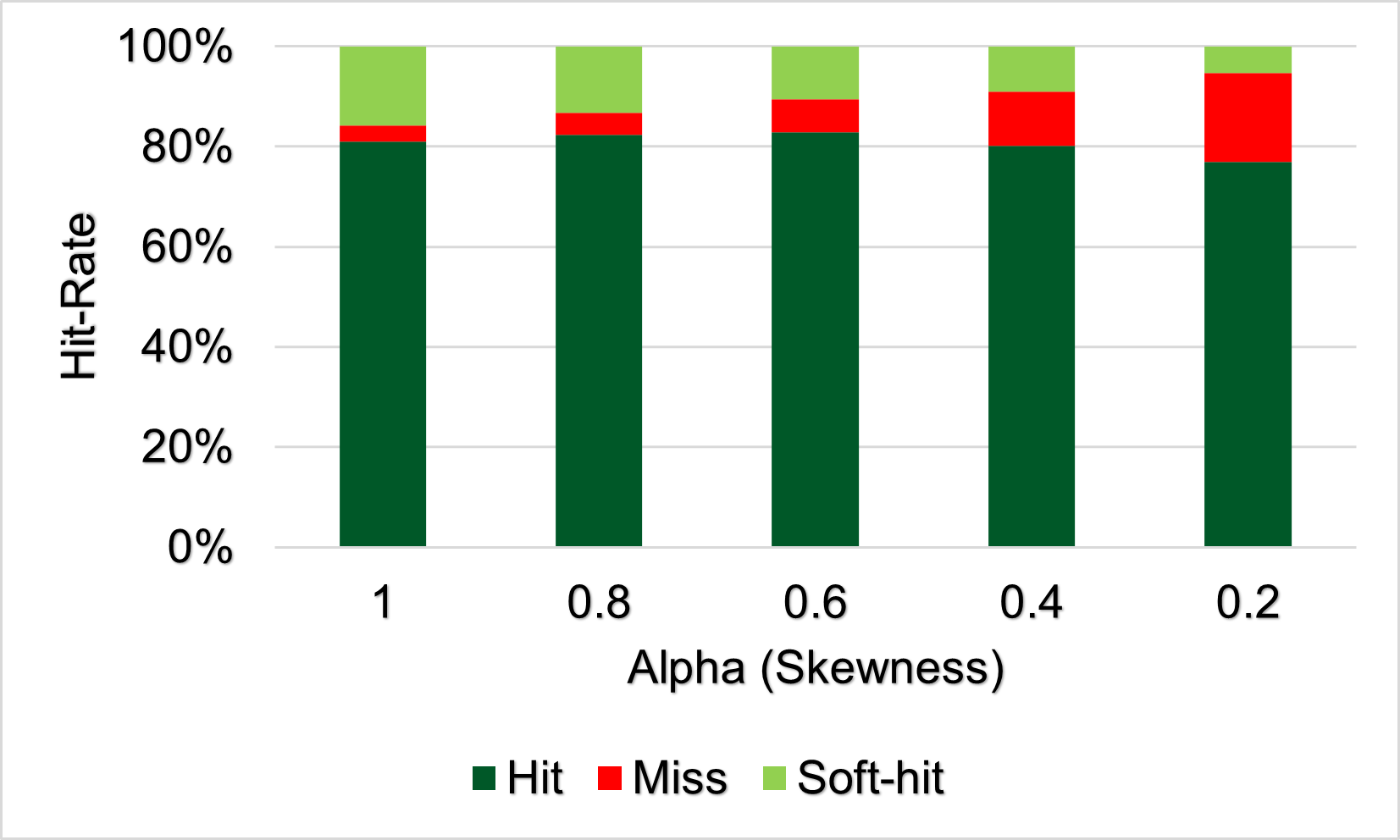}%
\label{fig:popularity_80}}
\hfil
\subfloat[]{\includegraphics[width=2in]{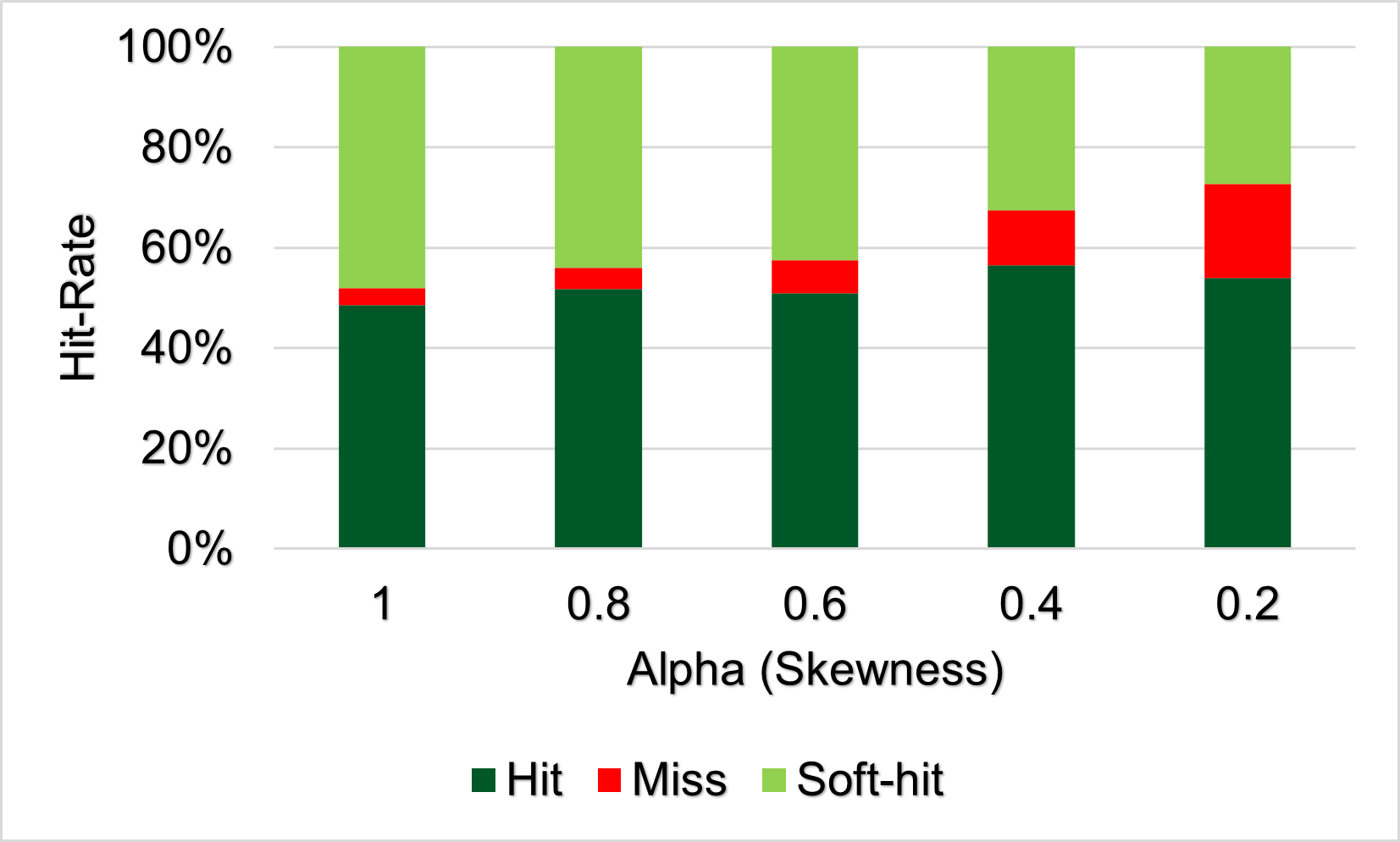}%
\label{fig:popularity_50}}
\hfil
\subfloat[]{\includegraphics[width=2in]{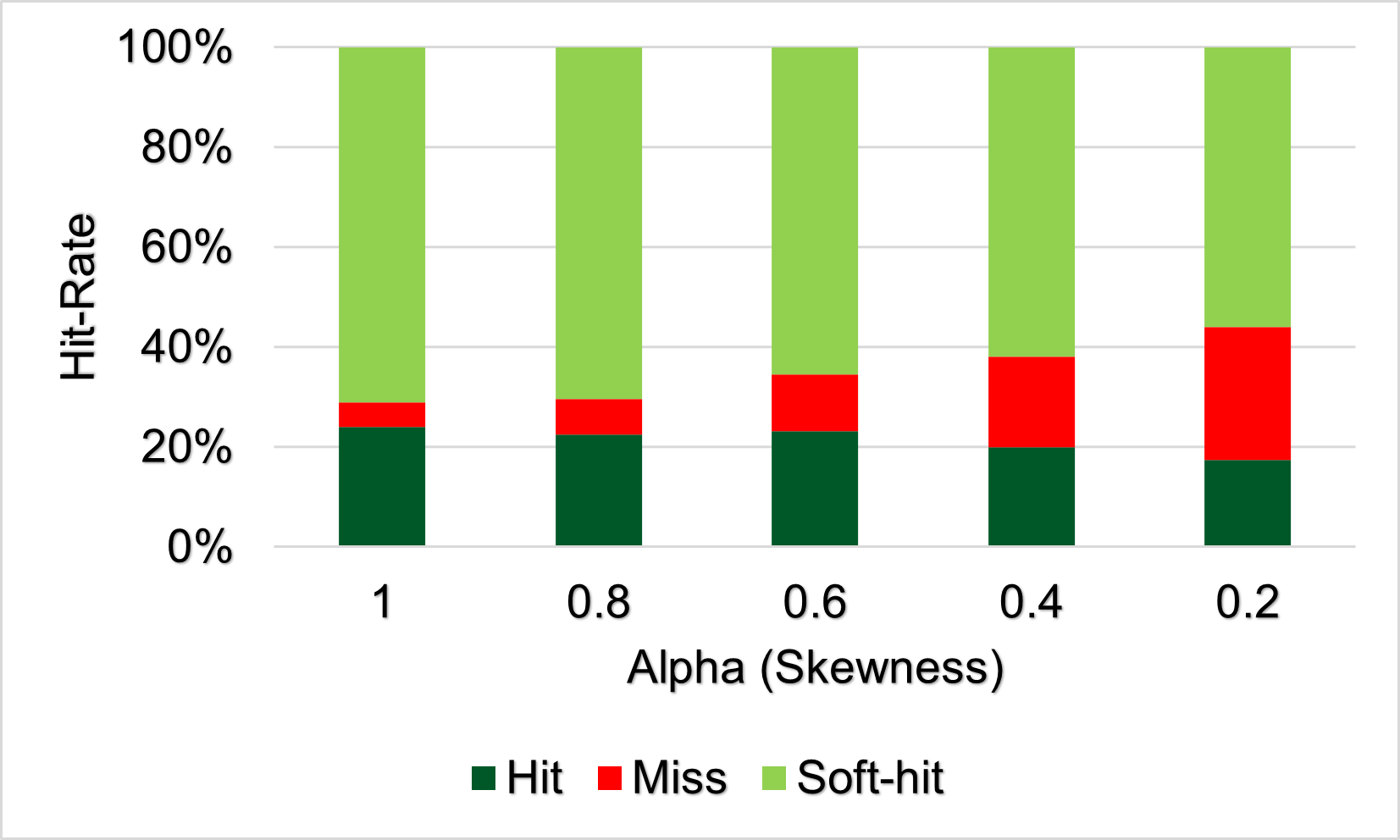}%
\label{fig:popularity_20}}
\caption{Analyzing cache hit rate in steady state for \textit{marking attendance} scenario across Alpha Values (a) Cache Size \textbf{80\%} and (a) Cache Size \textbf{50\%} (c) Cache Size \textbf{20\%}.}
\label{fig:populaity-hitrate}
\end{figure*}

Additionally, we evaluated the impact of different GE replacement methods \(\mathbf{B_1,B_2,O_1,O_2}\). This experiment aims to identify the most efficient refresh strategy for different workload configurations. 
We analyzed the frequency of updates, regenerations, and hit-and-miss percentages to assess their effectiveness. Soft-hits caused by outdated cache entries are handled via regeneration or update, depending on policy mergeability. Figure~\ref{fig:merge} shows that update frequency rises as policy insertions increase, which can reduce future merges' success, while on the other hand, repeated updates cause the GE to grow more fragmented, with disjointed conditions accumulating over time. Consequently, merging becomes harder, leading to inefficiencies in policy enforcement. While updates are faster initially, regeneration becomes preferable as policy volume grows, since it resets the GE to a streamlined form that better supports future merges.

\vspace{0.1cm}
\subsubsection*{\textbf{Experiment~6:} Querier Popularity and Cache Efficiency}  

This experiment analyzes the impact of querier popularity on cache performance using a Zipfian workload distribution. Policies and queries are distributed among 388 queriers with a consistent \textbf{10P5Q} configuration. The Zipfian parameter \(\alpha\) is varied to simulate different popularity skews, and results are observed for 80\%, 50\%, and 20\% cache sizes.

\begin{figure}[!t]
\centering
\includegraphics[width=\linewidth]{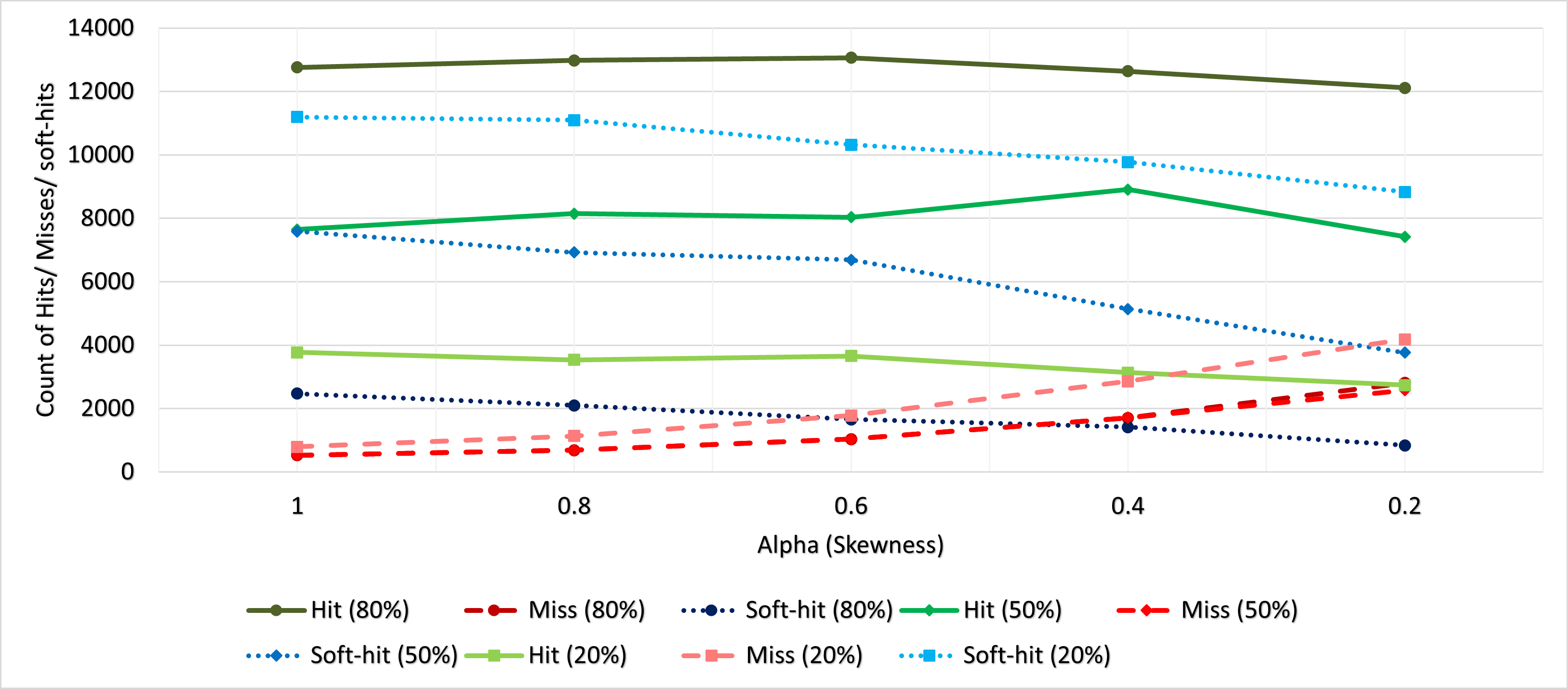}
\caption{Cache performance across different Alpha values.}
\label{fig:popularity-cache}
\end{figure}

Our evaluation highlights the impact of Zipfian-distributed queries and cache size on hit, miss and soft-hit rates. As shown in Figure \ref{fig:populaity-hitrate}, hit rates generally decline with decreasing alpha as query access patterns become more uniform, leading to fewer frequently reused queries. The soft-hit rate remains high, particularly for smaller cache sizes (e.g., 20\%), where frequent policy insertions result in outdated cached entries being accessed before regeneration. This aligns with our expectation that smaller caches struggle to retain fresh data, leading to higher soft-hit occurrences rather than outright misses. In contrast, larger caches (80\%) exhibit a lower soft-hit rate since they can store more queries, reducing the chance of outdated entries. Figure \ref{fig:popularity-cache} further illustrates the trends across cache sizes, showing a clear trade-off between cache efficiency and policy update frequency. Notably, for higher alpha values (i.e., more skewed workloads), hit rates remain relatively stable in larger caches, reinforcing the effectiveness of caching in Zipfian-heavy query distributions. These findings emphasize the need for adaptive caching strategies to mitigate soft-hit overhead in dynamic FGAC workloads.

\begin{figure}[!t]
\centering
\includegraphics[width=2.2in]{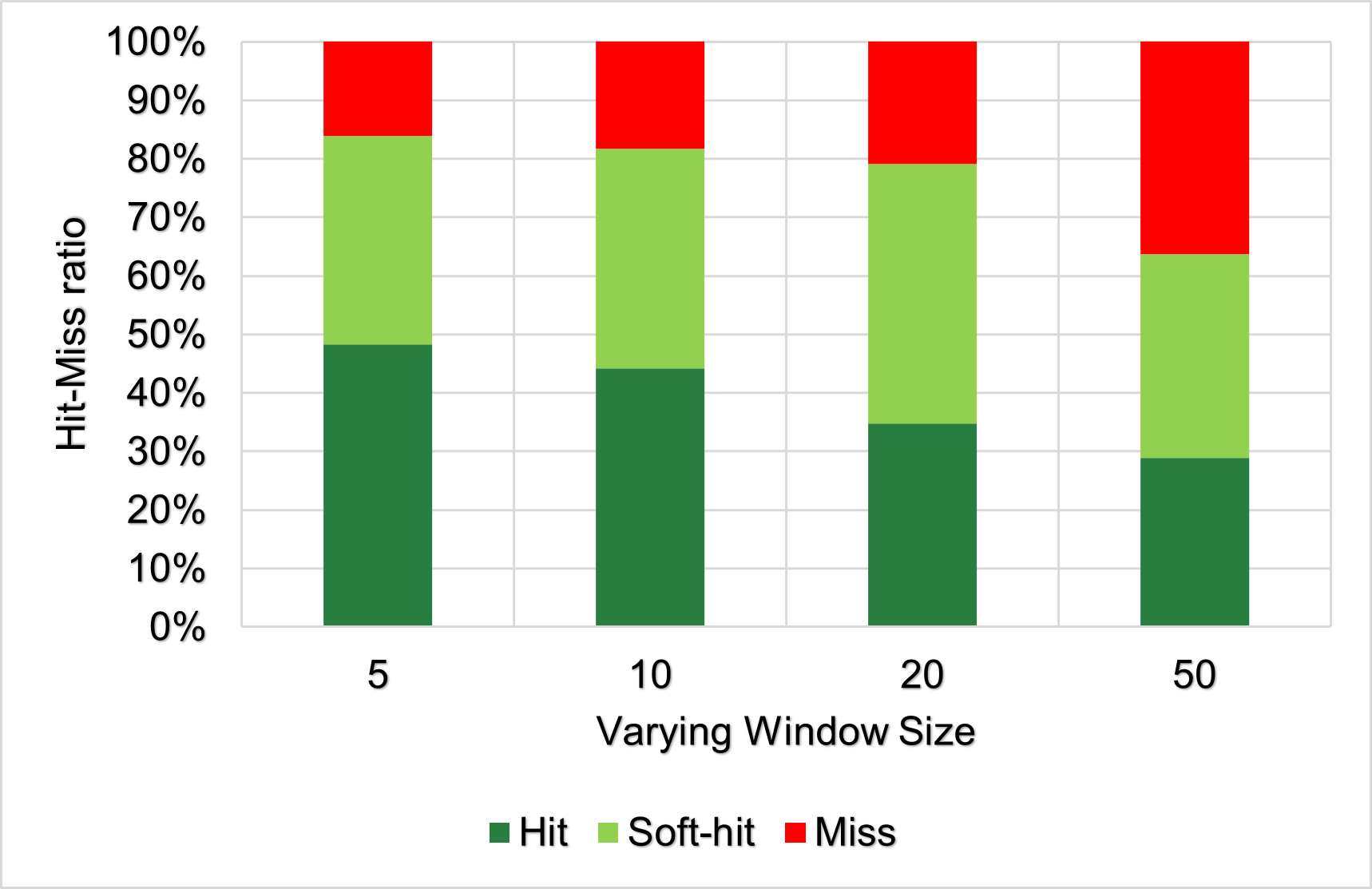}
\caption{Cache hit rate with varying window sizes (steady state).}
\label{fig:window-size-80}
\end{figure}

\vspace{0.1cm}
\subsubsection*{\textbf{Experiment~7:} Impact of Window Sizes on Cache Performance}  
This experiment evaluates the effect of varying window sizes on cache efficiency using the \textbf{5P10Q} configuration. Different window sizes (5, 10, 20, 50) were tested in a steady-state workload, focusing on cache hit, soft-hit, and miss ratios. 

\noindent   
\begin{itemize}
    \item \textbf{Small Window Sizes:} Higher hit rates and lower miss rates due to repeated queries benefiting from cached entries.
    \item \textbf{Large Window Sizes:} Increased miss rates as query diversity widened, reducing the likelihood of repeated queries within the cache.
    \item \textbf{Soft-Hits:} Observed consistently across all window sizes, attributed to stale entries in the cache.
\end{itemize}

Figure~\ref{fig:window-size-80} depicts the cache performance across window sizes. The results highlight the importance of balancing window size to optimize cache utility while minimizing stale data.

\vspace{0.1cm}
\subsubsection*{\textbf{Experiment~8:}~Performance Evaluation in Dynamic Setting}  

\begin{figure*}[ht]
\centering
\subfloat[]{\includegraphics[width=2.2in]{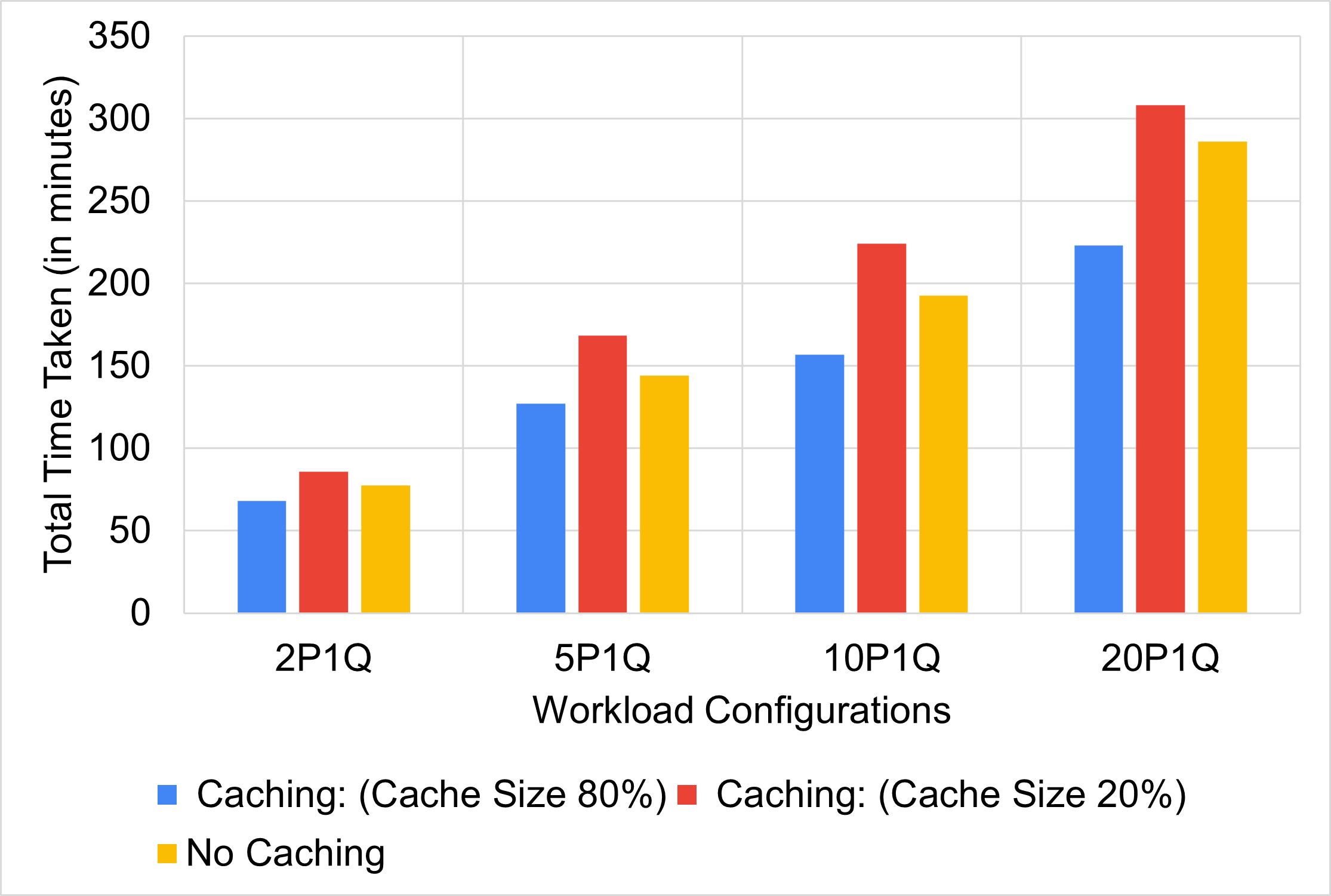}%
\label{perf-ac-xp}}
\hfil
\subfloat[]{\includegraphics[width=2.2in]{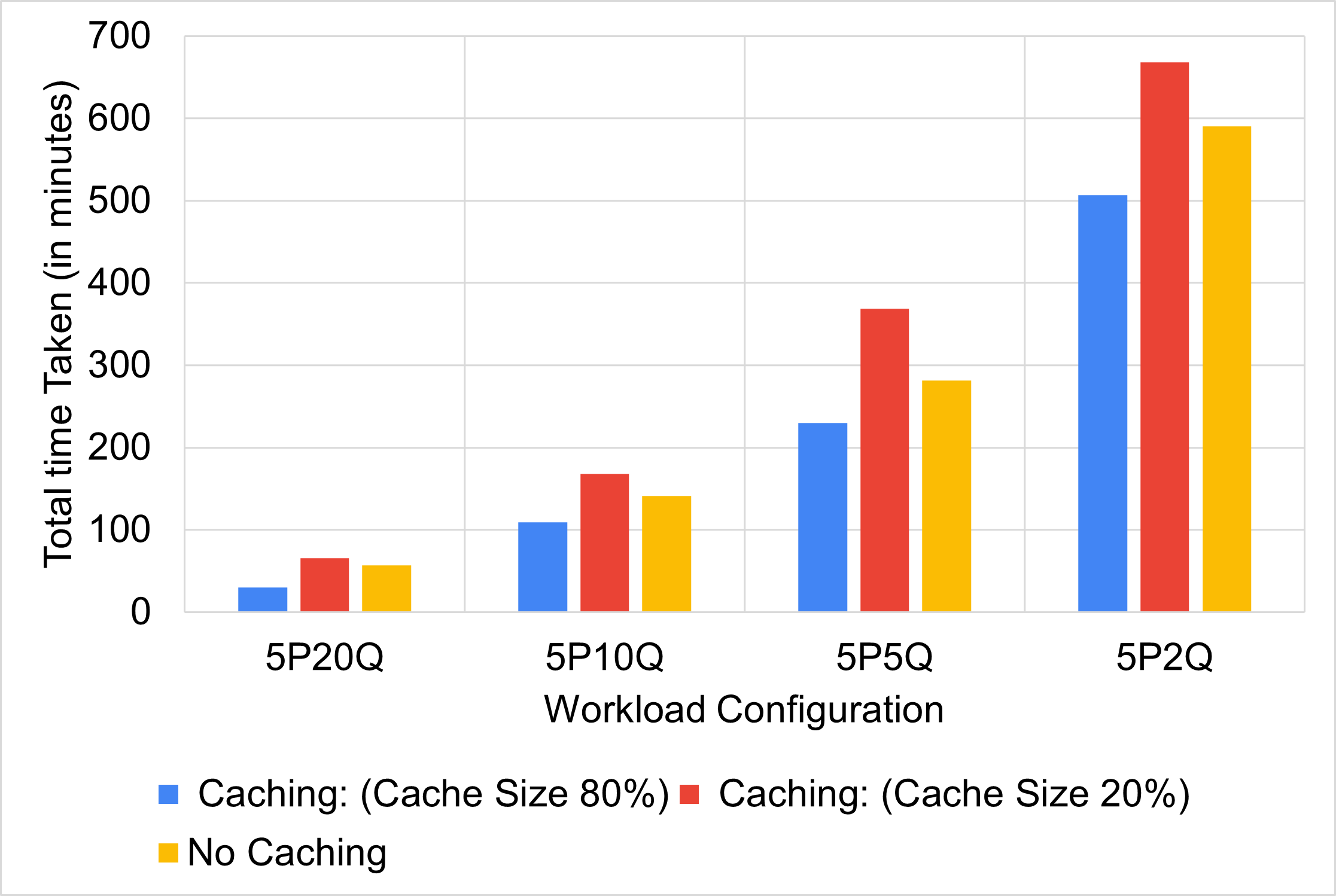}%
\label{perf-ac-yq}}
\hfil
\subfloat[]{\includegraphics[width=2.2in]{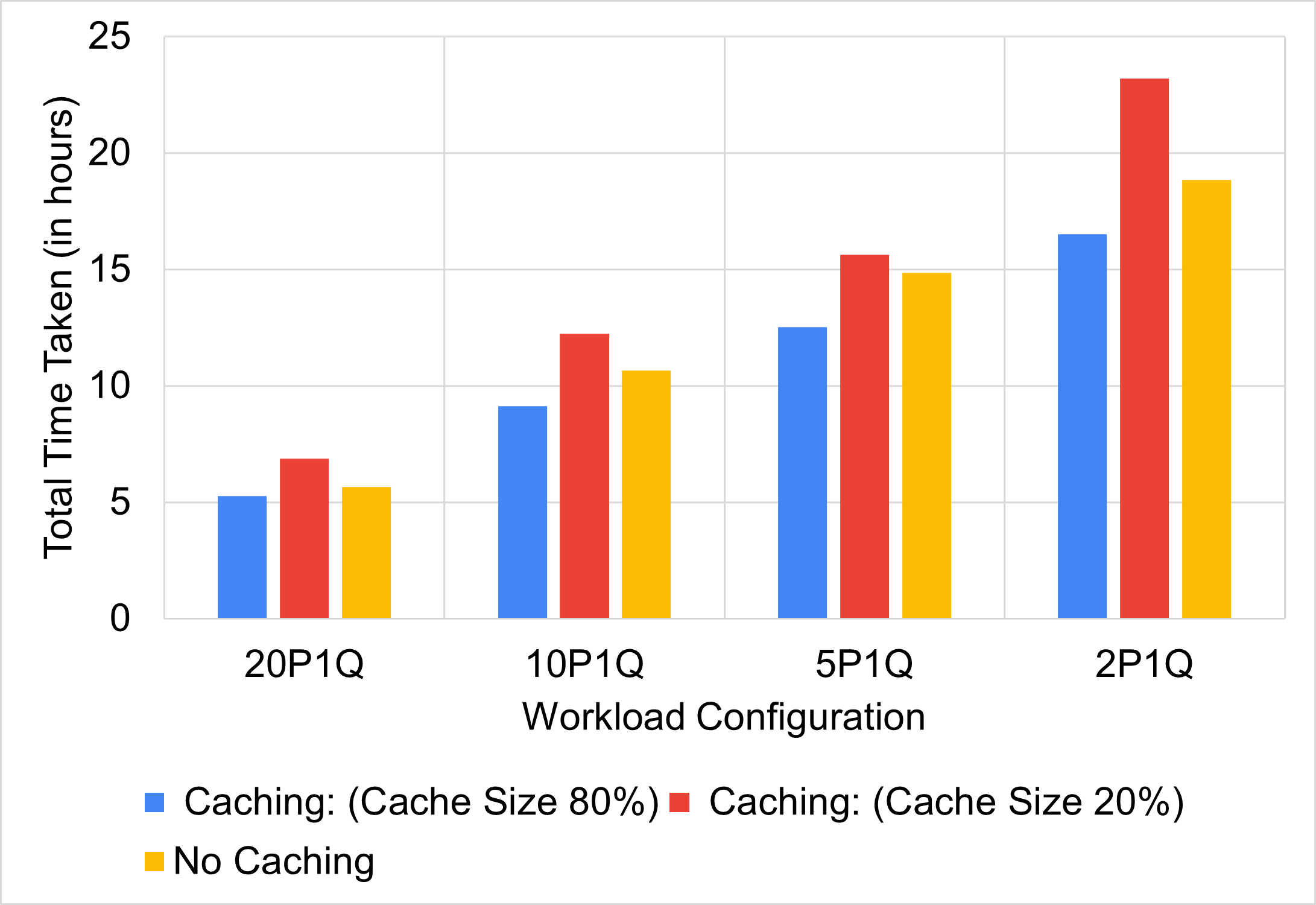}%
\label{fig:perf-eval-su}}
\caption{Performance analysis for \textit{marking attendance} scenario (a) Varying policy insertion \textbf{xP1Q} and (b) Varying query insertion \textbf{5PyQ}: Total Runtime Across Workloads (in minutes). (c) Performance analysis - for \textit{space-usage monitoring} scenario varying policy insertion \textbf{xP1Q}: Total Runtime Across Workloads (in hours).}
\label{fig:perf-graph}
\end{figure*}

In this experiment, we evaluate the performance implications of employing \oursystem under dynamic IoT scenarios: without caching and with caching for 80\% and 20\% cache size in the Steady State Scenario. Cache size is proportional to the number of queriers for the given scenario. We aim to understand how caching influences the efficiency of \oursystem across varied workload configurations. We present our findings in Figure \ref{fig:perf-graph} during steady-state scenarios, focusing on the \textit{marking attendance} and \textit{space usage monitoring} use cases for cache sizes 80\% and 20\% versus total no. of queriers. 

We investigate the impact of two distinct workload configurations: one where the number of policies varies while keeping query volume constant \textbf{xP1Q}, as presented in Figure \ref{perf-ac-xp}, and another where the number of queries increases for a fixed policy volume \textbf{5PyQ}, as illustrated in Figure \ref{perf-ac-yq}. In the first set, 3,152 queries are executed under increasing policy insertions ranging from 6,304 to 63,040. 39,400 policies are inserted consistently in the second set, while queries increase from 3,152 to 15,760. This design allows us to evaluate how policy volume and query load independently influence cache behavior and system performance.

We conduct the experiments using two cache sizes: 310 slots and 77 slots, which correspond to 80\% and 20\% of the total number of queriers, respectively. The larger cache allows room for regeneration and updates, highlighting the effectiveness of the refresh strategy. In contrast, the smaller cache emphasizes the role of the replacement policy under tighter memory constraints. Since policy insertion time is negligible across configurations, differences in execution time directly reflect the impact of caching. On the x-axis, we depict the workload configurations, while the y-axis shows the total execution time (in minutes).


Our results demonstrate that the Sieve algorithm with caching consistently outperforms Sieve without caching at a higher cache size. However, the 20\% cache size performs worse than \oursystem without caching with varying policy insertions. This is because the smaller cache size reduces the chance of achieving a high hit rate, and items in the cache can quickly become stale, leading to frequent regeneration of GE. Additionally, the overhead of maintaining a small cache with frequent evictions and regenerations adds extra processing time, which outweighs the benefits of caching in this scenario.\\

Building on the attendance control scenario results, we further evaluate the performance of \oursystem under a different workload: the \textit{space-usage monitoring} use case. This scenario includes many queriers, enabling a large-scale experiment closely reflecting real-world settings where the system must manage extensive policies and queries concurrently. We use a fixed query load of 18,218 queries and vary policy insertions across four configurations: \textbf{2P1Q}, \textbf{5P1Q}, \textbf{10P1Q}, and \textbf{20P1Q}, which correspond to 36,436, 91,090, 182,180, and 364,360 policies, respectively. We compare performance without caching and with 80\% and 20\% cache sizes, which correspond to 1,136 and 284 entries based on the total number of queriers. As shown in Figure~\ref{fig:perf-eval-su}, the results for both cache sizes are consistent with trends observed in the marking attendance scenario.



Additional results on workload-driven cache behavior, including steady, bursty, and deletion-based workloads, are available in Appendix~\ref{appendix:workload-cache-exp}.


\subsection{Summary of Experimental Findings}
Our evaluation of \oursystem in static settings demonstrates significant performance gains over traditional FGAC enforcement methods. By generating compact guarded expressions, \oursystem significantly reduces the enforcement time for policies. 
On benchmark queries ($Q_1$–$Q_3$), \oursystem consistently outperforms baselines across different query selectivities, achieving 1.5$\times$ to over 70$\times$ faster execution times. Moreover, \oursystem performs robustly across DBMSs. On PostgreSQL, it benefits from bitmap index scans to outperform both the native baseline and its own MySQL counterpart, achieving up to 5.6$\times$ speedup in larger policy sets (e.g., 1200 policies) on synthetic datasets. These results affirm \oursystem’s efficiency, scalability, and generalizability.

In dynamic settings, \oursystem enhanced with caching improves performance and scalability. Using a mergeability-aware refresh strategy (\(O_1\)), we achieved the best runtime results, minimizing unnecessary regenerations while maintaining correctness. Across workloads with 31,520 policies and up to 15,760 queries (e.g., 10P1Q), caching reduced total execution time from over 90 minutes to under 70 minutes. In Zipfian-distributed workloads, we observe that cache hit rates exceed 60\% when the cache size reaches 80\%, showing that larger caches effectively retain relevant guarded expressions. \oursystem with small cache size (e.g., 20\%) performed worse than \oursystem with no caching as frequent policy insertions lead to outdated entries being accessed before regeneration, resulting in high soft-hit rates. 
As guarded expressions are relatively small in size (few KBs), it is reasonable to cache most of the guarded expressions.
These results highlight the importance of caching for improving performance and addressing memory hierarchy limitations. Although systems can store all guarded expressions in memory, they cannot fit them into faster but much smaller tiers like CPU registers, making efficient caching essential. 
In bursty workloads, caching adapted dynamically, improving hit rates from 15\% to over 70\% and halving query latency. We saw the most significant improvements during query-heavy phases in bursty workloads and in Zipfian settings with high query reuse, where caching helped avoid repeated computation and saved time. However, heavy policy deletions (e.g., 10P5Q10D) increased regeneration overhead due to outdated cached entries. Overall, \oursystem with caching achieves up to 6-22\% runtime improvement over the non-caching version and scales effectively across diverse and evolving FGAC workloads.

\section{Related Work}
\label{sect:relWork}

\subsection{Fine-Grained Access Control Enforcement}
Access control mechanisms in databases, particularly those supporting Fine-Grained Access Control (FGAC), have been extensively explored. Techniques for FGAC enforcement can be broadly categorized as either \textit{authorization views} or \textit{policy-as-data}. Authorization views, exemplified by Oracle Virtual Private Database~\cite{loney2008oracle}, rely on creating materialized views tailored to specific access permissions. This approach is effective when the number of access control policies is small and does not depend on the size of the underlying database. However, it becomes prohibitive at scale due to the overhead of maintaining and updating many views per querier, making it less suitable for dynamic or policy-rich environments.

The policy-as-data approach, seen in Hippocratic databases~\cite{agrawal2002hippocratic} and their successors~ \cite{lefevre2004limiting,agrawal2005extending}, stores policies alongside tuples and enforces access control by appending policy predicates to queries. This strategy, however, suffers from scalability issues when dealing with large policy sets, as appending numerous conditions in disjunctive normal form significantly degrades query performance. 

Distributed and scalable authorization systems have been proposed as access control needs grow in complexity, particularly in large-scale, multi-tenant environments having a large number of policies. Zanzibar~\cite{zanzibar2019}, developed by Google, provides a global authorization service that ensures consistent access control across distributed applications. It employs a uniform data model and configuration language to express a wide range of access control policies, efficiently processing millions of authorization requests per second. However, Zanzibar is tightly integrated into Google's infrastructure, relying on Spanner, and may not generalize well to other environments. Additionally, Zanzibar’s policy model is based on relationship-based access control (ReBAC), which defines access through resource-user relationships. In contrast, \oursystem adopts an attribute-based access control (ABAC) model that supports fine-grained, context-aware policy evaluations based on user, resource, and environment attributes---better suited for dynamic IoT and query-driven scenarios.

In the context of IoT and smart spaces, prior work has focused on policy-based access control tailored to sensor data. For example, FENCE~\cite{nehme2013fence} processes data and associated policies based on incoming queries, enabling continuous access control enforcement in dynamic data stream environments. However, this method may face challenges with analytical queries involving large policy sets and might require modifications to the database management system (DBMS) operators to ensure they are security aware. Similarly, Colombo et al.~\cite{colombo2018access} introduced a Message Queuing Telemetry Transport (MQTT)-based architecture for secure data transmission in IoT ecosystems, focusing on secure data transmission rather than managing extensive policy sets at runtime. While augmenting tuples with access purposes during data ingestion reduces real-time policy enforcement overhead, it introduces substantial overhead during data ingestion. It fails to address dynamic policies that change post-ingestion. 

Recent advancements have attempted to overcome these limitations by incorporating flexible and scalable access control mechanisms. The Flexible Access Control Technique (FACT)~\cite{FACT} enhanced RBAC and ABAC by introducing a dual-hierarchy structure, including Spatial-Device and Temporal-Policy Trees to manage policies for millions of IoT devices and users efficiently. Additionally, studies have explored integrating ABAC into NoSQL databases~\cite{ABC-1, ABC-2} to handle environmental attributes and support dynamic access for ad-hoc users. While these methods address specific challenges, they struggle to support workloads with frequent policy changes or repeated queries, common in IoT environments.

\oursystem tightly couples access control enforcement with the database management system by dynamically rewriting queries based on access policies. By generating guarded expressions (GEs) at query time, \oursystem reduces enforcement overhead while remaining compatible with existing DBMSs. It requires no changes to the query engine or internals, making it easy to deploy alongside standard systems. To further improve efficiency, \oursystem reuses cached GEs to avoid recomputation for recurring queries, enhancing performance across both static and dynamic workloads. As a result, \oursystem supports scalable and responsive policy enforcement.

\subsection{Caching Mechanisms}
Caching mechanisms have been widely explored in database systems and access control enforcement. 
It has been extensively studied in database query optimization. Yang \etal~\cite{yang2023fifo} examined traditional adaptive caching strategies, such as Adaptive Replacement Cache (ARC), and found that simpler First-In, First-Out (FIFO)-based approaches can achieve comparable or even better performance in specific workload conditions, mainly when cache sizes are sufficiently large. Their study demonstrates that complex eviction policies, such as Least Recently Used (LRU), do not always offer significant benefits over First-In-First-Out (FIFO) in large-scale caching scenarios. However, these findings focus on general caching workloads and do not specifically address FGAC, where caching must efficiently handle frequent policy changes and repeated queries based on evolving access control rules.

Despite these advancements, many existing approaches defer FGAC enforcement to query time, as seen in works like Colombo \etal~\cite{6767117,DBLP:journals/dase/ColomboF16}, and other database policy enforcement models. While these methods address some scalability concerns, they fail to handle the complexities introduced by dynamic policy updates and repeated queries. Our system overcomes these limitations by combining guarded expressions with an adaptive caching mechanism, offering a robust and scalable solution for managing extensive policy sets in dynamic environments.

Qapla~\cite{qapla2017} presents a policy-compliance middleware that transparently enforces fine-grained access control through query rewriting. To mitigate the policy enforcement overhead, Qapla incorporates a query template cache that stores rewritten queries, reducing the cost of repeated parsing and rewriting. While effective for amortizing query transformation costs, Qapla's caching approach does not address the challenges introduced by frequent policy updates, where cached query templates may quickly become invalid. Additionally, Qapla’s reliance on static query rewriting means it does not dynamically adapt caching strategies in response to policy evolution or workload changes.

To bridge this gap, we enhance FGAC enforcement by introducing an adaptive caching mechanism designed to efficiently reuse guarded expressions (GEs). Unlike traditional caching strategies that optimize general query workloads, our approach accounts for policy evolution by balancing cache replacement decisions versus cache refresh depending upon the policies inserted. This ensures scalability and responsiveness in dynamic FGAC environments, where queries and policies are constantly changing.

\section{Conclusion and Future Work}
\label{sect:conclusion}

In this paper, we presented \oursystem, an advanced framework for enforcing a large number of fine-grained access control (FGAC) policies during query execution, with extensions to support dynamic IoT scenarios through caching. \oursystem employs a layered approach to optimize query processing by reducing the number of policy evaluations and tuple checks. The addition of a caching mechanism, with a replacement policy based on the Clock algorithm, further improves query latency and system load under dynamic workloads. 

Our experimental evaluation demonstrated that \oursystem significantly outperforms baseline approaches, maintaining low query processing times even with thousands of policies per query. The caching mechanism effectively increased cache hit rates and reduced the overhead of guarded expression regeneration, particularly under larger cache sizes. These results highlight \oursystem's scalability and efficiency, making it well-suited for dynamic and complex environments.

Future work will focus on extending \oursystem's applicability to diverse datasets and domains, optimizing caching strategies for improved resource utilization, and enhancing the processing of disjunctive query expressions. We also aim to explore tighter integration with database query optimizers to further streamline query and policy workload management. These directions will strengthen \oursystem’s capabilities as a robust and scalable solution for enforcing FGAC policies across a wide range of real-world scenarios.


\bibliographystyle{IEEEtran}
\bibliography{references}

@article{DBLP:journals/toit/YusBMV22,
  author       = {Roberto Yus and
                  Georgios Bouloukakis and
                  Sharad Mehrotra and
                  Nalini Venkatasubramanian},
  title        = {The SemIoTic Ecosystem: {A} Semantic Bridge between IoT Devices and
                  Smart Spaces},
  journal      = {{ACM} Trans. Internet Techn.},
  volume       = {22},
  number       = {3},
  pages        = {76:1--76:33},
  year         = {2022},
  url          = {https://doi.org/10.1145/3527241},
  doi          = {10.1145/3527241},
  bibsource    = {dblp computer science bibliography, https://dblp.org}
}

@INPROCEEDINGS{TIPPERS2016,
  author={Mehrotra, Sharad and Kobsa, Alfred and Venkatasubramanian, Nalini and Rajagopalan, Siva Raj},
  booktitle={2016 IEEE International Conference on Pervasive Computing and Communication Workshops (PerCom Workshops)}, 
  title={TIPPERS: A privacy cognizant IoT environment}, 
  year={2016},
  volume={},
  number={},
  pages={1-6},
  doi={10.1109/PERCOMW.2016.7457158}
}

@inproceedings {qapla2017,
author = {Aastha Mehta and Eslam Elnikety and Katura Harvey and Deepak Garg and Peter Druschel},
title = {Qapla: Policy compliance for database-backed systems},
booktitle = {26th USENIX Security Symposium (USENIX Security 17)},
year = {2017},
isbn = {978-1-931971-40-9},
address = {Vancouver, BC},
pages = {1463--1479},
url = {https://www.usenix.org/conference/usenixsecurity17/technical-sessions/presentation/mehta},
publisher = {USENIX Association},
month = aug
}

@inproceedings {zanzibar2019,
author = {Ruoming Pang and Ramon Caceres and Mike Burrows and Zhifeng Chen and Pratik Dave and Nathan Germer and Alexander Golynski and Kevin Graney and Nina Kang and Lea Kissner and Jeffrey L. Korn and Abhishek Parmar and Christina D. Richards and Mengzhi Wang},
title = {Zanzibar: {Google{\textquoteright}s} Consistent, Global Authorization System},
booktitle = {2019 USENIX Annual Technical Conference (USENIX ATC 19)},
year = {2019},
isbn = {978-1-939133-03-8},
address = {Renton, WA},
pages = {33--46},
url = {https://www.usenix.org/conference/atc19/presentation/pang},
publisher = {USENIX Association},
month = jul
}

@INPROCEEDINGS{FACT,
  author={Yu, Alian and Kang, Jian and Jiang, Wei and Lin, Dan},
  booktitle={2023 IEEE International Conference on Pervasive Computing and Communications Workshops and other Affiliated Events (PerCom Workshops)}, 
  title={FACT: A Flexible Access Control Technique for Very Large Scale Public IoT Services}, 
  year={2023},
  volume={},
  number={},
  pages={386-391},
  doi={10.1109/PerComWorkshops56833.2023.10150257}}

@ARTICLE{ABC-2,
  author={Gupta, Eeshan and Sural, Shamik and Vaidya, Jaideep and Atluri, Vijayalakshmi},
  journal={IEEE Transactions on Emerging Topics in Computing}, 
  title={Enabling Attribute-Based Access Control in {N}o{SQL} Databases}, 
  year={2023},
  volume={11},
  number={1},
  pages={208-223},
  doi={10.1109/TETC.2022.3193577}}

@INPROCEEDINGS {ABC-1,
author = {G. Meena and P. Paul and S. Sural},
booktitle = {2023 5th IEEE International Conference on Trust, Privacy and Security in Intelligent Systems and Applications (TPS-ISA)},
title = {Efficiently Supporting Attribute-Based Access Control in Relational Databases},
year = {2023},
volume = {},
issn = {},
pages = {230-239},
doi = {10.1109/TPS-ISA58951.2023.00037},
url = {https://doi.ieeecomputersociety.org/10.1109/TPS-ISA58951.2023.00037},
publisher = {IEEE Computer Society},
address = {Los Alamitos, CA, USA},
month = {nov}
}

@article{DBLP:journals/pvldb/ShastriBWKC20,
  author       = {Supreeth Shastri and
                  Vinay Banakar and
                  Melissa Wasserman and
                  Arun Kumar and
                  Vijay Chidambaram},
  title        = {Understanding and Benchmarking the Impact of {GDPR} on Database Systems},
  journal      = {Proc. {VLDB} Endow.},
  volume       = {13},
  number       = {7},
  pages        = {1064--1077},
  year         = {2020},
  url          = {http://www.vldb.org/pvldb/vol13/p1064-shastri.pdf},
  doi          = {10.14778/3384345.3384354},
  bibsource    = {dblp computer science bibliography, https://dblp.org}
}

@inproceedings {Shastri2019,
author = {Supreeth Shastri and Melissa Wasserman and Vijay Chidambaram},
title = {The Seven Sins of {Personal-Data} Processing Systems under {GDPR}},
booktitle = {11th USENIX Workshop on Hot Topics in Cloud Computing (HotCloud 19)},
year = {2019},
address = {Renton, WA},
url = {https://www.usenix.org/conference/hotcloud19/presentation/shastri},
publisher = {USENIX Association},
month = jul
}

@inproceedings{michelakaki2023unlocking,
  title={Unlocking data protection by design \& by default: Lessons from the enforcement of article 25 GDPR},
  author={Michelakaki, Christina and Vale, Sebasti{\~a}o Barros},
  year={2023},
  organization={< bound method Organization. get\_name\_with\_acronym of< Organization: Future~…}
}

@article{SIEVEPappachanYMF20,
 author = {Primal Pappachan and
Roberto Yus and
Sharad Mehrotra and
Johann-Christoph Freytag},
 bibsource = {dblp computer science bibliography, https://dblp.org},
 journal = {PVLDB Endow.},
 number = {11},
 pages = {2424--2437},
 title = {Sieve: A Middleware Approach to Scalable Access Control for Database
Management Systems},
 url = {http://www.vldb.org/pvldb/vol13/p2424-pappachan.pdf},
 volume = {13},
 year = {2020}
}

@article{CLOCKcorbato1968paging,
  title={A paging experiment with the {m}ultics system},
  author={Corbato, Fernando J and others},
  year={1969},
  publisher={Massachusetts Institute of Technology}
}

@inproceedings{yang2023fifo,
  title={{FIFO} can be Better than {LRU}: the Power of Lazy Promotion and Quick Demotion},
  author={Yang, Juncheng and Qiu, Ziyue and Zhang, Yazhuo and Yue, Yao and Rashmi, KV},
  booktitle={Proceedings of the 19th Workshop on Hot Topics in Operating Systems},
  pages={70--79},
  year={2023}
}

@article{kim,
author = {Kim, Albert and Madden, Samuel},
title = {Optimizing Disjunctive Queries with Tagged Execution},
year = {2024},
issue_date = {June 2024},
publisher = {Association for Computing Machinery},
address = {New York, NY, USA},
volume = {2},
number = {3},
url = {https://doi.org/10.1145/3654961},
doi = {10.1145/3654961},
journal = {Proc. ACM Manag. Data},
month = may,
articleno = {158},
numpages = {25}
}

@misc{GDPR,
  title = {General data protection regulation {GPDR}.},
  author = {},
  howpublished = {\url{https://gdpr.eu}},
  note = {Accessed: 2025-02-22}
}

@misc{CPRA,
  title = {California {P}rivacy {R}ights {A}ct {CPRA}.},
  author = {},
  howpublished = {\url{https://thecpra.org/}},
  note = {Accessed: 2025-02-22}
}

@misc{OCPA,
  title = {Oregon {C}onsumer {P}rivacy {A}ct {OCPA}.},
  author = {},
  howpublished = {\url{https://www.doj.state.or.us/consumer-protection/id-theft-data-breaches/privacy/}},
  note = {Accessed: 2025-02-22}
}

@article{SeattleDBReport2022,
author = {Abadi, Daniel and Ailamaki, Anastasia and Andersen, David and Bailis, Peter and Balazinska, Magdalena and Bernstein, Philip A. and Boncz, Peter and Chaudhuri, Surajit and Cheung, Alvin and Doan, Anhai and Dong, Luna and Franklin, Michael J. and Freire, Juliana and Halevy, Alon and Hellerstein, Joseph M. and Idreos, Stratos and Kossmann, Donald and Kraska, Tim and Krishnamurthy, Sailesh and Markl, Volker and Melnik, Sergey and Milo, Tova and Mohan, C. and Neumann, Thomas and Ooi, Beng Chin and Ozcan, Fatma and Patel, Jignesh and Pavlo, Andrew and Popa, Raluca and Ramakrishnan, Raghu and Re, Christopher and Stonebraker, Michael and Suciu, Dan},
title = {The Seattle report on database research},
year = {2022},
issue_date = {August 2022},
publisher = {Association for Computing Machinery},
address = {New York, NY, USA},
volume = {65},
number = {8},
issn = {0001-0782},
url = {https://doi.org/10.1145/3524284},
doi = {10.1145/3524284},
journal = {Commun. ACM},
month = jul,
pages = {72–79},
numpages = {8}
}

@inproceedings{byun2005purpose,
  title={Purpose based access control of complex data for privacy protection},
  author={Byun, Ji-Won and Bertino, Elisa and Li, Ninghui},
  booktitle={10th ACM Symposium on Access Control Models and Technologies},
  pages={102--110},
  year={2005},
}

@inproceedings{chaudhuri2003factorizing,
  title={Factorizing complex predicates in queries to exploit indexes},
  author={Chaudhuri, Surajit and Ganesan, Prasanna and Sarawagi, Sunita},
  booktitle={ACM SIGMOD Int. Conf. on Management of data},
  pages={361--372},
  year={2003},
}

@inproceedings{rizvi2004extending,
  title={Extending query rewriting techniques for fine-grained access control},
  author={Rizvi, Shariq and Mendelzon, Alberto and Sudarshan, Sundararajarao and Roy, Prasan},
  booktitle={ACM SIGMOD Int. Conf. on Management of data},
  pages={551--562},
  year={2004},
}

@inproceedings{nehme2013fence,
  title={Fence: Continuous access control enforcement in dynamic data stream environments},
  author={Nehme, Rimma V and Lim, Hyo-Sang and Bertino, Elisa},
  booktitle={3rd ACM Conf. on Data and Application Security and Privacy},
  pages={243--254},
  year={2013},
}

@inproceedings{agrawal2002hippocratic,
  title={Hippocratic databases},
  author={Agrawal, Rakesh and Kiernan, Jerry and Srikant, Ramakrishnan and Xu, Yirong},
  booktitle={PVLDB},
  pages={143--154},
  year={2002},
}

@inproceedings{lefevre2004limiting,
  title={Limiting disclosure in {h}ippocratic databases},
  author={LeFevre, Kristen and Agrawal, Rakesh and Ercegovac, Vuk and Ramakrishnan, Raghu and Xu, Yirong and DeWitt, David},
  booktitle={PVLDB},
  pages={108--119},
  year={2004},
}

@inproceedings{agrawal2005extending,
  title={Extending relational database systems to automatically enforce privacy policies},
  author={Agrawal, Rakesh and Bird, Paul and Grandison, Tyrone and Kiernan, Jerry and Logan, Scott and Rjaibi, Walid},
  booktitle={21st Int. Conf. on Data Engineering},
  pages={1013--1022},
  year={2005},
}

@book{loney2008oracle,
  title={Oracle Database 11g The Complete Reference},
  author={Loney, Kevin},
  year={2008},
  publisher={McGraw-Hill, Inc.}
}

@inproceedings{stonebraker1974access,
  title={Access control in a relational data base management system by query modification},
  author={Stonebraker, Michael and Wong, Eugene},
  booktitle={1974 Annual Conf.},
  pages={180--186},
  year={1974},
}

@inproceedings{colombo2018access,
  title={Access Control Enforcement within {MQTT}-based Internet of Things Ecosystems},
  author={Colombo, Pietro and Ferrari, Elena},
  booktitle={23nd ACM Symposium on Access Control Models and Technologies},
  pages={223--234},
  year={2018},
}

@article{bertino2011access,
  title={Access control for databases: concepts and systems},
  author={Bertino, Elisa and Ghinita, Gabriel and Kamra, Ashish and others},
  journal={Foundations and Trends in Databases},
  volume={3},
  number={1--2},
  pages={1--148},
  year={2011},
}

@inproceedings{lin2014privacy,
 author = {Lin, Jialiu and Liu, Bin and Sadeh, Norman and Hong, Jason I.},
 title = {Modeling Users' Mobile App Privacy Preferences: Restoring Usability in a Sea of Permission Settings},
 booktitle = {10th USENIX Conf. on Usable Privacy and Security},
 year = {2014},
 pages = {199--212}
}

@inproceedings{wang2007correctness,
  title={On the correctness criteria of fine-grained access control in relational databases},
  author={Wang, Qihua and Yu, Ting and Li, Ninghui and Lobo, Jorge and Bertino, Elisa and Irwin, Keith and Byun, Ji-Won},
  booktitle={PVLDB},
  pages={555--566},
  year={2007},
}

@article{wang2019idea,
  title={An IDEA: an ingestion framework for data enrichment in {AsterixDB}},
  author={Wang, Xikui and Carey, Michael J},
  journal={PVLDB},
  volume={12},
  number={11},
  pages={1485--1498},
  year={2019},
}

@article{FARAHANI2018659,
title = {Towards fog-driven IoT eHealth: Promises and challenges of IoT in medicine and healthcare},
journal = {Future Generation Computer Systems},
volume = {78},
pages = {659-676},
year = {2018},
issn = {0167-739X},
doi = {https://doi.org/10.1016/j.future.2017.04.036},
url = {https://www.sciencedirect.com/science/article/pii/S0167739X17307677},
author = {Bahar Farahani and Farshad Firouzi and Victor Chang and Mustafa Badaroglu and Nicholas Constant and Kunal Mankodiya}
}

@inproceedings{Acar2020,
author = {Acar, Abbas and Fereidooni, Hossein and Abera, Tigist and Sikder, Amit Kumar and Miettinen, Markus and Aksu, Hidayet and Conti, Mauro and Sadeghi, Ahmad-Reza and Uluagac, Selcuk},
title = {Peek-a-boo: i see your smart home activities, even encrypted!},
year = {2020},
isbn = {9781450380065},
publisher = {Association for Computing Machinery},
address = {New York, NY, USA},
url = {https://doi.org/10.1145/3395351.3399421},
doi = {10.1145/3395351.3399421},
booktitle = {Proceedings of the 13th ACM Conference on Security and Privacy in Wireless and Mobile Networks},
pages = {207–218},
numpages = {12},
location = {Linz, Austria},
series = {WiSec '20}
}

@article{Kroger2022,
author = {Kröger, Jacob},
year = {2022},
month = {06},
pages = {},
title = {The Privacy-Invading Potential of Sensor Data},
journal = {SSRN Electronic Journal},
doi = {10.2139/ssrn.4362987}
}

@inproceedings{mehrotra2016tippers,
  title={{TIPPERS}: A privacy cognizant IoT environment},
  author={Mehrotra, Sharad and Kobsa, Alfred and Venkatasubramanian, Nalini and Rajagopalan, Siva Raj},
  booktitle={2016 IEEE Int. Conf. on Pervasive Computing and Communication Workshops (PerCom Workshops)},
  pages={1--6},
  year={2016}
}

@article{hellerstein1998optimization,
  title={Optimization techniques for queries with expensive methods},
  author={Hellerstein, Joseph M},
  journal={ACM Transactions on Database Systems (TODS)},
  volume={23},
  number={2},
  pages={113--157},
  year={1998},
  publisher={ACM New York, NY, USA}
}

@inproceedings{pappachan2017towards,
  title={Towards privacy-aware smart buildings: Capturing, communicating, and enforcing privacy policies and preferences},
  author={Pappachan, Primal and Degeling, Martin and Yus, Roberto and Das, Anupam and Bhagavatula, Sruti and Melicher, William and Naeini, Pardis Emami and Zhang, Shikun and Bauer, Lujo and Kobsa, Alfred and others},
  booktitle={IEEE 37th Int. Conf. on Distributed Computing Systems Workshops (ICDCSW)},
  pages={193--198},
  year={2017}
}

@inproceedings{joon,
author = {Heo, Joon and Lim, Hyoungjoon and Yun, Sung Bum and Ju, Sungha and Park, Sangyoon and Lee, Rebekah},
title = {Descriptive and Predictive Modeling of Student Achievement, Satisfaction, and Mental Health for Data-Driven Smart Connected Campus Life Service},
year = {2019},
booktitle = {9th Int. Conf. on Learning Analytics \& Knowledge}
}

@inproceedings{DBLP:conf/percom/0001K17,
  author    = {Hosub Lee and
               Alfred Kobsa},
  title     = {Privacy preference modeling and prediction in a simulated campuswide
               IoT environment},
  booktitle = {{IEEE} Int. Conf. on Pervasive Computing and Communications,
               PerCom 2017},
  pages     = {276--285},
  year      = {2017}
}

@article{DBLP:journals/dase/ColomboF16,
  author    = {Pietro Colombo and
               Elena Ferrari},
  title     = {Fine-Grained Access Control Within {N}o{SQL} Document-Oriented Datastores},
  journal   = {Data Science and Engineering},
  volume    = {1},
  number    = {3},
  pages     = {127--138},
  year      = {2016},
}

@ARTICLE{6767117,  
author={P. {Colombo} and E. {Ferrari}},  
journal={IEEE Transactions on Knowledge and Data Engineering},   
title={Enforcement of Purpose Based Access Control within Relational Database Management Systems},   
year={2014},  
volume={26},  
number={11},  
pages={2703-2716}
}

@inproceedings{iotbenchmark,
  author    = {Peeyush Gupta and Michael J. Carey and Sharad Mehrotra and Roberto Yus},
  title     = {SmartBench: A Benchmark For Data Management In Smart Spaces},
  booktitle = {PVLDB},
  volume = {13},
  number = {11},
  year      = {2020},
}

\vspace{-2.5cm}

\begin{IEEEbiography}[{\includegraphics[width=1in,height=1.25in,clip,keepaspectratio]{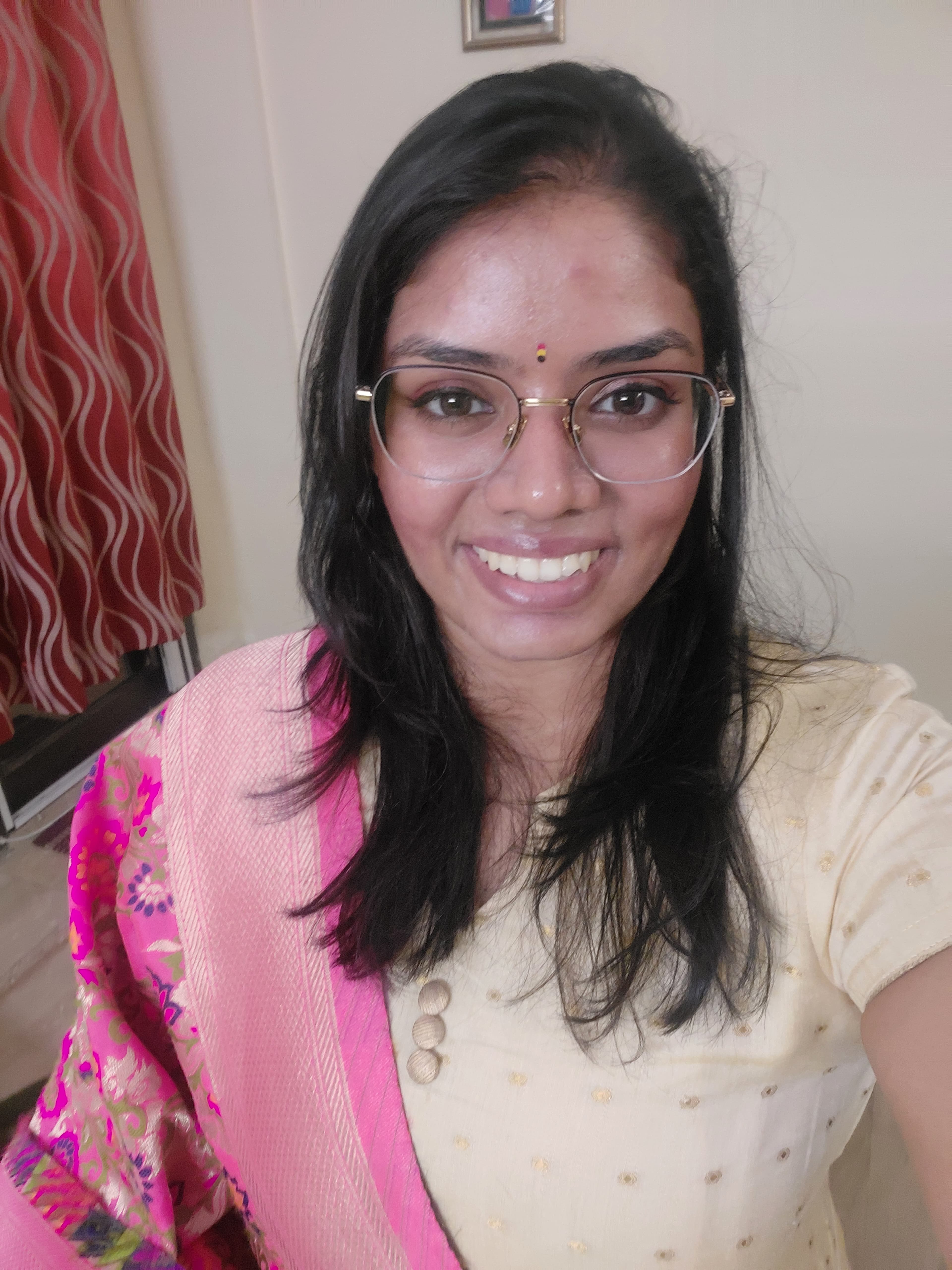}}]{Anadi Shakya}
received an MS degree in Software Engineering from Delhi Technological University, Delhi, in 2019, and a second MS degree in Computer Science from Portland State University, Oregon, in 2024. She is a Ph.D. student in Computer Science at Portland State University. Her research centers on data management, privacy, and enhancing operational efficiency in contemporary database environments.
\end{IEEEbiography}

\vspace{-2cm}

\begin{IEEEbiography}[{\includegraphics[width=1in,height=1.25in,clip,keepaspectratio]{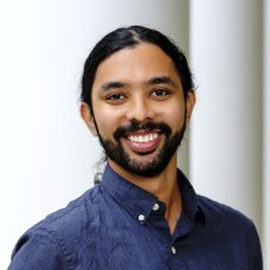}}]{Primal Pappachan}
(Member, IEEE) received the Ph.D.
degree in computer science from the University of
California, Irvine, in 2021. He is an assistant professor
with the Department of Computer Science, Portland
State University. Afterwards, he was a postdoctoral
scholar with the College of Information Sciences
and Technology at Pennsylvania State University.
His research interests are in the intersection of data
management and privacy, particularly data protection
methods such as access control, differential privacy, and privacy policies.\end{IEEEbiography}

\vspace{-2cm}

\begin{IEEEbiography}[{\includegraphics[width=1in,height=1.25in,clip,keepaspectratio]{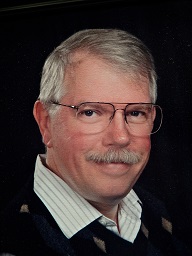}}]{David Maier}
(Senior Member, IEEE) received his Ph.D. in Electrical Engineering
and Computer Science from Princeton University. 
He is Maseeh Professor Emeritus of
Emerging Technologies in the Department of
Computer Science at Portland State University.
He is an ACM Fellow and an IEEE Senior
Member. He received an NSF Young Investigator Award in 1984, was awarded the 1997
ACM SIGMOD Innovations Award for his contributions in objects and databases and received
a Microsoft Research Outstanding Collaborator
Award in 2016.\end{IEEEbiography}
\vspace{-2cm}

\begin{IEEEbiography}[{\includegraphics[width=1in,height=1.25in,clip,keepaspectratio]{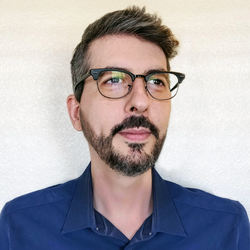}}]{Roberto Yus}
received the Ph.D. degree in computer science from the University of Zaragoza,
Spain, in 2016, researching on issues related
to semantic data management in distributed and
mobile environments. He is currently an assistant professor at the Department of Computer Science and Electrical Engineering at the University of Maryland, Baltimore County. His current research
interest includes privacy issues in data management on the Internet of Things.\end{IEEEbiography}
\vspace{-2cm}

\begin{IEEEbiography}[{\includegraphics[width=1in,height=1.25in,clip,keepaspectratio]{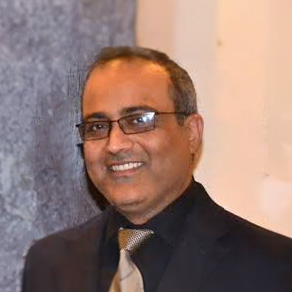}}]{Sharad Mehrotra} (Fellow, ACM \&  IEEE) received the PhD
degree in computer science from the University of Texas, Austin, Austin, Texas, in 1993. He is currently a Distinguished Professor of Computer Science
at the  University of California, Irvine. He has received numerous awards
and honors, including the 2011 SIGMOD Best Paper
Award, 2007 DASFAA Best Paper Award, SIGMOD
Test of Time Award, 2012, DASFAA ten year best
paper awards for 2013 and 2014, 1998 CAREER Award from the US National
Science Foundation (NSF), and ACM ICMR Best Paper Award for 2013.
His primary research interests include the area of database management, distributed systems, secure databases, and Internet of Things.
\end{IEEEbiography}

\vspace{-2cm}

\begin{IEEEbiography}[{\includegraphics[width=1in,height=1.25in,clip,keepaspectratio]{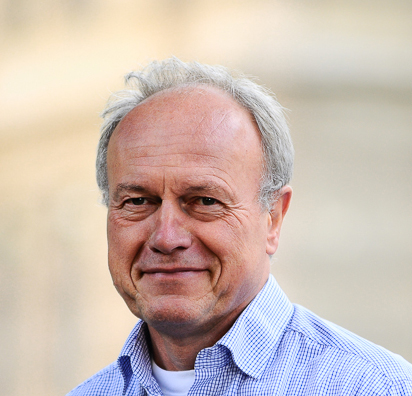}}]{Johann-Christoph Freytag}
(Member, IEEE) received the PhD degree in applied mathematics and computer science from Harvard University in 1985. He is a professor emeritus with the Institut für Informatik at Humboldt-Universität zu Berlin, Germany, where he was professor for Databases and Information Systems (DBIS) since 1994. From 1985 to 1994, before joining the department, Freytag spent nine years in industrial research at the IBM Almaden Research Center (1985-1987), at the European Computer Industry Research Centre (ECRC), Munich (1987-1989), and at DEC’s (Digital Equipment) Database System Research Center, Munich (1990-1993). 
His research spans different areas: query processing in scalable data management systems, information integration, data and information quality, privacy and data systems, information as a service, and new hardware architectures for data and information management.
He is a member of the ACM and a member of the Gesellschaft für Informatik (GI), Germany.
 TBA.\end{IEEEbiography}

\appendices

\section{Implementing Operator $\Delta$ and Its Integration with Guards}
\label{appendix:delta-guard}

This appendix provides additional implementation details and performance evaluation for the $\Delta$ operator used in \oursystem. The $\Delta$ operator enables tuple-aware policy evaluation and is an optimization strategy in combination with guarded expressions(GEs). We describe how $\Delta$ is implemented using User Defined Functions (UDFs), explain the conditions under which it is selected over inlined policy evaluation, and present an experiment, which empirically compares the cost trade-offs between inlining and invoking $\Delta$, as well as the choice between query-based and guard-based indexing strategies.

\subsection{Implementing Operator \texorpdfstring{$\Delta$}{Delta}}
\label{appendex:UDF}

We implement the policy evaluation operation $\Delta$ (see Section~\ref{sect:ourApproach}) by User Defined Functions (UDFs) on top of a DBMS. 
\ifextended
Consider a set of policies $\vPolicySet{}$ and the query metadata $\vQueryMetadata{i}{}$ and a tuple $\vTuple{}{t}$ belonging to relation $\vRelation{}{j}$. $\Delta(\vPolicySet{}, \vQueryMetadata{i}{}, \vTuple{}{t})$ is implemented as the following UDF
\begin{lstlisting}[style=mystyle]
CREATE FUNCTION delta($[policy]$, $querier$, $purpose$, $[attrs]$) 
{BEGIN
 Cursor c = 
  SELECT $\vRelation{}{OC}.attr$ as attr, $\vRelation{}{OC}.op$ as op, $\vRelation{}{OC}.val$ as val
  FROM $\vRelation{}{P}, \vRelation{}{OC}$ 
  WHERE $\vRelation{}{P}.querier = querier$ AND $\vRelation{}{P}.purpose = purpose$ AND $\vRelation{}{P}.id$ IN $[policy]$ AND $\vRelation{}{P}.owner = [attrs].owner$ AND $\vRelation{}{P}.id = \vRelation{}{OC}.policy-id$
  LET satisfied_flag = true
  READ UNTIL c.isNext() = false:
     FETCH c INTO p_attr, p_op, p_val
      FOR each t_attr in $[attrs]$
        IF t_attr = p_attr THEN
          satisfied_flag = satisfied_flag AND /*Check whether t_val satisfies p_op p_val*/
  return satisfied_flag
END}
\end{lstlisting}

The UDF above performs two operations: 1) It takes a set of policies and retrieves a subset $\hat{\vPolicySet{}}$ which contains the relevant policies to be evaluated based on the query metadata $\vQueryMetadata{i}{}$ and the tuple $\vTuple{}{t}$; 2) It performs the evaluation of each policy $\vPolicy{}{i} \in \hat{\vPolicySet{}}$ on $\vTuple{}{t}$. 
\else
Given a set of policies $\vPolicySet{}$, the query metadata $\vQueryMetadata{i}{}$, and a tuple $\vTuple{}{t}$ belonging to relation $\vRelation{}{j}$. $\Delta(\vPolicySet{}, \vQueryMetadata{i}{}, \vTuple{}{t})$ is implemented as a UDF which performs two operations: 1) It takes a set of policies and retrieves a subset $\hat{\vPolicySet{}}$ which contains the relevant policies to be evaluated based on the query metadata $\vQueryMetadata{i}{}$ and the tuple $\vTuple{}{t}$. This way, it retrieves the policies that are defined by the owner of the tuple for the specific querier and her purpose. 2) It performs the evaluation of each policy $\vPolicy{}{i} \in \hat{\vPolicySet{}}$ on $\vTuple{}{t}$. 
\fi

\subsection{Combining \texorpdfstring{$\Delta$}{Delta} with Guards}
\label{appendex:selectingGuardStrategy}

Depending upon the number of policies in the associated guard partition (i.e., $\vSetCardinality{\vPolicySet{\vGuard{}{i}}}$), we could rewrite the policy partition part using the $\Delta$ operator as $\Delta(\vPolicySet{\vGuard{}{i}}, \vQueryMetadata{i}{}, \vTuple{}{t})$ if it reduces the execution cost,  instead of checking the polices inline as shown in Section~\ref{sect:implementingGuards}. 
The $\Delta$ operator has an associated cost due to the invocation and execution of a UDF. For each guarded expression $\vGuard{}{i}$ in a guarded policy expression $\vPolicyGuardedExpression{}$ for a relation $\vRelation{}{i}$ and a specific querier and purpose, we check the overhead of using the $\Delta$ operator (to which we will refer to as $Guard\&\Delta$) versus not using it ($Guard\&Inlining$) and use $\Delta$ if $\vCostMethod{Guard\&\Delta}<\vCostMethod{Guard\&Inlining}$.

We model the cost of each strategy by computing the cost of evaluating policies per tuple since the number of tuples to check are the same in both cases. As modeled in Equation~\ref{eq:costEvalTuplePolicies},  $\vCostMethod{Guard\&Inlining}=\vShortCircuit.\vSetCardinality{\vPolicySet{\vGuard{}{i}}}.\vEvalCost$ where the values of $\vShortCircuit$, the percentage of policies that have to be checked before one returns true, and $\vEvalCost$, the cost of evaluating a policy against a single tuple, are obtained experimentally. We compute $\vShortCircuit$ by executing a query which counts the number of policy checks done over $\vPolicySet{\vGuard{}{i}}$ before a tuple either satisfies one of the policies or is discarded (does not satisfy any policy) and averaging the number of policy checks across all tuples. 
We estimate $\vEvalCost$ by computing the difference of the read cost per tuple without policies (estimated by dividing the time it takes to perform a table scan by the total number of tuples) and the average cost per tuple with policies. The former is estimated by executing a table scan with different number of policies with different selectivities (number of tuples) and averaging the cost per tuple per policy.
$\vCostMethod{Guard\&\Delta} = UDF_{inv} + UDF_{exec}$ where the two factors represent the cost of invocation and execution of the UDF, respectively\footnote{Recent work, such as~\cite{wang2019idea}, shows that in some situations batching of UDF operations might be possible to save the overhead of $UDF_{inv}$ per tuple. While current DBMSs generally lack support for this optimization, our model could be easily adapted to consider such cost amortizations.}. We obtain this cost experimentally by executing $\Delta$ with varying number of tuples (by changing the selection predicate before $\Delta$) and the number of policies to be checked against (by changing the guard associated with invocation of operator).

Most of the terms in both cost models are constants, the term that varies depending on the specific guard is $\vSetCardinality{\vPolicySet{\vGuard{}{i}}}$. Our experiments (see Section~\ref{sect:expSieve}) indicate that the usage of the $Guard\&\Delta$ strategy is beneficial if $\vSetCardinality{\vPolicySet{\vGuard{}{i}}}>120$.

When \oursystem selects the $\Delta$ strategy for a specific guard $\vGuard{}{i}$, the query rewrite includes a function call to the $\Delta$ UDF instead of expanding the policy expression inline. This allows the system to defer policy evaluation to the tuple runtime while using guard-level selectivity for efficient indexing. For example, consider the following rewritten query using $\Delta$ for the final guard partition:

\begin{lstlisting}[style=mystyle]
WITH WiFiDatasetPol AS (
 SELECT * FROM WiFiDataset as W FORCE INDEX($\vObjectCondition{1}{g} \cdots \vObjectCondition{n}{g}$)
 WHERE ($\vObjectCondition{1}{g}$ AND W.ts-date between "9/25/19" AND "12/12/19" AND ($\vObjectCondition{1}{1}$ AND $\cdots$ AND $\vObjectCondition{1}{n}$))
        OR $\cdots$ OR 
       ($\vObjectCondition{n}{g}$ AND W.ts-date between "9/25/19" AND "12/12/19" AND delta(32,"Prof.Smith", "Analysis","owner","ts-date", "ts-time", "wifiAP")=true)
) StudentPerf(WifiDatasetPol, Enrollment, Grades)
 \end{lstlisting}

In this example, the policy partition associated with $\vObjectCondition{n}{g}$ is evaluated by the $\Delta$ operator using the attributes of each tuple in the $\texttt{WiFiDataset}$. This UDF-based evaluation is preferred when the number of policies in the guard partition is large and tuple-specific filtering leads to better performance.

\vspace{0.1cm}
\subsection{\textbf{Experiment: }~Inline vs. Operator $\Delta$ \& Query Index vs. Guard Index}\label{sect:exp-inlive vs delta}

\oursystem uses a cost model to determine for each guard whether to inline the policies or to evaluate the policies using the $\Delta$ operator. The $\Delta$ operator has an associated overhead of UDF invocation, but it can utilize the tuple context to reduce the number of policies that need to be checked per tuple. To study this tradeoff in both inlining and using the $\Delta$ operator, we gradually increased the number of policies that are part of the partition of a guard and observed the cost of policy evaluation. As expected, we observed that when the number of policies is about 120, the cost of UDF invocations is amortized by the savings from filtering policies by the $\Delta$ operator (see Figure~\ref{fig:designChoice1}). 

\begin{center}
\begin{minipage}{\linewidth}
    \begin{minipage}[b]{0.49\linewidth}
        \vspace{0pt}
        \centering
        \includegraphics[width=\columnwidth]{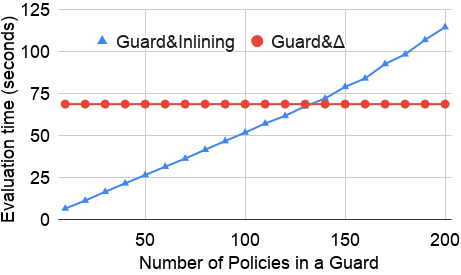}
        \captionof{figure}{$Inlining$ vs. $\Delta$.}
        \label{fig:designChoice1}
    \end{minipage}
    \hfill
    \begin{minipage}[b]{0.5\linewidth}
        \vspace{0pt}
        \centering
        \includegraphics[width=\columnwidth]{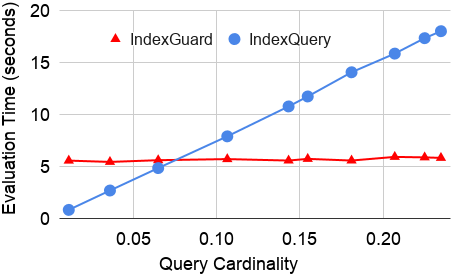}
        \captionof{figure}{Index choice.}
        \label{fig:designChoice2}
  \end{minipage}
\end{minipage}
\end{center}


In \oursystem, we use a cost model to choose between using the  \textit{IndexQuery} and \textit{IndexGuards} as explained in Section~\ref{sect:implementingSieve}. We evaluated this cost model by analyzing the evaluation cost against increasing query cardinality for three guard cardinalities (low, medium, high). Figure~\ref{fig:designChoice2} shows the results averaged across these three guard cardinalities. As expected, at low query cardinality, it is better to utilize \textit{IndexQuery}, while at medium and high query cardinalities ($>0.07$), \textit{IndexGuards} are the better choice. In both these options, guarded expressions are used as filters on top of the results from Index Scan.


\section{Workload Generation Methodology}
\label{appendix:workload}

The workload generation process involves several steps to ensure that the workload accurately mimics real-world conditions, such as Sampling Users, Policy Creation, Query Creation, and Workload Generation. Each of these steps is described in further detail below.

\begin{table*}[!ht]
\caption{Users for the two scenarios\label{tab:user-sampling}}
\centering
\begin{tabular}{l l l l l l l}
\multirow{2}{*}{Scenarios} & \multicolumn{6}{c}{User Profiles}\\
\cline{2-7}
 & visitor & staff & graduate & undergrad & faculty & total \\
\hline
Marking of Attendance & N.A & N.A & 1,394 & 1,758 & 388 & 3,540\\
\hline
Monitoring of Space Usage & 31,796 & 1,029 & 1,428 & 1,795 & 388 & 36,436\\
\end{tabular}
\end{table*}

\subsection{Sampling Users:}
First, we sample the users to filter the ones that are relevant to the given scenario. For the \textit{marking attendance} scenario, we include users with roles defined as \textit{faculty}, \textit{undergraduate students} (undergrad) and \textit{graduate students} (graduate) in \textit{User Sampling}. Similarly, for \textit{marking space usage} scenario, we include all the roles: \textit{visitor} as well as non-visitor, which encompass \textit{staff}, \textit{faculty}, \textit{undergraduate students} (undergrad) and \textit{graduate students} (graduate) as shown in Table \ref{tab:user-sampling}. Each user is associated with a unique user ID.

\subsection{Policy Creation: }We define 10 policies for each user, ensuring they have a substantial set of policies to govern their interactions in the smart space.

The policies consist of query conditions, object conditions, and policy actions, as discussed in Section \ref{sect:modeling}.
Depending upon the scenario selected, their values may vary.
For the \textit{marking attendance} scenario, students define policies for faculty members. 

Furthermore, in each scenario, we include additional checks to enhance the realism of the generated workload. For the \textit{marking attendance} scenario, we ensure that each student defines a policy for a querier (\textit{faculty}) in a way that they share the same location. This approach guarantees that the policy is relevant to classrooms, as the querier is always a faculty member. Each policy is defined by the student for the entire three-month term, with some policies having a late start date to indicate late enrollment in the class and others having an earlier end date to indicate early or mid-term dropouts. We also have policies defined for a day or two, reflecting guest lectures. The querier is consistently set to a faculty member, and the purpose and action are set to \textit{marking attendance} and allow, respectively. This strategy avoids other locations on campus, such as labs and study rooms, where students might be present but are not relevant for \textit{marking attendance}. Consequently, policies for \textit{marking attendance} are not generated in such cases, as given in Listing~\ref{list:attendance}.

\begin{lstlisting}[style=mystyle, caption= Sample Policy for Marking Attendance Scenario., label=list:attendance]
object_conditions=[
    user_id = 177, 
    user_profile = graduate, 
    date = 2018-02-01>=, 2018-04-30<=,
    time = 08:00:00>=, 10:00:00<=] 
querier_conditions=[
    querier = 51, 
    purpose = marking attendance]
action = allow
\end{lstlisting}

For the \textit{space usage monitoring} scenario, we include all roles, such as \textit{visitors}, \textit{staff}, \textit{faculty}, \textit{undergraduate students}, and \textit{graduate students}. Each user defines 10 policies to ensure comprehensive governance of their interactions in the smart space. Any user, either a staff member or a faculty member, can define policies.

The policies cover a three-month period with varying date and time possibilities, reflecting different usage patterns. For instance, policies for PhD spaces where graduates might spend more time include late hours, while lab policies account for varying durations of stay. Additionally, this scenario can extend to recording space usage for events, allowing an event manager, either a staff or faculty member, to check the number of people present, as depicted in the Listing~\ref{list:space}.

\begin{lstlisting}[style=mystyle, caption= Sample Policy for Space Utilization Scenario., label=list:space]
object_conditions=[
    user_id = 76, 
    user_profile = staff, 
    date = 2018-02-01>=, 2018-04-30<=,
    time = 08:00:00>=, 17:00:00<=]   
querier_conditions=[
    querier = 2576, 
    purpose = $\texttt{space-utilization}$]
action = allow
    \end{lstlisting}

\subsection{Query Creation: }Queries are generated based on the selected scenario and query templates, which ensure consistency and realism. These templates include location-based queries, user-specific queries, and aggregated presence counts by location. We ensure that each querier has policies defined over it, though the selectivity of queries may vary. 

The templates over which we generate queries are as follows:

\begin{enumerate}[label=\arabic*.]
    \item \textbf{Query 1: location-based}
    \begin{lstlisting}[style=mystyle]
    SELECT ...
    FROM ...
    WHERE date >= "..." AND date <= "..."
        AND time >= "..." AND time <= "..."
        AND location_id IN ("loc1", "loc2", ...)
    \end{lstlisting}

    \item \textbf{Query 2: user-specific}
    \begin{lstlisting}[style=mystyle]
    SELECT ...
    FROM ...
    WHERE date >= "..." AND date <= "..."
        AND time >= "..." AND time <= "..."
        AND user_id IN (id1, id2, ...)
    \end{lstlisting}

    \item \textbf{Query 3: aggregated presence counts by locations}
    \begin{lstlisting}[style=mystyle]
    SELECT location_id, COUNT(*)
    FROM PRESENCE
    WHERE time >= "..." AND time <= "..."
    GROUP BY location_id
    \end{lstlisting}
\end{enumerate}

In the \textit{marking attendance} scenario, a faculty member acts as the querier, querying information about students present in their classroom. The predicates involved in the query are time, date, location ID, and user ID. By varying these predicates, we formulate queries according to the indicated templates. For this scenario, the querier is always selected from the pool of faculty, and the location ID maps to designated classrooms. 

Relevant queries generated based on the given templates are as follows:

\begin{enumerate}
    \item \textbf{Q1}: 
    \begin{lstlisting}[style=mystyle]
    SELECT * FROM WiFi_Dataset AS W 
    WHERE W.location_id IN $locations$ 
        AND W.time BETWEEN t1 AND t2 
        AND W.date BETWEEN d1 AND d2
    \end{lstlisting}
                                           
    \begin{itemize}
        \item The faculty member needs to check which students were present during a particular class session.
        \item The query helps track attendance by filtering WiFi connections in designated classroom areas within a specified time and date range.
    \end{itemize}
    \item \textbf{Q2}: 
    \begin{lstlisting}[style=mystyle]
    SELECT * FROM WiFi_Dataset AS W 
    WHERE W.USER_id IN $users$ 
        AND W.time BETWEEN t1 AND t2 
        AND W.date BETWEEN d1 AND d2
    \end{lstlisting}  
    
    \begin{itemize}
        \item The faculty member wants to verify the attendance of specific students.
        \item This query allows checking which individual students were connected to the WiFi during a particular class period, aiding in attendance verification.
    \end{itemize}
    \item \textbf{Q3}: 
    \begin{lstlisting}[style=mystyle]
    SELECT W.location_id, COUNT(*) 
    FROM WiFi_Dataset AS W 
    WHERE W.time BETWEEN t1 AND t2 
        AND W.date BETWEEN d1 AND d2
    GROUP BY W.location_id
    \end{lstlisting}
    
    \begin{itemize}
        \item Assessing the overall occupancy or attendance in different classroom areas over a specified period.
        \item This query helps determine how many devices (students) were connected to the WiFi network in each designated room during the specified time and date range, providing insights into room usage and attendance patterns.
    \end{itemize}
\end{enumerate}

These templates allow comprehensive attendance tracking, ensuring queries are accurately targeted to relevant classroom locations and time frames.

In the \textit{space usage monitoring} scenario, a staff member or a faculty is selected as the querier. The queries are designed to track users and non-visitors visited areas, similar to the one described in \textit{marking attendance} scenario. 

\subsection{Workload Generation: }\label{sec:wg-gen}
In our implementation, we follow a structured approach of user sampling, policy creation, query formation, and workload generation.
We previously explained the first three steps, and now we will explain how the workload generator combines them to create a workload.

To reflect real-world scenarios, the workload includes a mix of previously \textit{seen} and \textit{unseen} queries, i.e., sampling with selective replacement. It mimics the situation where the same query is repeatedly posed to the system.
To implement the specified queries, we employ a sliding window mechanism that retains the most recent 10 queries from the workload, categorizing any query outside this window as \textit{unseen}. We alternate between picking an unseen query from the query set and a seen query from the sliding window. This simulates the scenario where the same query might be executed multiple times over a period of time. For example, an administrator might check the usage of a meeting room to determine if it has become available. 


We studied two distinct operational states: the steady state and the bursty state. In the steady state, workload configurations involve a consistent rate, where either the number of policy insertions (\textbf{xP1Q}) or query insertions (\textbf{1PyQ}) is varied but remains steady throughout the workload. On the other hand, the bursty state is characterized by fluctuating workload patterns, where the rate of policy and query insertions changes from one epoch to the next. This approach allows us to capture query execution rates and cache hit-and-miss ratios effectively, providing an evaluation of the policy models under different workload conditions. 

Throughout this paper, we discuss different workload configurations in terms of the rate of policy insertions and queries posed to the DBMS, expressed as \textit{x}\textbf{P}\textit{y}\textbf{Q}, where \textit{x} and \textit{y} represent the number of policies and queries within a unit of time. For example, \textbf{2P1Q} indicates that in each epoch—representing a fixed cycle of operations—two policies are inserted into the system, followed by one query. This means that two new policies are added between each query, defining the operations structure within the epoch.

For a comprehensive understanding, refer to Table \ref{tab:ac-stats} and Table \ref{tab:su-stats} for the statistics of the data generated for the experiment. Table \ref{tab:ac-stats} shows that the total users include faculty, undergraduate students, and graduate students. However, the number of policyholders is smaller because policies are not defined for faculty members. In contrast, in the space-usage scenario, anyone can define a policy for the queriers (faculty and staff members). The number of queries generated is half the number of policies to maintain a balanced workload. It ensured we would not run out of questions before inserting all the policies into the system.

\begin{table}[h]
\caption{Marking of Attendance Scenario Statistics.\label{tab:ac-stats}}
\centering
\scriptsize
\begin{tabular}{l l l}
\textbf{Users} & $\vUserSet{ac}$ & 3540\\
\hline
\textbf{Policy Holder} & $\vUserSet{p}$ & 3152\\
\hline
\textbf{No. of policies per user} & $x$ & 10\\
\hline
\textbf{Total no. of Policies} & $\vPolicySet{ac}$ = $x \times \vUserSet{p}$ & 31520\\
\hline
\textbf{Total no. of Queries} & $\vQuery{}{ac}$ = $\frac{\vPolicySet{ac}}{2}$ & 15760\\
\hline
\textbf{Unseen - Seen Queries} & $\frac{\vQuery{}{ac}}{2}$ & 7880\\
\hline
\textbf{Queriers} & $\vUserSet{q}$ & 388\\
\end{tabular}
\end{table}

\begin{table}[h]
\caption{Space Usage Monitoring Scenario Statistics.\label{tab:su-stats}}
\centering
\scriptsize
\begin{tabular}{l l l}
\textbf{Users} & $\vUserSet{su}$ & 36436\\
\hline
\textbf{Policy Holder} & $\vUserSet{p}$ & 36436\\
\hline
\textbf{No. of policies per user} & $x$ & 10\\
\hline
\textbf{Total no. of Policies} & $\vPolicySet{su}$ = $x \times \vUserSet{p}$& 364360\\
\hline
\textbf{Total no. of Queries} & $\vQuery{}{su}$ = $\frac{\vPolicySet{su}}{2}$  & 182180\\
\hline
\textbf{Unseen - Seen Queries} & $\frac{\vQuery{}{su}}{2}$ & 91090\\
\hline
\textbf{Queriers} & $\vUserSet{q}$ & 1417\\
\end{tabular}
\end{table}


\section{Additional Experiments on Workload-Driven Cache Behavior}
\label{appendix:workload-cache-exp}




This appendix presents additional experiments that extend our caching evaluation to more dynamic workloads. We examine cache behavior across three dimensions: (1) varying policy and query insertion rates, (2) bursty workloads with shifting ratios, and (3) the impact of policy deletions. We measure cache effectiveness by tracking hit, miss, and soft-hit rates, offering insights into cache reuse and data staleness under different conditions.

\subsection{Experiment: Assessing Cache Effectiveness in Steady State}

\begin{figure*}[!b]
\centering
\subfloat[]{\includegraphics[width=3.5in]{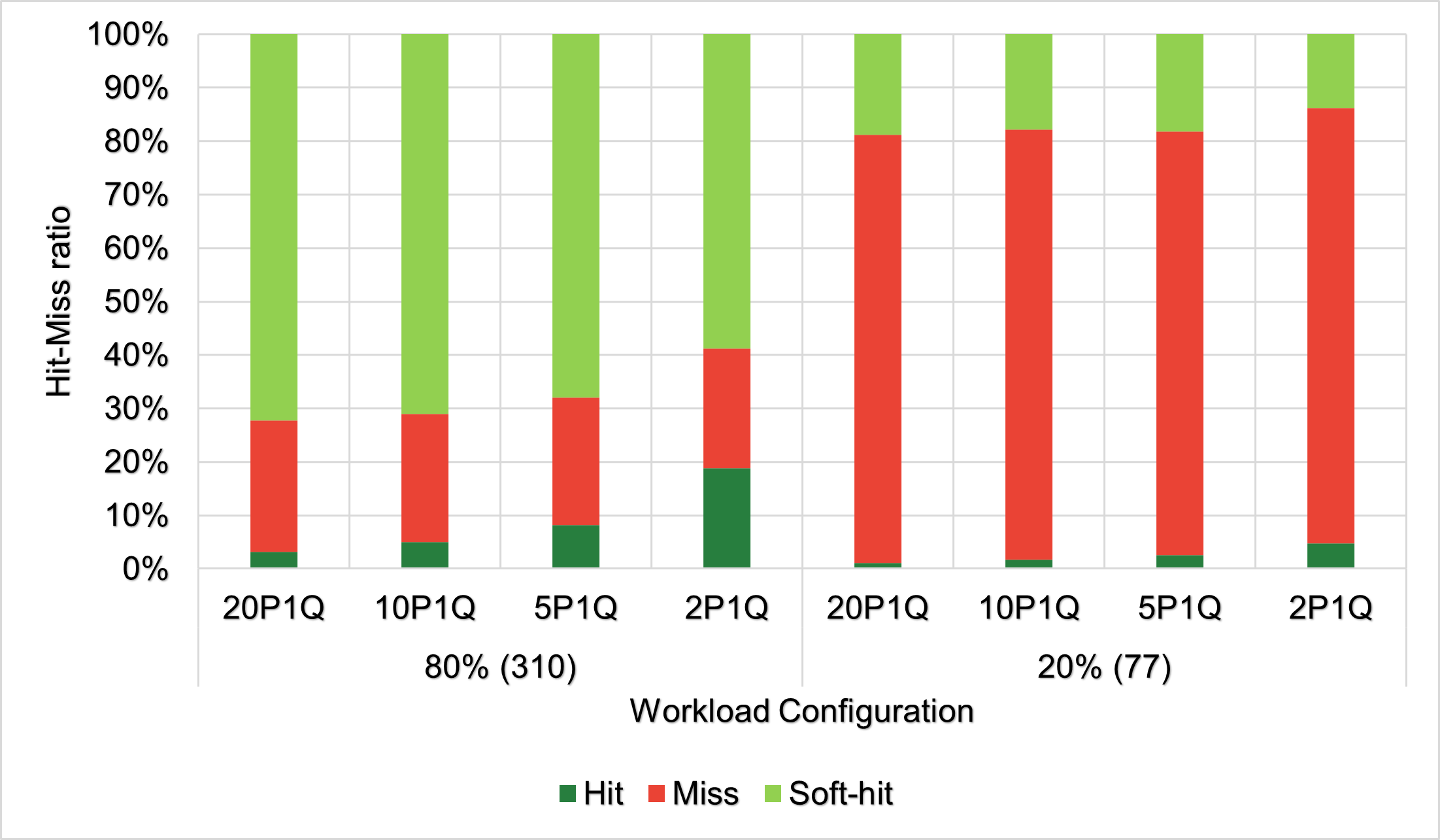}%
\label{fig:S-cache-xp}}
\hfil
\subfloat[]{\includegraphics[width=3.5in]{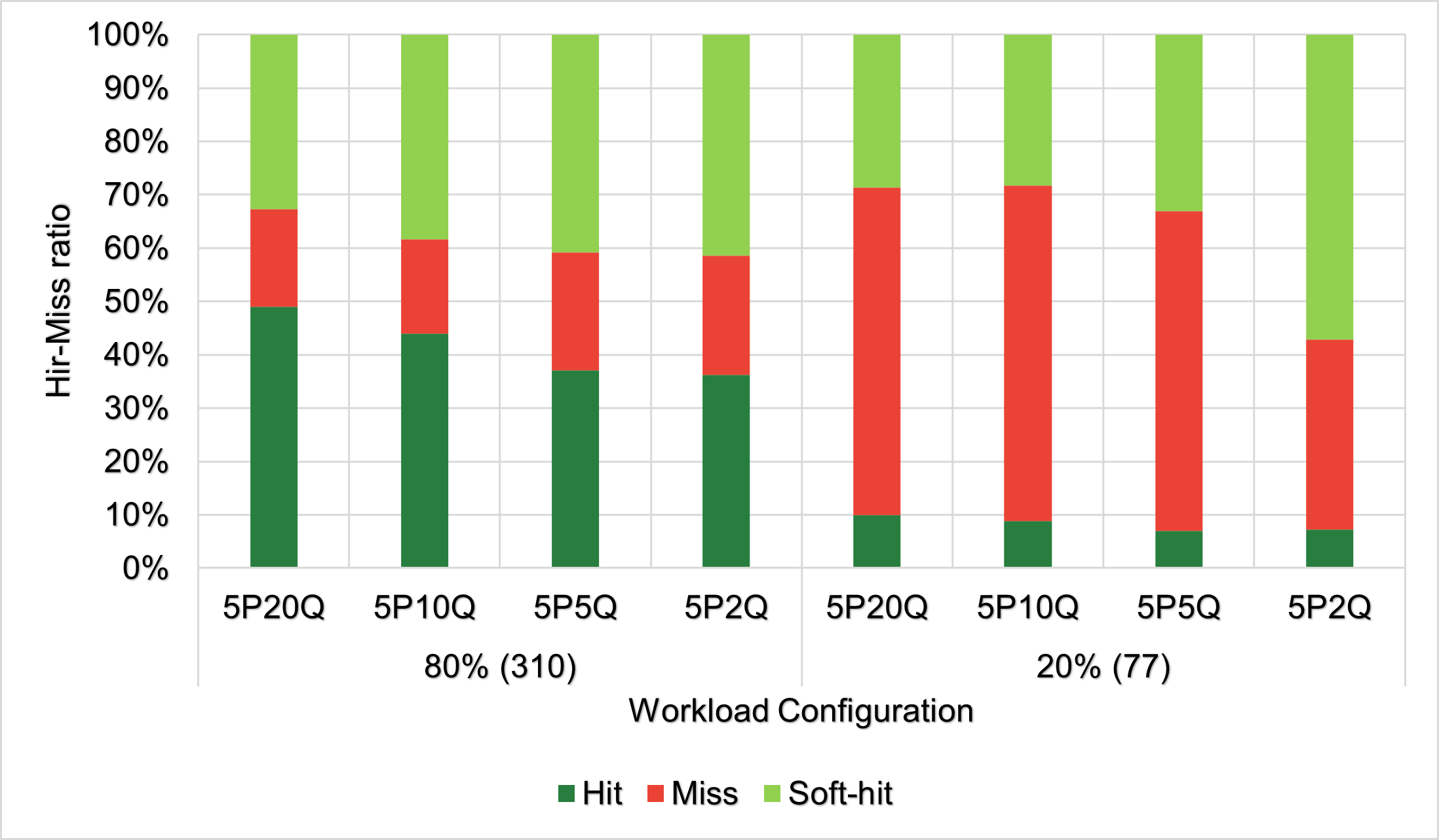}%
\label{fig:S-cache-yq}}\\
\caption{Analyzing cache hit rate in steady state for \textit{marking attendance} scenario :(a) Varying policy insertion \textbf{xP1Q} and (b) Varying query insertion \textbf{5PyQ}.}
\label{fig:S-cache}
\end{figure*}

\begin{table}[h]
\caption{Workload configuration and statistics: Marking Attendance Scenario in a steady state.\label{tab:ac-steady-stats}}
\centering
\scriptsize
\begin{tabular}{l l l}
Workload & Total Policy Inserted & Total Query Executed\\
    \hline
    2P1Q & 6304 & 3152\\
    \hline
    5P1Q & 15760 & 3152\\
    \hline
    10P1Q & 31520 & 3152\\
    20P1Q & 63040 & 3152\\
    \hline
    \hline
    5P20Q & 3940 & 15760\\
    \hline
    5P10Q & 7880 & 15760\\
    \hline
    5P5Q & 15760 & 15760\\
    \hline
    5P2Q & 39400 & 15760\\
\end{tabular}
\end{table}

Our investigation has unveiled intriguing trends as shown in Figure \ref{fig:S-cache-xp} when performed over \textit{marking of attendance} scenario for varying policy insertion workload configuration \textbf{2P1Q - 20P1Q} and varying query insertion \textbf{5P2Q-5P20Q}.  The dataset details can be found in Table \ref{tab:ac-stats}, while Table \ref{tab:ac-steady-stats} summarizes the workload configuration and overall query statistics. As expected, with each increment in the number of inserted policies, we observed a corresponding increase in soft hits, indicating a proliferation of outdated data within the cache.

\begin{figure}[!t]
\centering
\includegraphics[width=\linewidth]{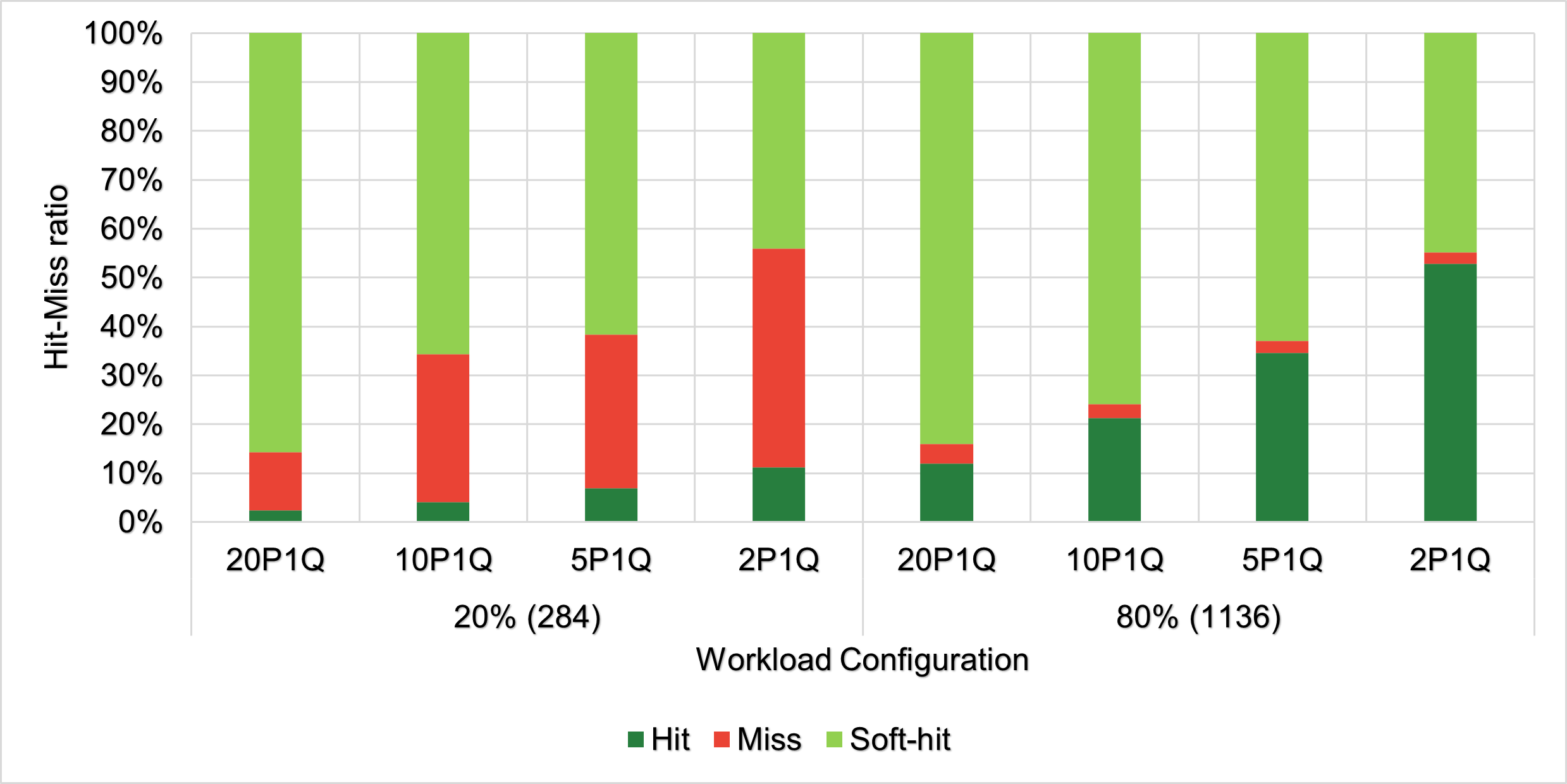}
\caption{Analyzing cache hit rate in steady state for \textit{space usage monitoring} scenario in a varying policy insertion \textbf{xP1Q} workload.}
\label{fig:S-cache-su}
\end{figure}

As we reduce the cache size from 80\% to 20\%, we observe a significant increase in the miss rate, indicating that the cache becomes too small and inefficient, performing similarly to \oursystem without caching. These results suggest that \oursystem benefits from a higher cache size. In the space usage scenario, as shown in Figure \ref{fig:S-cache-su}, reducing the cache size from 80\% to 20\% leads to a noticeable rise in miss rates, similar to our previous results. 

\subsection{Experiment: Assessing Cache Effectiveness under Bursty Workloads}  

\begin{figure}[!t]
    \centering
    \includegraphics[width=\linewidth]{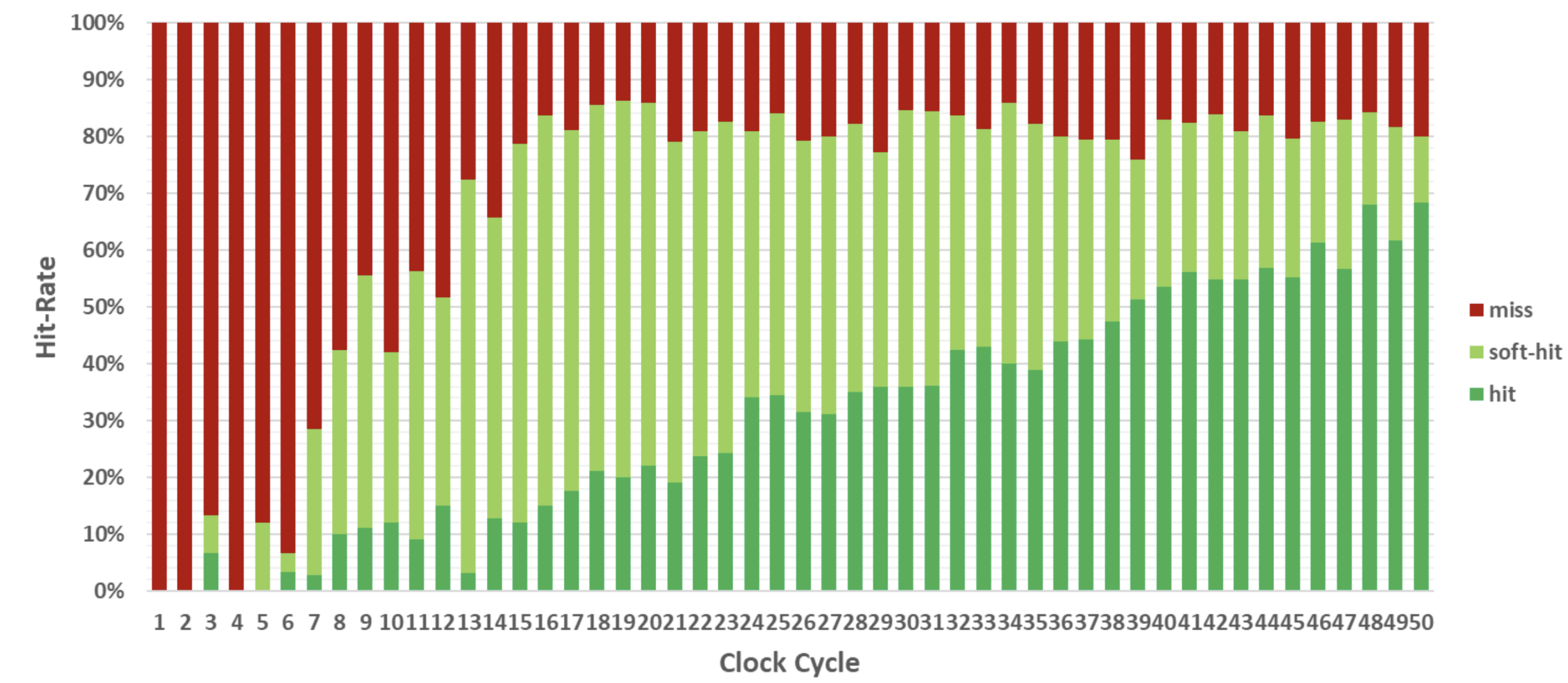}
    \caption{Cache hit-miss and soft hit counts during bursty workload.}
    \label{fig:bursty-hit-miss-soft}
\end{figure}

\begin{figure}[!t]
    \centering
    \includegraphics[width=\linewidth]{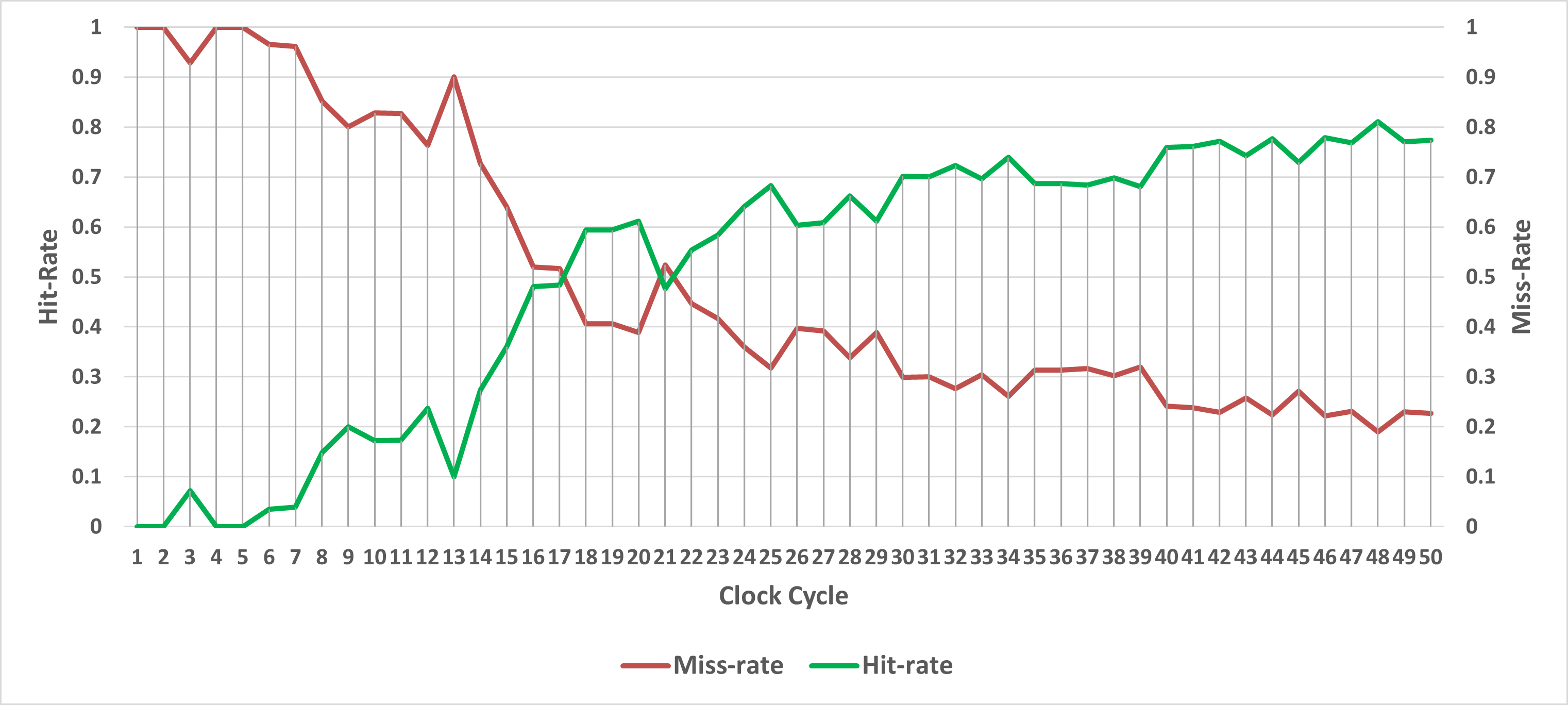}
    \caption{Hit and miss ratios per clock cycle during bursty workload.}
    \label{fig:bursty-hit-miss-ratio}
\end{figure}

This experiment models a bursty workload scenario, starting with high policy insertion rates that taper off while query rates gradually increase, mimicking dynamic *Attendance Control* scenarios. The workload progression involves a shift from configurations with high policy insertions and low query rates (e.g., \textbf{500P1Q}, representing 500 policies per 1 query) to configurations with low policy insertions and high query rates (e.g., \textbf{1P250Q}, representing 1 policy per 250 queries). This gradual transition, achieved by decreasing policies by 10 and increasing queries by 5 at each clock cycle, allows us to observe cache performance trends and compare cache hit-and-miss rates across different workload phases.

The experiment was conducted with a cache size of 80\%, a window size of 10, and a total workload of 6,375 queries and 12,750 policies. Three graphs illustrate the results, each providing insights into different aspects of the caching performance.

We observe a significant improvement in cache hit rates as the workload transitions from policy-dominant to query-dominant phases, with a corresponding decrease in miss rates. Initially, frequent policy insertions lead to instability and lower hit rates. As insertions taper off and query rates increase, the cache stabilizes, adapting effectively to the query-heavy workload. The graph shows this trend, where each clock cycle represents a workload configuration (e.g., 100P1Q, 90P5Q, 80P10Q). The y-axis shows the hit rate, illustrating the cache’s dynamic adaptation. Similarly, the hit and miss ratios per clock cycle demonstrate a steady rise in hit ratio and decline in miss ratio as queries dominate, effectively highlighting the cache’s stabilization during query-heavy phases as depicted in Figures~\ref{fig:bursty-hit-miss-soft} and~\ref{fig:bursty-hit-miss-ratio}, respectively. This trend in hit rate improvement directly translates to performance gains, where \oursystem with caching completes execution significantly faster than without caching, reducing total runtime by over 35\% under bursty configurations.

\subsection{Experiment: Impact of Policy Deletions on Cache Performance}  
In this experiment, we analyze the effects of policy deletions on cache performance using configurations like \textbf{10P5Q2D}, \textbf{10P5Q5D}, and \textbf{10P5Q10D}. Each configuration involves inserting policies, executing queries, and deleting policies in varying proportions.

As the number of deletions increases, we observe higher soft-hit rates caused by outdated cached entries, which necessitate more frequent regenerations of GEs. These regenerations, triggered by deletions, tend to result in more permissive GEs, which may reduce the precision of policy enforcement. Additionally, higher deletion rates introduce increased processing overhead due to the repeated need for GE regeneration. Figure~\ref{fig:deletion} shows the cache hit rates for varying deletion workloads, while Figure~\ref{fig:deletion-perf} highlights the corresponding performance impact.

\begin{table}[!ht]
\caption{Workload configuration and statistics for deletion experiments.}
\label{tab:deletion}
\centering
\scriptsize
\resizebox{\columnwidth}{!}{
\begin{tabular}{l l l l}
Workload & Total Policies Inserted & Total Queries Executed & Total Policies Deleted \\
\hline
10P5Q2D & 31,520 & 15,760 & 6,304 \\
\hline
10P5Q5D & 31,520 & 15,760 & 15,760 \\
\hline
10P5Q10D & 31,520 & 15,760 & 31,520 \\
\end{tabular}}
\end{table}

\vspace{1cm}
\begin{figure}[!ht]
\centering
\includegraphics[width=3in]{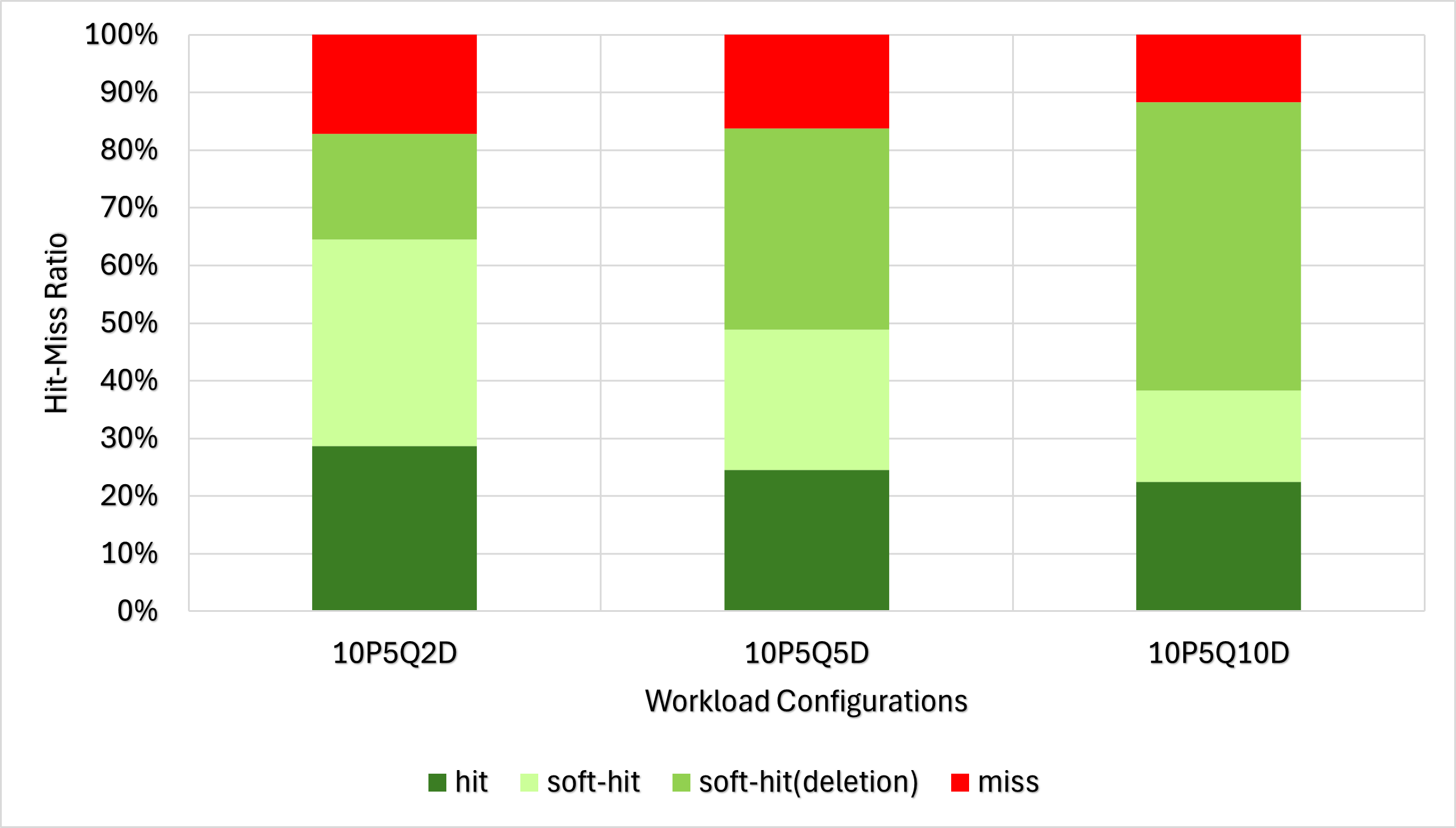}
\caption{Cache hit rate with varying policy deletion workloads.}
\label{fig:deletion}
\end{figure}

\begin{figure}[!ht]
\centering
\includegraphics[width=3in]{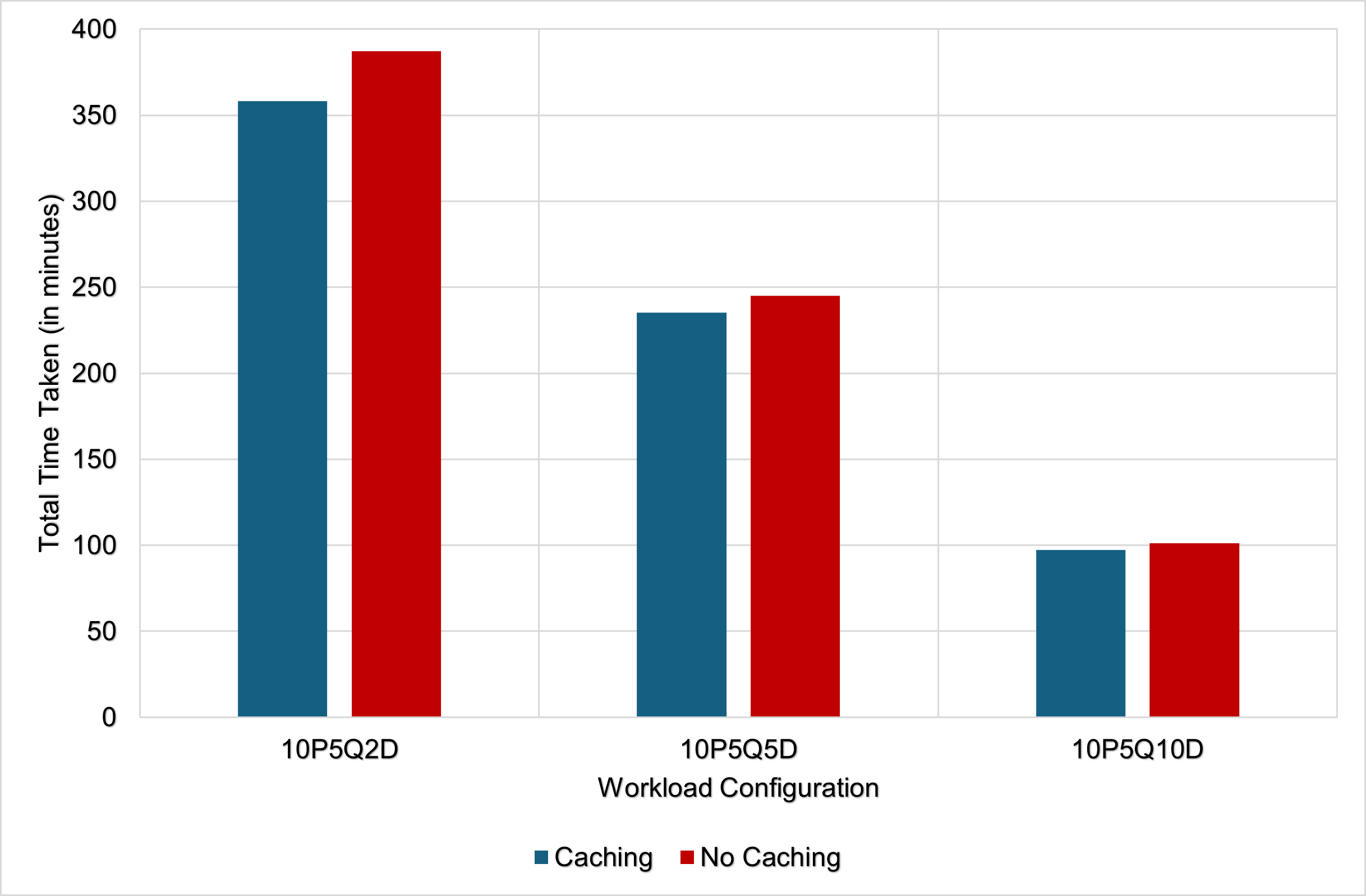}
\caption{Performance with varying policy deletion workloads.}
\label{fig:deletion-perf}
\end{figure}

\vspace{11pt}

\end{document}